\documentclass{article}

\usepackage{arxiv}

\usepackage[utf8]{inputenc} 
\usepackage[T1]{fontenc}    
\usepackage{hyperref}       
\usepackage{url}            
\usepackage{booktabs}       
\usepackage{amsfonts}       
\usepackage{amsmath}
\usepackage{stmaryrd}       

\usepackage{float}
\usepackage{siunitx}
\usepackage{nicefrac}       
\usepackage{microtype}      
\usepackage{lipsum}
\usepackage{graphicx}
\usepackage{subcaption} 

\graphicspath{ {./images/} }

\title{A Monolithic Computational Homogenization Framework for Nearly Incompressible Magnetoelastic Composites}

\author{
L. River Spencer \\
Department of Aerospace Engineering \& Engineering Mechanics,\\
The University of Texas at Austin,\\
Austin, TX 78712 \\
\And
Manuel K. Rausch\\
Department of Aerospace Engineering \& Engineering Mechanics,\\
The Oden Institute of Computational Science and Engineering,\\
Department of Biomedical Engineering,\\
Department of Mechanical Engineering,
The University of Texas at Austin,\\
Austin, TX 78712 \\
\And
Chad M. Landis\\
Department of Aerospace Engineering \& Engineering Mechanics,\\
The Oden Institute of Computational Science and Engineering,\\
The University of Texas at Austin,\\
Austin, TX 78712 \\
\And
Jan N. Fuhg \\
Department of Aerospace Engineering \& Engineering Mechanics,\\
The Oden Institute of Computational Science and Engineering,\\
The University of Texas at Austin,\\
Austin, TX 78712 \\
}

\begin{document}
\maketitle
\begin{abstract}
Magneto-active elastomers exhibit large, nonlinear deformations under combined mechanical loading and magnetic fields, and their effective behavior is strongly governed by microstructural heterogeneity. Predictive modeling of these materials is challenging because their response involves strong magneto-mechanical coupling, large deformations, and the nearly incompressible behavior of elastomeric matrices. Existing multiscale approaches often rely on staggered solution strategies or formulations that struggle to robustly treat near-incompressibility in strongly coupled settings.
This work presents a fully coupled computational homogenization framework for nearly incompressible magnetoelastic composites in which the mechanical deformation and magnetostatic fields are solved monolithically on a representative volume element (RVE). The microscale problem employs a mixed finite-element discretization with Lagrangian displacement degrees of freedom and a N\'ed\'elec-based magnetic vector potential, enabling a curl-conforming representation of the magnetic induction together with periodic boundary constraints for both mechanical and magnetic fields. Near-incompressibility is treated using a J-bar stabilization in which the volumetric response is controlled by the cell-averaged dilatation, while the isochoric response is evaluated using a scaled deformation gradient. The constitutive behavior is derived from an additive free-energy decomposition consisting of a hyperelastic contribution, a vacuum magnetic energy term, and a saturation-type magnetization potential that is active within the particle phase. 
The resulting formulation enables robust three-dimensional RVE simulations of heterogeneous magneto-elastic composites containing complex particle distributions under large deformations and strong magneto-mechanical coupling. Numerical examples demonstrate how particle interactions, microstructural arrangement, and inclusion compressibility influence deformation patterns and the effective magneto-mechanical response. In particular, the simulations highlight the role of particle compressibility in governing local deformation modes and the resulting homogenized response under magnetic loading. 
\end{abstract}


\section{Introduction}
\label{sec:introduction}

Magneto-active elastomers (MAEs) and related magneto-responsive soft composites combine a compliant non-magnetic polymer matrix with magnetic inclusions, enabling field-controlled actuation, tunable stiffness, and programmable shape change \cite{Farshad2004Magnetoactive,Kang2020MRE}. These materials are attractive for applications such as soft robotics, adaptive vibration isolation, reconfigurable metamaterials, and sensing/energy harvesting devices \cite{Farshad2004Magnetoactive,Aghamiri2025Magnetoactive,Becker2022Magnetoactive}.
The macroscopic response of MAEs is governed by strongly nonlinear, coupled magneto--mechanical physics and by the underlying microstructure (e.g., particle volume fraction, distribution, and anisotropy) \cite{Galipeau2014Magnetoactive}. Large strains, geometric nonlinearity, and near-incompressibility of elastomeric matrices challenge standard displacement-based finite elements, while magnetic saturation and material property contrast between phases can introduce additional nonlinearity and strong field localization. As a result, predictive modeling requires constitutive descriptions that are both thermodynamically consistent and capable of resolving microstructure-induced interactions between the mechanical and magnetic fields, including field-driven particle reorientation and rotation under finite deformation. \cite{Kalina2020MAE_Multiscale, Kalina2024NN_Multiscale_MagnetoElasticity}.

Computational homogenization offers a systematic route to bridge the microscale physics to effective macroscopic behavior by solving a representative volume element (RVE) boundary-value problem and extracting homogenized mechanical and magnetic quantities for use at the macroscale \cite{Geers2010MultiscaleCH}. In this setting, periodic boundary conditions are commonly adopted to reduce boundary artifacts and to represent an infinite periodic medium \cite{TaheriBehrooz2019RVE_PBC}, while consistent linearization of the coupled residual is essential for robust Newton-type solvers in multiscale contexts \cite{Tikarrouchine2018FE2CompositeStructures}.
A persistent practical difficulty in RVE simulations of elastomeric composites is enforcing (near-)incompressibility without sacrificing numerical stability. Mixed formulations with additional pressure-like fields can be effective \cite{ValverdeGonzalez2023LockingTreatment}, but they increase complexity and saddle-point structure, particularly when combined with additional electromagnetic unknowns \cite{schrautzer2025identification}. An alternative is J-bar type stabilization \cite{saliji2025mfe2}, where the volumetric response is driven by a cell-averaged dilatation while the isochoric response is evaluated using a scaled deformation gradient, mitigating volumetric locking while retaining a displacement-based representation \cite{saliji2025mfe2}.
On the magnetic side, vector-potential formulations provide a convenient framework for magnetostatics by enforcing $\nabla\cdot\mathbf{B}=0$ identically and enabling curl-conforming discretizations \cite{biro1996use}. In a finite element setting, N\'ed\'elec elements are a natural choice for representing the vector potential and its curl \cite{kanayama2002three}, and they integrate well with periodic constraints and gauge-fixing strategies needed to remove null modes \cite{chari2003three}.

Despite substantial progress, existing coupled magneto-elastic RVE formulations often (i) adopt staggered or weakly coupled solution strategies that can become fragile for strong coupling, large strain, or near-saturation regimes \cite{liu2020finite}, (ii) use magnetic scalar-potential approaches that complicate periodic enforcement and/or the representation of discontinuous material coefficients at phase boundaries \cite{lahart2004electromagnetic}, or (iii) rely on discretizations and incompressibility treatments that can suffer from locking or deteriorated Newton convergence in nearly incompressible settings \cite{coombs2018overcoming}. These limitations motivate a monolithic, energy-based formulation with a curl-conforming magnetic discretization, robust near-incompressibility stabilization, and a consistent coupled tangent tailored for use in computational homogenization.
In addition, many microstructure-resolved studies are restricted to idealized lattice geometries, two-dimensional settings, small numbers of inclusions, or moderate deformation regimes. Robust three-dimensional RVE simulations of random particle distributions at large strain and near-incompressible matrix response remain comparatively limited, particularly under strong magneto–mechanical coupling \cite{saber2023modeling}.

This work develops a fully coupled, energy-based computational homogenization framework for nearly incompressible magnetoelastic composites in which the mechanical displacement field and magnetic vector potential are solved monolithically on three-dimensional RVEs. The formulation enables large-deformation simulations of random particle distributions containing $O(10^{2})$ inclusions under periodic boundary conditions. Field-induced particle reorientation emerges naturally from the coupled solution without imposing prescribed particle motions, allowing the study of rotation-driven anisotropy evolution in strongly nonlinear regimes.
The framework employs (i) periodic boundary constraints for both mechanical and magnetic fields, (ii) a J-bar stabilization to robustly treat near-incompressibility at finite strain, and (iii) automatic differentiation of the free energy to obtain a consistent coupled tangent operator including magneto–mechanical coupling terms. The constitutive model is defined through an additive free-energy decomposition comprising a hyperelastic contribution, a vacuum magnetic energy term, and a saturation-type magnetization potential active in the particle phase.

The remainder of the paper is organized as follows. Section~\ref{sec:theoretical_background} introduces the governing equations, energetic setting, and homogenization framework. Section~\ref{sec:theoretical_background} also details the near-incompressibility treatment and magnetic vector-potential formulation. Section~\ref{sec:numerical_method} describes the finite element discretization, periodic constraints, and the consistent tangent construction. Section~\ref{sec:examples} presents representative numerical examples, and Section~\ref{sec:conclusion} concludes with a discussion of limitations and future directions.

\section{Theoretical Background}
\label{sec:theoretical_background}

\subsection{Kinematics}
\label{sec:kinematics}

Consider the motion of a body whose reference configuration at time $t_0 \in \mathbb{R}_{\geq0}$ is given by $B_0 \subset \mathbb{R}^d$ with the spatial dimension $d$. The current configuration of the body at time $t \in \mathcal{I} := \{\, t \in \mathbb{R} \mid t \ge t_0 \,\} $ is defined as $B \subset \mathbb{R}^d3$. To describe the motion of this body, a smooth bijective mapping $\boldsymbol{\varphi}: B_0 \times \mathcal{I} \to B$, which links material points $\mathbf{X} \in B_0 \;\mapsto\; \mathbf{x} = \varphi(\mathbf{X}, t) \in B$, is introduced. The displacement $\mathbf{u}$ of each material point is given by $\mathbf{u}(\mathbf{X}, t) := \boldsymbol{\varphi}(\mathbf{X}, t) - \mathbf{X}$. The deformation gradient and its determinate are defined as $\mathbf{F} := \left(\frac{\partial \boldsymbol{\varphi}}{\partial \mathbf{X}}\right)^{T}$ and $J := \det \mathbf{F}.$ We further define the right Cauchy--Green tensor $\mathbf{C} := \mathbf{F}^T\mathbf{F}$. For more details on kinematics, refer to Ref. \cite{dym1973solid}.  

\subsection{Volumetric--isochoric split}
\label{sec:vol_split}
To robustly enforce near-incompressibility, we employ an element-wise volumetric projection (the $J$-bar method) within a mixed formulation following \cite{saliji2025mfe2}. This choice is motivated by the periodic boundary conditions of the RVE problem and by the need to represent a nearly incompressible matrix response without introducing additional volumetric unknowns. In a full mixed incompressible formulation, such extra fields would also need to satisfy admissible microscale constraints, which considerably complicates the periodic homogenization setting. By contrast, the $J$-bar approach replaces the pointwise volumetric response with an averaged measure, which is consistent with the present homogenization framework and also provides a practical basis for studying cases in which different phases exhibit different degrees of compressibility. For each element domain $\Omega_e$ with volume $V_e$, we define the element-averaged Jacobian
\begin{equation}
\label{eq:cell_j}
  \overline{J}^{(e)} := \frac{1}{V_e}\int_{\Omega_e} J \,\mathrm{d}V \, .
  \qquad
  V_e := \int_{\Omega_e} 1 \,\mathrm{d}V,
\end{equation}
Using $\overline{J}^{(e)}$, we introduce a volumetric--isochoric split by defining the scaling factor
\begin{equation}
\label{eq:vol_split}
  s := \left(\frac{\overline{J}^{(e)}}{J}\right)^{1/d},
  \qquad
  \overline{\mathbf{F}} := s\,\mathbf{F},
\end{equation}
where $d$ denotes the number of spatial dimensions. By construction, the modified deformation gradient satisfies
$\det\overline{\mathbf{F}}=\overline{J}^{(e)}$, so the volumetric deformation is assumed to be constant over each element. We then define the modified right Cauchy--Green tensor and its first invariant as
\begin{equation}
\label{eq:iso_C}
  \overline{\mathbf{C}} := \overline{\mathbf{F}}^{T}\overline{\mathbf{F}},
  \qquad
  \overline{I}_1 := \operatorname{tr}(\overline{\mathbf{C}}).
\end{equation}

\subsection{Magnetostatic formulation}
\label{sec:magnetics}
We consider quasi-stationary magnetostatics in the absence of free current densities. In the current configuration, Maxwell's equations reduce to \cite{huray2009maxwell}
\begin{equation}
\label{eq:maxwell}
  \nabla \cdot \mathbf{b} = 0,
  \qquad
  \nabla \times \mathbf{h} = \mathbf{0},
\end{equation}
supplemented by the standard interface conditions $\llbracket \mathbf{b}\rrbracket\cdot\mathbf{n}=0$ and $\llbracket \mathbf{h}\rrbracket\times\mathbf{n}=\mathbf{0}$.

For constitutive modeling and coupling to deformation, we employ a Lagrangian description in which the spatial fields are related by $\mathbf{b}=\mu_0(\mathbf{h}+\mathbf{m})$ and pulled back to the reference configuration via
\begin{equation}
\label{eq:lag_maxwell_quantities}
  \mathbf{B} := (\text{cof}\, \mathbf{F})\,\mathbf{b} = J\mathbf{F}^{-1}\mathbf{b},
  \qquad
  \mathbf{H} := \mathbf{F}^{T}\mathbf{h},
\end{equation}
so that the magnetostatic equations become $\nabla_{\!X}\cdot\mathbf{B}=0$ and $\nabla_{\!X}\times\mathbf{H}=\mathbf{0}$ in the reference configuration \cite{dorfmann2005some}.

To enforce $\nabla_{\!X}\cdot\mathbf{B}=0$ identically, we introduce a (fluctuation) magnetic vector potential $\mathbf{A}$ and decompose the induction as
\begin{equation}
\label{eq:mag_from_applied}
  \mathbf{B} = \mathbf{B}_M + \nabla_{\!X}\times\mathbf{A},
\end{equation}
where $\mathbf{B}_M$ denotes a prescribed macroscopic induction. The uniqueness of the discrete magnetic vector potential $\mathbf{A}$ is ensured by removing the remaining gauge freedom, for example, by fixing a single degree of freedom \cite{li2015vectorial} and weakly enforcing the Coulomb gauge condition $\nabla \cdot \mathbf{A} = 0$. In the present formulation, this weak gauge enforcement is provided through the boundary conditions described in Section~\ref{sec:pbc_constraints}.

\subsection{Free energy density}
\label{sec:SEF}

We assume an additive decomposition of the Helmholtz free energy density into mechanical and magnetic parts,
\begin{equation}
\label{eq:SEF}
\psi(\mathbf{F},\mathbf{B})
=
\psi_{\mathrm{mech}}(\overline{\mathbf{F}},\overline{J}^{(e)})
+
\psi_{\mathrm{vac}}(\mathbf{F},\mathbf{B})
+
\psi_{\mathrm{mag}}(\mathbf{F},\mathbf{B}),
\end{equation}
where $\overline{\mathbf{F}}$ and $\overline{J}^{(e)}$ are the element-wise $J$-bar projected measures defined in Section~\ref{sec:vol_split}.

The $J$-bar projection is applied only to the mechanical volumetric response to alleviate locking; magnetic terms are evaluated using the physical deformation gradient $\mathbf F$ to preserve the exact referential–spatial transformation of the induction.

\subsubsection{Mechanical contribution}
\label{sec:mech_contribution}

The mechanical response is modeled using a Yeoh-type potential written in terms of the averaged dilatation $\overline{J}^{(e)}$ and the first invariant $\overline{I}_1$ of the modified right Cauchy--Green tensor,
\begin{equation}
\label{eq:SEF_mech}
\psi_{\mathrm{mech}}
=
\frac{K}{2}\big(\overline{J}^{(e)}-1\big)^{2}
+ C_{1}\big(\overline{I}_{1}-3\big)
+ C_{2}\big(\overline{I}_{1}-3\big)^{2}
+ C_{3}\big(\overline{I}_{1}-3\big)^{3},
\end{equation}
where $K$ is the bulk modulus and $C_{1}$, $C_{2}$, and $C_{3}$ are Yeoh material parameters \cite{holzapfel2002nonlinear}. 

\subsubsection{Vacuum-field contribution}
\label{sec:vacuum_contribution}

The vacuum-field contribution accounts for the magnetic energy stored in free space and is expressed in referential form as
\begin{equation}
\label{eq:SEF_vac}
\psi_{\mathrm{vac}}
=
\frac{1}{2\mu_0} J\, \mathbf{b}\cdot \mathbf{b}
=
\frac{1}{2\mu_0 J} (\mathbf{F}\mathbf{B})\cdot(\mathbf{F}\mathbf{B}),
\end{equation}
where $\mu_{0}$ is the permeability of free space, and the spatial magnetic induction $\mathbf{b}=J^{-1}\mathbf{F}\mathbf{B}$.

\subsubsection{Particle magnetization contribution }
\label{sec:mag_contribution}

To model the additional energetic contribution associated with particle magnetization, we adopt a Legendre-based magnetic free energy depending on the spatial induction magnitude.
Assuming the magnetic field is aligned with $\mathbf{b}$, we write $\mathbf{h}=h\,\mathbf{n}$ with $\mathbf{n}:=\mathbf{b}/b$ where  $ b= \|\mathbf{b}\|$ , and determine the field magnitude $h$ implicitly from
\begin{equation}
\label{eq:h_from_b}
\mu_0\Big(h + m_s^{\mathrm{leg}}\,\mathcal{L}(\alpha^{\mathrm{leg}} h)\Big) = b,
\qquad
\mathcal{L}(x) := \coth(x) - \frac{1}{x},
\end{equation}
where $\mu_0$ is the permeability of free space, $m_s^{\mathrm{leg}}$ is the (Legendre-model) saturation magnetization, and $\alpha^{\mathrm{leg}}$ is the Langevin parameter \cite{metsch2019two}. The magnetization energy density per \emph{current} volume is then defined as
\begin{equation}
\label{eq:w_mag_legendre}
w_{\mathrm{mag}}(b)
=
b\,h - \frac{\mu_0}{2}h^2 - W^\star(h),
\qquad
W^\star(h)=\frac{\mu_0\,m_s^{\mathrm{leg}}}{\alpha^{\mathrm{leg}}}\ln\!\left(\frac{\sinh(\alpha^{\mathrm{leg}} h)}{\alpha^{\mathrm{leg}} h}\right).
\end{equation}
Finally, the corresponding \emph{referential} contribution is
\begin{equation}
\label{eq:psi_mag_legendre}
\psi_{\mathrm{mag}}(\mathbf{F},\mathbf{B})
=
\eta\,J\,w_{\mathrm{mag}}(b),
\qquad
\eta=
\begin{cases}
0, & \text{matrix},\\
1, & \text{particle}.
\end{cases}
\end{equation}
The parameters $(m_s^{\mathrm{leg}},\alpha^{\mathrm{leg}})$ are prescribed material constants for the particle magnetization model.

\subsection{Derived quantities}
\label{sec:derived_quantities}

The constitutive responses follow from the free energy density $\psi(\mathbf{F},\mathbf{B})$ by energetic conjugacy. In the reference configuration, the first Piola--Kirchhoff stress tensor and the Lagrangian magnetic field are defined as
\begin{equation}
\label{eq:derived_quantities}
\mathbf{P} = \frac{\partial \psi}{\partial \mathbf{F}},
\qquad
\mathbf{H} = \frac{\partial \psi}{\partial \mathbf{B}}.
\end{equation}
In the present setting, $\psi$ depends on $\mathbf{F}$ both explicitly and implicitly through the volumetric--isochoric split, since $\overline{\mathbf{F}}=\overline{\mathbf{F}}(\mathbf{F},\overline{J}^{(e)})$ with $\overline{J}^{(e)}$ treated as an element-wise constant. Consequently, the consistent linearization required for Newton iterations involves the mixed tangent blocks
\begin{equation}
\label{eq:consistent_tangents}
\mathbb{A} := \frac{\partial \mathbf{P}}{\partial \mathbf{F}}, \qquad
\mathbb{C} := \frac{\partial \mathbf{P}}{\partial \mathbf{B}}, \qquad
\mathbb{D} := \frac{\partial \mathbf{H}}{\partial \mathbf{F}}, \qquad
\mathbb{E} := \frac{\partial \mathbf{H}}{\partial \mathbf{B}}.
\end{equation}
All first- and second-order derivatives are evaluated algorithmically via automatic differentiation \cite{rall1981automatic} of $\psi$ with respect to the independent components of $(\mathbf{F},\mathbf{B})$, yielding a monolithic magneto-mechanical tangent consistent with the chosen energy density \cite{Kalina2024NN_Multiscale_MagnetoElasticity}. 

\subsection{Homogenization}
\label{sec:homogenization}

Let $\Omega \subset \mathbb{R}^d$ denote the macroscopic body and let $\Omega_m \subset \mathbb{R}^d$ be a representative volume element (RVE) associated with a macroscopic material point $\mathbf{X}\in\Omega$. At the macroscale, the prescribed loading measures are the macroscopic deformation gradient $\overline{\mathbf{F}}(\mathbf{X})$ and the macroscopic Lagrangian magnetic induction $\overline{\mathbf{B}}(\mathbf{X})$. The objective of computational homogenization is to define an effective response such that the macroscopic constitutive quantities are consistent with the underlying heterogeneous microscale fields on $\Omega_m$.

\paragraph{Kinematic and magnetic scale transition.}
The microscale motion is decomposed into an affine part and a periodic fluctuation,
\begin{equation}
\label{eq:rve_motion}
\mathbf{x}_m(\mathbf{X}_m)=\overline{\mathbf{F}}\,\mathbf{X}_m+\widetilde{\mathbf{u}}_m(\mathbf{X}_m),
\qquad
\widetilde{\mathbf{u}}_m \ \text{periodic on } \partial\Omega_m,
\end{equation}
together with an anchoring condition to eliminate rigid body modes. The corresponding deformation gradient reads
\begin{equation}
\label{eq:rve_F}
\mathbf{F}_m=\nabla_m \mathbf{x}_m=\overline{\mathbf{F}}+\nabla_m \widetilde{\mathbf{u}}_m.
\end{equation}
The macroscopic deformation gradient is identified with the RVE volume average,
\begin{equation}
\label{eq:avg_F}
\overline{\mathbf{F}}=\langle \mathbf{F}_m\rangle
:=\frac{1}{|\Omega_m|}\int_{\Omega_m}\mathbf{F}_m\,\mathrm{d}V .
\end{equation}

For the magnetic problem, we introduce a vector potential $\mathbf{A}_m$ such that the microscale Lagrangian magnetic induction is divergence-free by construction, i.e.,
\begin{equation}
\label{eq:Bm_curlA}
\mathbf{B}_m'=\nabla_m\times\mathbf{A}_m .
\end{equation}
Accordingly, we decompose the total induction into a prescribed macroscopic part and a fluctuation field,
\begin{equation}
\label{eq:rve_B}
\mathbf{B}_m = \overline{\mathbf{B}} + \mathbf{B}_m'
= \overline{\mathbf{B}} + \nabla_m \times \mathbf{A}_m,
\end{equation}
together with periodic boundary conditions on $\mathbf{A}_m$ (up to gauge) and a gauge-fixing constraint to ensure uniqueness. The macroscopic induction is defined as the RVE volume average \cite{castaneda2011homogenization},
\begin{equation}
\label{eq:avg_B}
\overline{\mathbf{B}}=\langle \mathbf{B}_m\rangle
:=\frac{1}{|\Omega_m|}\int_{\Omega_m}\mathbf{B}_m\,\mathrm{d}V .
\end{equation}

\paragraph{Energetic consistency (Hill--Mandel condition).}
The multiscale coupling is enforced through the Hill--Mandel macro-homogeneity condition, which requires equality of macroscopic and volume-averaged microscopic power expenditures. In the present setting with work-conjugate pairs $(\mathbf{P},\mathbf{F})$ and $(\mathbf{H},\mathbf{B})$, the condition is stated as \cite{hill1963elastic}
\begin{equation}
\label{eq:hill_mandel}
\overline{\mathbf{P}} : \dot{\overline{\mathbf{F}}}
+ \overline{\mathbf{H}} \cdot \dot{\overline{\mathbf{B}}}
=
\left\langle
\mathbf{P}_m : \dot{\mathbf{F}}_m
+ \mathbf{H}_m \cdot \dot{\mathbf{B}}_m
\right\rangle ,
\end{equation}
where $\langle \bullet \rangle := |\Omega_m|^{-1}\int_{\Omega_m}(\bullet)\,\mathrm{d}V$ denotes volume averaging. Under the assumed periodicity of the fluctuation fields and the corresponding anti-periodicity of microscopic tractions and magnetic flux terms on $\partial\Omega_m$, Eq. \eqref{eq:hill_mandel} holds and implies the standard averaging relations for the macroscopic constitutive quantities,
\begin{equation}
\label{eq:avg_PH}
\overline{\mathbf{P}}=\langle \mathbf{P}_m\rangle
=\frac{1}{|\Omega_m|}\int_{\Omega_m}\mathbf{P}_m\,\mathrm{d}V,
\qquad
\overline{\mathbf{H}}=\langle \mathbf{H}_m\rangle
=\frac{1}{|\Omega_m|}\int_{\Omega_m}\mathbf{H}_m\,\mathrm{d}V .
\end{equation}

\paragraph{Effective energy density and macroscopic response.}
With the microscale free energy density $\psi_m(\mathbf{F}_m,\mathbf{B}_m)$ and the derived quantities $\mathbf{P}_m=\partial\psi_m/\partial\mathbf{F}_m$ and $\mathbf{H}_m=\partial\psi_m/\partial\mathbf{B}_m$, the effective stored energy density is defined by volume averaging,
\begin{equation}
\label{eq:avg_energy}
\overline{\psi}(\overline{\mathbf{F}},\overline{\mathbf{B}})
:= \left\langle \psi_m(\mathbf{F}_m,\mathbf{B}_m)\right\rangle
=\frac{1}{|\Omega_m|}\int_{\Omega_m}\psi_m(\mathbf{F}_m,\mathbf{B}_m)\,\mathrm{d}V .
\end{equation}
The macroscopic work-conjugate quantities then follow as \cite{coleman1974thermodynamics}
\begin{equation}
\label{eq:macro_work_conjugates}
\overline{\mathbf{P}}=\frac{\partial \overline{\psi}}{\partial \overline{\mathbf{F}}},
\qquad
\overline{\mathbf{H}}=\frac{\partial \overline{\psi}}{\partial \overline{\mathbf{B}}},
\end{equation}
and the consistent macroscopic tangents are obtained by linearization of the RVE equilibrium with respect to $(\overline{\mathbf{F}},\overline{\mathbf{B}})$.

\section{Numerical Methods}
\label{sec:numerical_method}

This section summarizes the numerical formulation used to solve the fully coupled magneto--mechanical RVE problem
and to extract homogenized quantities. The numerical scheme was implemented in \textit{deal.II} \cite{arndt2025deal}. The microscale boundary-value problem is posed in the reference
configuration $\Omega_0^m$ and is discretized with a conforming finite element method. The primary unknowns are the
microscale displacement field $\mathbf{u}_m$ and the magnetic vector potential $\mathbf{A}_m$, from which the
deformation gradient $\mathbf{F}_m$ and magnetic induction $\mathbf{B}_m$ are obtained as \cite{zabihyan2018aspects}
\begin{equation}
  \mathbf{F}_m = \mathbf{I} + \nabla_{X_m}\mathbf{u}_m,
  \qquad
  \mathbf{B}_m = \mathbf{B}_M + \nabla_{X_m}\times \mathbf{A}_m  \,.
\end{equation}
Here, $\mathbf{B}_M$ denotes the prescribed macroscopic (uniform) magnetic induction applied to the RVE. The
constitutive response follows from a Helmholtz-type free energy density $\psi_m(\mathbf{F}_m,\mathbf{B}_m)$, yielding the work-conjugate relations \cite{coleman1974thermodynamics}
\begin{equation}
  \mathbf{P}_m = \frac{\partial \psi_m}{\partial \mathbf{F}_m},
  \qquad
  \mathbf{H}_m = \frac{\partial \psi_m}{\partial \mathbf{B}_m}.
\end{equation}
Accordingly, the residual statements correspond to mechanical equilibrium and magnetostatic balance (in the
absence of free currents) and are solved using a monolithic Newton scheme.

\subsection{Finite element discretization}
\label{sec:fe_discretization}

To balance accuracy and meshing flexibility, we consider two complementary discretization pipelines. Hexahedral meshes enable a
standard $H^1$--$H(\mathrm{curl})$ pairing with tensor-product Lagrange elements for mechanics and N\'ed\'elec elements for the magnetic
vector potential, which is particularly convenient for enforcing curl conformity. However, resolving complex particle geometries and
performing local refinement is often more straightforward with unstructured tetrahedral meshes. For this reason, we also
implement a simplex-based discretization on tetrahedra, allowing robust local refinement near inclusions while maintaining the same
monolithic coupling strategy and averaging procedures used in the hexahedral setting.

\subsubsection{Hex Mesh}
\label{sec:fe_discretization_hex}

The weak form is discretized using a mixed finite element space consisting of $H^1$-conforming Lagrange shape functions for the displacement field and $H(\mathrm{curl})$-conforming N\'ed\'elec elements for the vector potential \cite{reynolds2024h1, schoberl2005high}. Let $\mathcal{T}_h$ denote a conforming partition of $\Omega_0^m$ into hexahedral cells. The discrete trial and test spaces are chosen as
\begin{equation}
  \mathbf{u}_m^h \in [\mathcal{Q}_{p}]^d, \qquad
  \mathbf{A}_m^h \in \mathcal{N}_{p-1},
\end{equation}
where $\mathcal{Q}_{p}$ denotes tensor-product Lagrange polynomials of degree $p$ (with $d\in\{2,3\}$ the spatial dimension) and $\mathcal{N}_{p-1}$ denotes the first-family N\'ed\'elec space of degree $p-1$. This choice ensures conformity of the displacement gradient and a curl-conforming representation for $\mathbf{A}_m$, such that $\nabla_{X_m}\times \mathbf{A}_m^h$ is well-defined element-wise and $\nabla_{X_m}\cdot \mathbf{B}_m = 0$ is weakly enforced.

The coupled discrete unknown vector collects both fields in a monolithic form,
\begin{equation}
  \mathbf{x}
  =
  \begin{bmatrix}
    \mathbf{u}_m^h \\
    \mathbf{A}_m^h
  \end{bmatrix},
\end{equation}
leading to a block-structured nonlinear system
\begin{equation}
  \mathbf{R}(\mathbf{x})=
  \begin{bmatrix}
    \mathbf{R}_u(\mathbf{u}_m^h,\mathbf{A}_m^h)\\
    \mathbf{R}_A(\mathbf{u}_m^h,\mathbf{A}_m^h)
  \end{bmatrix}
  =\mathbf{0},
\end{equation}
with a consistent tangent operator $\mathbf{K}=\partial \mathbf{R}/\partial \mathbf{x}$ assembled from the algorithmic moduli $\partial \mathbf{P}_m/\partial \mathbf{F}_m$, $\partial \mathbf{P}_m/\partial \mathbf{B}_m$, $\partial \mathbf{H}_m/\partial \mathbf{F}_m$, and $\partial \mathbf{H}_m/\partial \mathbf{B}_m$.

Numerical integration is performed on each cell using Gauss quadrature of order $q$, and all macroscopic averages reported in this work are computed as volume averages over $\Omega_0^m$ using the same quadrature rule \cite{brass2011quadrature}.

\subsubsection{Tet Mesh}
\label{sec:fe_discretization_tet}

To enable unstructured meshing and local refinement around particle geometries, we also discretize the microscale domain $\Omega_0^m$ using a tetrahedral partition. Let $\mathcal{T}_h^\Delta$ denote a conforming triangulation of $\Omega_0^m$ into simplex (tetrahedral) cells. On $\mathcal{T}_h^\Delta$ we employ simplex polynomial finite elements for both the mechanical and magnetic unknowns \cite{delingette1999general}. The discrete trial and test spaces are chosen as
\begin{equation}
  \mathbf{u}_m^h \in [\mathcal{P}_{p}]^d, \qquad
  \mathbf{A}_m^h \in [\mathcal{P}_{p_A}]^d,
\end{equation}
where $\mathcal{P}_{p}$ denotes the space of complete polynomials of degree $p$ on tetrahedra (implemented with $H^1$-conforming simplex Lagrange elements). In all computations reported here, we take $p_A=p$ unless stated otherwise.

Compared to the hexahedral discretization, the key difference is the approximation of the magnetic vector potential. While N\'ed\'elec elements are a natural choice for $H(\mathrm{curl})$-conforming discretizations on general meshes, we adopt an $H^1$-conforming approximation for $\mathbf{A}_m$ on simplices to maintain a uniform simplex pipeline and to simplify the imposition of periodic constraints (which are enforced at nodal support points). Since $H^1(\Omega)\subset H(\mathrm{curl};\Omega)$, the quantity $\nabla_{X_m}\times\mathbf{A}_m^h$ remains well-defined and can be evaluated elementwise for constructing the magnetic induction $\mathbf{B}_m^h$.

As in the hexahedral case, the coupled discrete unknown vector is collected in monolithic form,
\begin{equation}
  \mathbf{x}
  =
  \begin{bmatrix}
    \mathbf{u}_m^h \\
    \mathbf{A}_m^h
  \end{bmatrix},
\end{equation}
yielding the same block-structured nonlinear residual $\mathbf{R}(\mathbf{x})=\mathbf{0}$ and consistent tangent $\mathbf{K}=\partial\mathbf{R}/\partial\mathbf{x}$. The algorithmic moduli $\partial \mathbf{P}_m/\partial \mathbf{F}_m$, $\partial \mathbf{P}_m/\partial \mathbf{B}_m$, $\partial \mathbf{H}_m/\partial \mathbf{F}_m$, and $\partial \mathbf{H}_m/\partial \mathbf{B}_m$ are assembled identically, with the only modification being the shape-function gradients/curls associated with the simplex basis.

Numerical integration on $\mathcal{T}_h^\Delta$ is performed using simplex Gauss quadrature of order $q$, and all macroscopic averages are computed as volume averages over $\Omega_0^m$ using the same quadrature rule \cite{brass2011quadrature}. Periodic boundary conditions are applied component-wise for both $\mathbf{u}_m^h$ and $\mathbf{A}_m^h$ by enforcing equality of nodal fluctuation values on paired periodic faces, resulting in a consistent periodicity of the discrete fields on the tetrahedral mesh.

\subsection{Periodic constraints and removal of null modes}
\label{sec:pbc_constraints}

To represent an infinite periodic medium and to suppress boundary artifacts, periodic boundary constraints are
enforced directly at the discrete degree-of-freedom level. Let $\partial\Omega_0^m$ be decomposed into three pairs of opposing faces $\Gamma_i^{-}$ and $\Gamma_i^{+}$ ($i\in\{1,2,3\}$) with outward normals $\mathbf{N}_i^{-}=-\mathbf{N}_i^{+}$. A one-to-one mapping $\mathcal{M}_i:\Gamma_i^{-}\rightarrow \Gamma_i^{+}$ is introduced such that $\mathbf{X}_m^{+}=\mathcal{M}_i(\mathbf{X}_m^{-})$ and $\mathbf{X}_m^{+}-\mathbf{X}_m^{-}=\mathbf{L}_i$, where $\mathbf{L}_i$ denotes the RVE period vector in direction $i$ \cite{zabihyan2018aspects}.

\paragraph{Mechanical periodicity.}
The microscopic displacement is decomposed into an affine part consistent with the prescribed macroscopic deformation gradient and a periodic fluctuation. In discrete form, the periodicity condition can be imposed as
\begin{equation}
\label{eq:pbc_u}
\mathbf{u}_m^h(\mathbf{X}_m^{+})-\mathbf{u}_m^h(\mathbf{X}_m^{-})
=
\big(\overline{\mathbf{F}}-\mathbf{I}\big)\big(\mathbf{X}_m^{+}-\mathbf{X}_m^{-}\big)
=
\big(\overline{\mathbf{F}}-\mathbf{I}\big)\mathbf{L}_i,
\qquad
\mathbf{X}_m^{-}\in\Gamma_i^{-},
\end{equation}
which enforces periodic fluctuations while admitting the macroscopic strain through the known jump on opposite faces. In practice, \eqref{eq:pbc_u} is implemented by selecting one side of each face pair as \emph{leader} and the opposing side as \emph{follower}, and constraining each follower displacement DOF as an affine function of the corresponding leader DOF plus the prescribed offset implied by $\overline{\mathbf{F}}$ \cite{Kalina2024NN_Multiscale_MagnetoElasticity}.

\paragraph{J-bar in periodic boundary conditions.}
The $J$-bar formulation is adopted in this work instead of a full mixed incompressible formulation, such as a $u$--$p$--$J$ approach, because it is much more straightforward to combine with periodic boundary conditions at the RVE level. In a mixed formulation, additional volumetric fields such as pressure and dilatation must also satisfy admissible microscale constraints, which complicates their treatment under periodic boundary conditions and the associated scale transition. By contrast, the $J$-bar approach replaces the pointwise volumetric response with an averaged volumetric measure, allowing the RVE problem to be formulated in terms of the displacement field alone. This makes the enforcement of periodicity substantially simpler while still providing a practical treatment of near-incompressibility. The tradeoff is that the volumetric response is represented in an averaged sense rather than enforced pointwise, which may smear local volume changes but greatly simplifies the periodic homogenization setting. In addition, retaining a displacement-based mechanical formulation is advantageous for the present coupled magneto-mechanical setting, since it avoids introducing extra volumetric unknowns alongside the magnetic field variables. This choice is also consistent with reduced mixed formulations in the literature, where replacing the pointwise dilatation by an averaged measure is used specifically to alleviate volumetric locking while retaining a formulation based on the modified deformation gradient alone, thereby avoiding static condensation and improving computational efficiency \cite{saliji2025mfe2}. This choice is also advantageous for the present class of heterogeneous RVEs, where the matrix is treated as nearly incompressible while the particle phase may be assigned a different, and potentially compressible, volumetric response. In such settings, the $J$-bar treatment provides a practical way to stabilize the nearly incompressible matrix without requiring a fully mixed incompressible formulation for all phases, thereby making it easier to study how inclusion compressibility influences the local and homogenized magneto-mechanical response.

\paragraph{Magnetic periodicity for the vector potential.}
The magnetic induction is represented as $\mathbf{B}_m=\overline{\mathbf{B}}+\nabla_{X_m}\times\mathbf{A}_m^h$, where $\overline{\mathbf{B}}$ is prescribed and uniform. Therefore, only the fluctuation field generated by $\nabla\times\mathbf{A}_m^h$ must be periodic \cite{Kalina2024NN_Multiscale_MagnetoElasticity}. This is achieved by enforcing periodicity of the vector potential in the $H(\mathrm{curl})$-conforming space,
\begin{equation}
\label{eq:pbc_A}
\mathbf{A}_m^h(\mathbf{X}_m^{+}) = \mathbf{A}_m^h(\mathbf{X}_m^{-}),
\qquad
\mathbf{X}_m^{-}\in\Gamma_i^{-},
\end{equation}
so that the tangential trace is periodic and the resulting curl field is compatible across opposite boundaries.

Because N\'ed\'elec basis functions are associated with oriented edges, periodic identification must respect the edge orientation induced by the face mapping $\mathcal{M}_i$. Specifically, if an edge on $\Gamma_i^{-}$ is mapped to an edge on $\Gamma_i^{+}$ with the same geometric support but opposite orientation, the associated DOFs must be related by a sign change. Denoting by $\alpha_e\in\{+1,-1\}$ the orientation factor for a periodic edge pairing, the discrete constraint for each paired edge DOF takes the form
\begin{equation}
\label{eq:pbc_A_sign}
a_e^{+} = \alpha_e\, a_e^{-},
\end{equation}
where $a_e^{\pm}$ are the scalar N\'ed\'elec DOFs on the paired edges. This preserves the conformity of the
$H(\mathrm{curl})$ space under periodic identification \cite{kikuchi2001theoretical}.

\paragraph{Elimination of rigid body and gauge modes.}
The periodic mechanical constraints admit rigid translations of the RVE. To ensure uniqueness of the displacement field, a minimal anchoring constraint is applied, e.g.,
\begin{equation}
\label{eq:anchor_u}
\mathbf{u}_m^h(\mathbf{X}_\mathrm{ref})=\mathbf{0},
\end{equation}
at a chosen reference point $\mathbf{X}_\mathrm{ref}$ (or, equivalently, by fixing a minimal set of displacement DOFs). This removes the null space associated with rigid body motion without affecting the periodic fluctuations \cite{okereke2018boundary}. 

Similarly, the magnetic vector potential is defined up to a gradient field (gauge freedom), which manifests as a near-null mode in the discrete system. Uniqueness is enforced by fixing one component of $\mathbf{A}_m^h$ at a single
node/edge location,
\begin{equation}
\label{eq:gauge_fix}
\mathbf{e}_k\cdot \mathbf{A}_m^h(\mathbf{X}_\mathrm{g}) = 0,
\end{equation}
for a selected component $k\in\{1,2,3\}$ and point $\mathbf{X}_\mathrm{g}\in\Omega_0^m$. This constraint removes the
gauge null mode while leaving the physically relevant quantity $\nabla\times\mathbf{A}_m^h$  \cite{stark2015boundary}.

\subsection{Stress Relaxed Boundary Conditions}
\label{sec:stress_relaxed_bc}

A limitation of the boundary conditions described above is that they may artificially restrict rigid translations of magnetically interacting particles in addition to the prescribed macroscopic deformation. In particular, when the macroscopic deformation is imposed strictly, magnetically induced tractions can be “locked” into the RVE response, suppressing the particle rearrangements that would otherwise occur under the same magnetic loading. To mitigate this effect, we introduce a stress-relaxed boundary condition that adjusts the macroscopic deformation to compensate for the magnetically induced stresses, thereby permitting particle translations while preserving the applied magnetic excitation.

In classical computational homogenization, stress relaxation is often realized by selecting a macroscopic deformation gradient such that one or more components of the homogenized stress vanish (e.g., traction-free or stress-free conditions along selected directions) \cite{miehe1999computational}. In the present setting, however, nonlocal magnetic interactions and strong magneto-mechanical coupling preclude a reliable closed-form determination of such a macroscopic deformation. We therefore employ an iterative stress-minimization procedure in which the macroscopic deformation gradient is treated as an optimization variable and updated until the targeted homogenized stress components satisfy a prescribed tolerance \cite{javili2017aspects}. The prescribed stress tolerance was set to $10^{-3}$. Further reduction of this tolerance led to negligible changes in the resulting macroscopic deformation gradient and the observed deformation patterns, while significantly increasing the computational cost.

\section{Numerical Examples}
\label{sec:examples}

This section presents a set of numerical examples designed to verify the proposed formulation and to illustrate key features of the fully coupled magneto--mechanical response in the nearly incompressible regime. We begin with a single-particle inclusion problem,
which serves as a controlled benchmark for assessing solution quality and consistency between the hexahedral (N\'ed\'elec-based) and tetrahedral (simplex-based) discretizations. We then consider a short chain of particles to highlight particle--particle interactions and the resulting localization of deformation and magnetic fields. Finally, we study RVEs containing a large number of randomly distributed particles to demonstrate robustness and scalability for realistic microstructures.

\subsection{Material properties}
\label{sec:material_properties}

Both the matrix and particle phases are modeled as nearly incompressible. Near-incompressibility is enforced through the $J$-bar projection in the mechanical energy, with the bulk modulus $K$ acting as a volumetric penalty parameter. The remaining mechanical response is governed by the Yeoh coefficients $(C_1,C_2,C_3)$. Magnetic behavior is described using the vacuum permeability $\mu_0$ and the Legendre-model parameters $(m_s^{\mathrm{leg}},\alpha^{\mathrm{leg}})$ used in the particle magnetization contribution. The parameters employed in all simulations are summarized in Table~\ref{tab:sef_parameters}.

\begin{table}[ht]
\centering
\caption{Strain-energy function (SEF) parameters used in the simulations.}
\label{tab:sef_parameters}
\begin{tabular}{lcc}
\toprule
Parameter (units) & Matrix & Particle \\
\midrule
Bulk modulus $K$ (Pa)   & \num{1.25e6}   & \num{2.5e8} \\
Yeoh $C_1$ (Pa)         & \num{1.2595e4} & \num{1.0e7} \\
Yeoh $C_2$ (Pa)         & \num{7.0244e1}      & \num{0.0} \\
Yeoh $C_3$ (Pa)         & \num{9.8177}      & \num{0.0} \\
\midrule
Vacuum permeability $\mu_0$ (H\,m$^{-1}$) & \multicolumn{2}{c}{\num{1.2564e-6}} \\
Legendre $m_s^{\mathrm{leg}}$ (A\,m$^{-1}$) & \multicolumn{2}{c}{\num{8.41e5}} \\
Legendre $\alpha^{\mathrm{leg}}$ (m\,A$^{-1}$) & \multicolumn{2}{c}{\num{2.18e-5}} \\
\bottomrule
\end{tabular}
\end{table}

For reference, the permeability of free space is $\mu_0 = 4\pi\times 10^{-7}\,\mathrm{H\,m^{-1}}$ (equivalently $\mathrm{N\,A^{-2}}$), which is consistent with the numerical value reported in Table~\ref{tab:sef_parameters}.
The material parameters used in this work are chosen to reflect experimentally calibrated constituent behavior following the multiscale strategy of Kalina \textit{et al.} \cite{kalina2023multiscale}. In their work, the microscopic constitutive response of each phase is identified separately from experimental data and then used as input for computational homogenization. In particular, the magnetic response of magnetically soft carbonyl-iron particles is represented by a saturation-type law based on the Langevin function, where the saturation magnetization $M_s$ and the scaling factor $\alpha$ are obtained by fitting the model to measured magnetization curves. The fitted values reported are $M_s=841~\mathrm{kA\,m^{-1}}$ and $\alpha=2.18\times10^{-5}~\mathrm{m\,A^{-1}}$, which directly motivate the Legendre-model parameters adopted here. 
For the polymer matrix, Kalina \textit{et al.} emphasize that a nearly incompressible mechanical response is essential for stable finite-element simulations and recommend enforcing it through a large volumetric penalty (equivalently, a high bulk/compression modulus), consistent with selecting a Poisson ratio close to $\nu=0.5$ (they use $\nu=0.49$) to obtain a nearly incompressible response. 
Accordingly, the bulk moduli $K$ and Yeoh coefficients $(C_1,C_2,C_3)$ in Table~\ref{tab:sef_parameters} are selected to (i) ensure near-incompressibility through the volumetric penalty and (ii) match the desired nonlinear deviatoric stiffness response of the matrix and particle phases within the strain range considered.

Table~\ref{tab:metamaterial_properties} summarizes the representative volume element (RVE) configurations used in the numerical examples.

\begin{table}[ht]
\centering
\caption{RVE microstructure configurations used in the numerical examples.}
\label{tab:metamaterial_properties}
\begin{tabular}{lccc}
\toprule
Property & Single particle & Particle chain & Random particles \\
\midrule
Number of particles & 1 & 6 & 100 \\
Particle radius $r$ & 0.2 & 0.1 & 0.2 \\
RVE size $L_x \times L_y \times L_z$ & $1\times1\times1$ & $1\times1\times1$ & $6\times6\times6$ \\
Particle volume fraction $\phi$ (\%) & 2.86 & 2.46 & 2.41 \\
\bottomrule
\end{tabular}
\end{table}

\subsection{Applied loading}
\label{sec:applied_loading}

The numerical experiments consider purely mechanical loading, purely magnetic loading, and combined magneto-mechanical loading. Table~\ref{tab:loads_for_experiments} summarizes the prescribed macroscopic deformation gradients and macroscopic Lagrangian magnetic inductions.
Magnetic loading is prescribed in terms of the macroscopic Lagrangian induction $\mathbf{B}_M$ in the reference configuration; we denote its $z$-component by $B_{M,z}$. For purely mechanical runs, we set $\mathbf{B}_M=\mathbf{0}$ so that the magnetic fields arise only from the coupled solution when applicable.
The prescribed deformation magnitudes are selected to (i) activate the nonlinear hyperelastic response and (ii) remain close to isochoric deformation, consistent with the nearly incompressible setting.

\newcommand{\Fstretch}{\mathrm{diag}\!\left(1.1,\,0.9,\,\frac{1}{1.1\cdot0.9}\right)}

\begin{table}[htbp]
\centering
\caption{Prescribed macroscopic loading cases.}
\label{tab:loads_for_experiments}
\begin{tabular}{lcc}
\toprule
Case & $\mathbf{F}_M$ & $\mathbf{B}_M$ \\
\midrule
Mechanical loading & $\Fstretch$ & $\mathbf{0}$ \\
Magnetic loading   & $\mathbf{I}$ & $250~\mathrm{mT}\,\mathbf{e}_z$ \\
Combined loading   & $\Fstretch$ & $250~\mathrm{mT}\,\mathbf{e}_z$ \\
\bottomrule
\end{tabular}
\end{table}

\begin{figure}[htbp]
\centering
\begin{subfigure}[t]{0.46\linewidth}
  \centering
  \includegraphics[width=\linewidth]{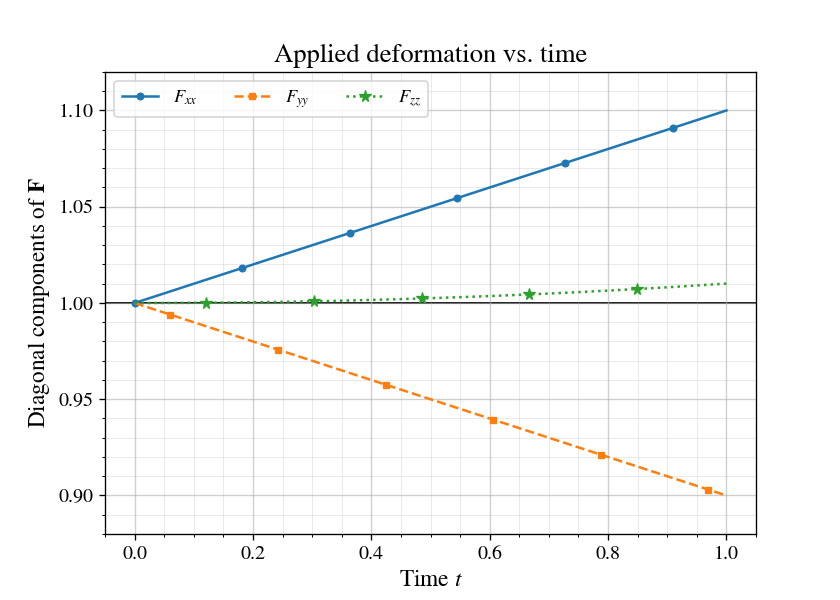}
  \caption{Prescribed components of $\mathbf{F}_M(t)$.}
  \label{fig:mech_loads}
\end{subfigure}
\begin{subfigure}[t]{0.46\linewidth}
  \centering
  \includegraphics[width=\linewidth]{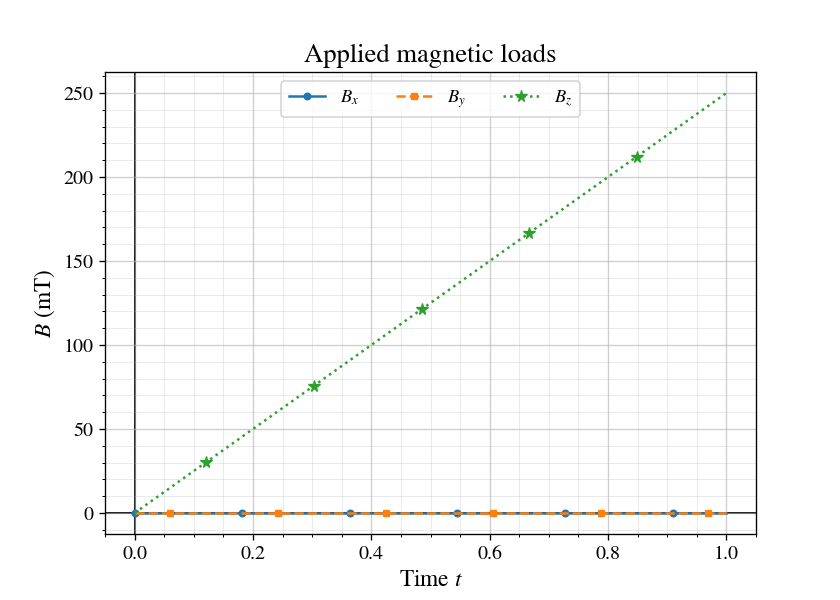}
  \caption{Prescribed $B_{M,z}(t)$.}
  \label{fig:mag_loads}
\end{subfigure}
\caption{Applied loading paths used in the simulations.}
\label{fig:applied_loading}
\end{figure}

All loads are applied quasi-statically through a linear ramp in pseudo-time $t\in[0,1]$. For the deformation gradient, we prescribe
\begin{equation}
\label{eq:loading_strain}
\mathbf{F}_M(t) = \mathbf{I} + t\big(\mathbf{F}_{M}^{\mathrm{final}}-\mathbf{I}\big),
\end{equation}
with the out-of-plane component chosen to enforce an isochoric macroscopic deformation,
\begin{equation}
\label{eq:loading_strain_zz}
F_{M,zz}(t) = \frac{1}{F_{M,xx}(t)\,F_{M,yy}(t)}.
\end{equation}
The magnetic induction is ramped as
\begin{equation}
\label{eq:loading_mag}
\mathbf{B}_M(t) = t\,\mathbf{B}_{M}^{\mathrm{final}},
\qquad \text{with } \mathbf{B}_{M}^{\mathrm{final}} = B_{M,z}^{\mathrm{final}}\mathbf{e}_z.
\end{equation}
Figure~\ref{fig:applied_loading} illustrates the applied loading paths.

For stress-relaxed boundary conditions, the macroscopic deformation $\mathbf{F}_M$ is not prescribed directly but is solved for iteratively to satisfy the chosen macroscopic stress constraint, while the magnetic loading continues to follow Eq.~\eqref{eq:loading_mag}.

\subsection{Simple Inclusion}
\label{sec:simple_inclusion}

\subsubsection{Meshes}
\label{sec:simple_inclusion_mesh}

We first consider a baseline RVE consisting of a single spherical inclusion embedded in an incompressible matrix. This example is intentionally simple so that differences between discretizations can be attributed primarily to the finite element spaces and mesh topology rather than microstructural complexity. Two geometrically equivalent meshes are generated: a structured hexahedral mesh and an unstructured tetrahedral mesh. The inclusion occupies approximately $10\%$ of the RVE volume, and the tetrahedral mesh employs local refinement in the vicinity of the inclusion--matrix interface to better resolve steep gradients in deformation and magnetic
fields.

\begin{figure}[htbp]
  \centering
  \begin{subfigure}[b]{0.48\linewidth}
    \centering
    \includegraphics[width=\linewidth]{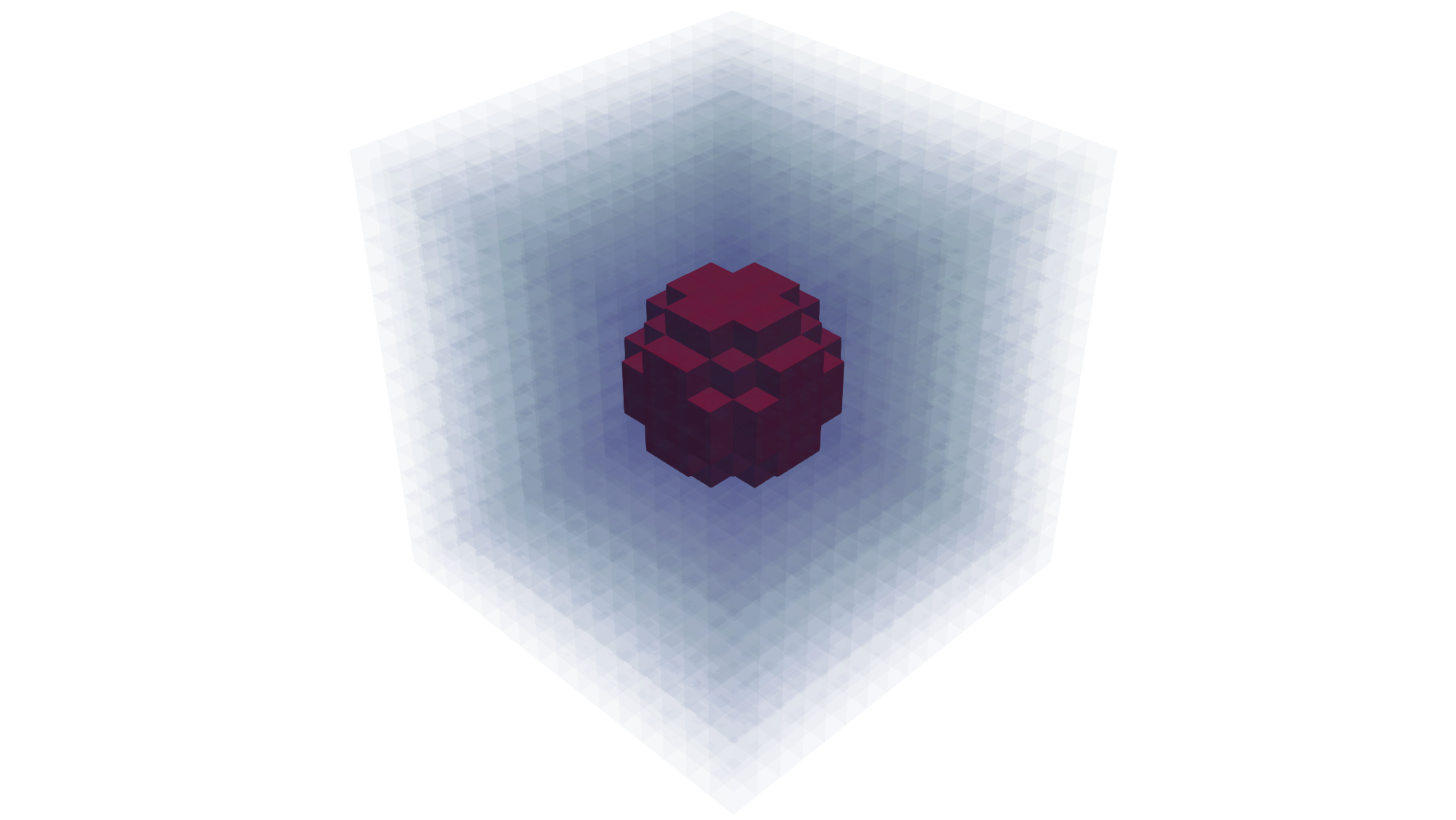}
    \caption{Hex Element Mesh}
    \label{fig:SI_hex_mesh}
  \end{subfigure}
  \hfill
  \begin{subfigure}[b]{0.48\linewidth}
    \centering
    \includegraphics[width=\linewidth]{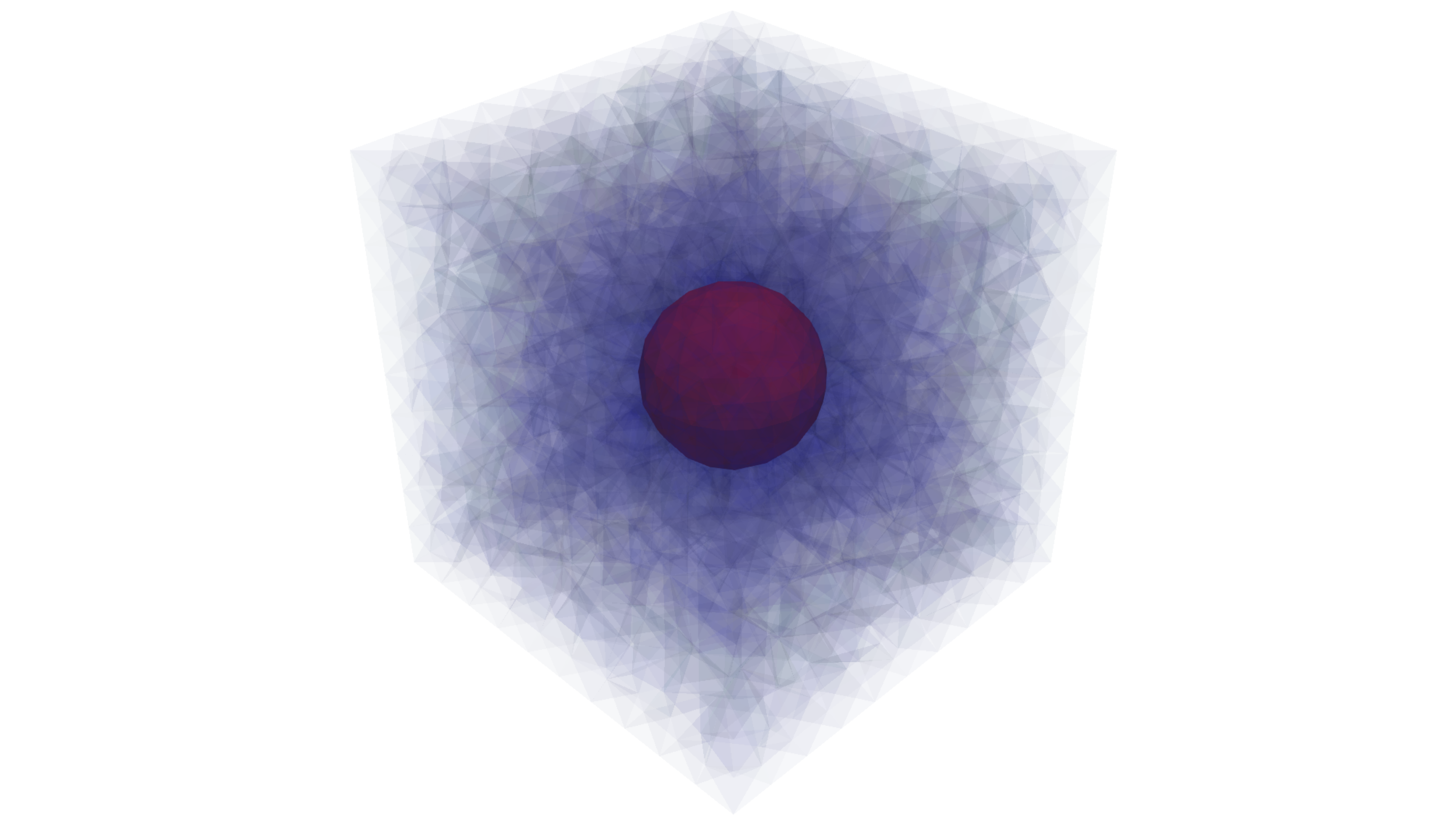}
    \caption{Tet Element Mesh}
    \label{fig:SI_tet_mesh}
  \end{subfigure}
  \caption{Simple Inclusion Meshes}
  \label{fig:pair}
\end{figure}

Figure~\ref{fig:pair} shows the two discretizations used for the comparison. The hexahedral mesh provides a tensor-product representation that is well-suited for the $H^1$--$H(\mathrm{curl})$ pairing used in Section~\ref{sec:fe_discretization_hex}, while the tetrahedral mesh enables flexible refinement and serves as a representative unstructured setting for complex particle geometries. Unless otherwise stated, both meshes use the same polynomial degree and comparable quadrature order, and periodic boundary conditions are enforced on all faces for both the mechanical displacement and magnetic vector potential.

\subsubsection{Homogenized Responses}
\label{sec:simple_inclusion_homo_responses}
This benchmark is used as a controlled verification case for the homogenized response obtained from the coupled RVE problem. For both discretizations, identical macroscopic loading paths are prescribed through $(\overline{\mathbf{F}},\overline{\mathbf{B}})$ as depicted in Table \ref{tab:loads_for_experiments} and the resulting homogenized quantities $(\overline{\boldsymbol{\sigma}},\overline{\mathbf{H}})$ are extracted using the averaging relations introduced in Section~\ref{sec:homogenization}. The objective is twofold: (i) to confirm that the tetrahedral simplex implementation reproduces the same effective response as the reference hexahedral formulation, and (ii) to verify that the near-incompressible stabilization remains effective under coupled magneto--mechanical loading.
We consider three classes of loading depicted in Section \ref{sec:applied_loading}. For each case, the macroscopic response is reported in terms of selected components of $\overline{\boldsymbol{\sigma}}$ and $\overline{\mathbf{H}}$ as functions of the applied loading parameter. In addition, the volumetric behavior is monitored via the RVE-averaged dilatation to confirm that the response remains essentially incompressible throughout the loading history.
Agreement between the hexahedral and tetrahedral results across these paths provides a direct consistency check of the simplex-based discretization and its periodic constraints. Any residual discrepancies are attributable primarily to discretization error and the different refinement patterns near the inclusion interface.

\paragraph{Purely mechanical loading:}
\label{par:sphere_mech_loading}
A purely mechanical loading path is prescribed via the macroscopic deformation gradient $\overline{\mathbf{F}}(t)$ while the macroscopic magnetic flux density is set to $\overline{\mathbf{B}}=\mathbf{0}$. The homogenized response is reported in terms of selected components of the macroscopic Cauchy stress $\overline{\boldsymbol{\sigma}}$ as a function of the loading parameter (e.g., stretch, shear magnitude, or time).
\begin{figure}[htbp]
  \centering

  \begin{subfigure}[t]{0.32\linewidth}
    \centering
    \includegraphics[width=\linewidth]{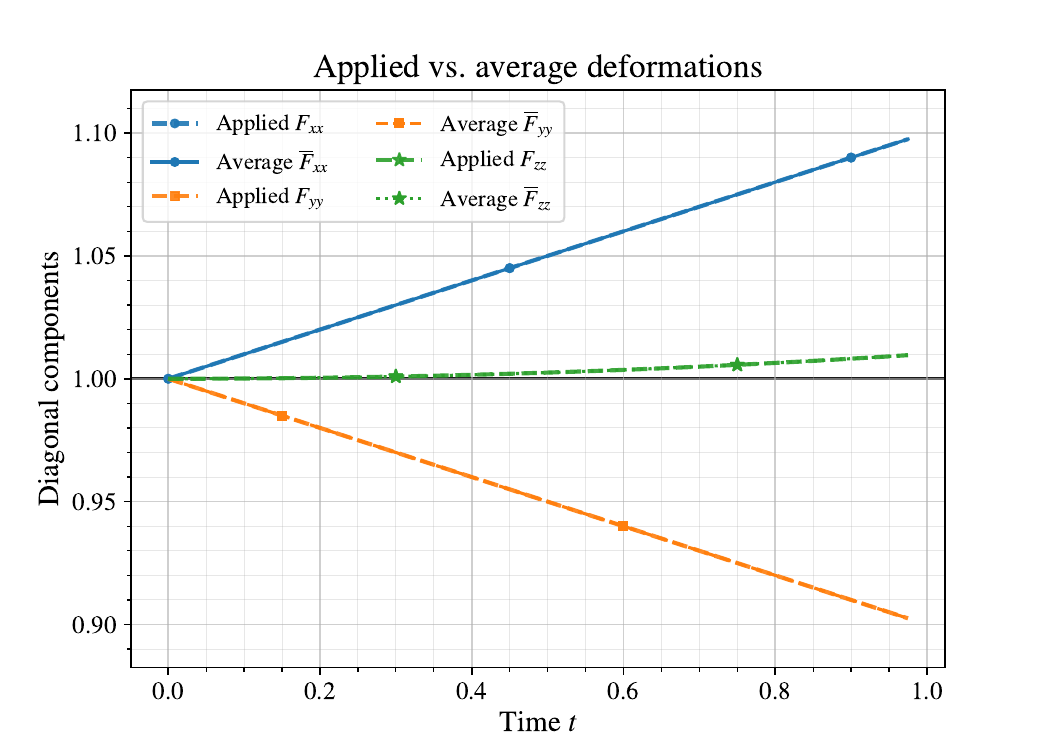}
    \caption{Homogenized deformation (average vs.\ applied)}
    \label{fig:tet_sphere_js_plots_homo_def}
  \end{subfigure}\hfill
  \begin{subfigure}[t]{0.32\linewidth}
    \centering
    \includegraphics[width=\linewidth]{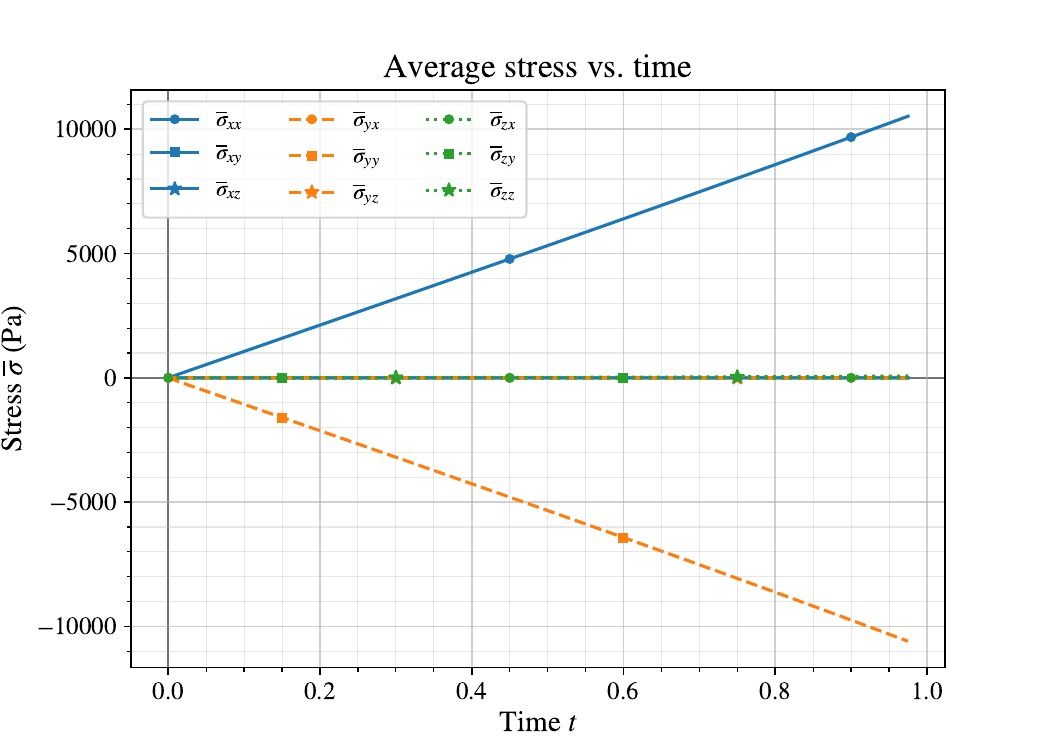}
    \caption{Homogenized stress response}
    \label{fig:tet_sphere_js_plots_homo_stress}
  \end{subfigure}\hfill
  \begin{subfigure}[t]{0.32\linewidth}
    \centering
    \includegraphics[width=\linewidth]{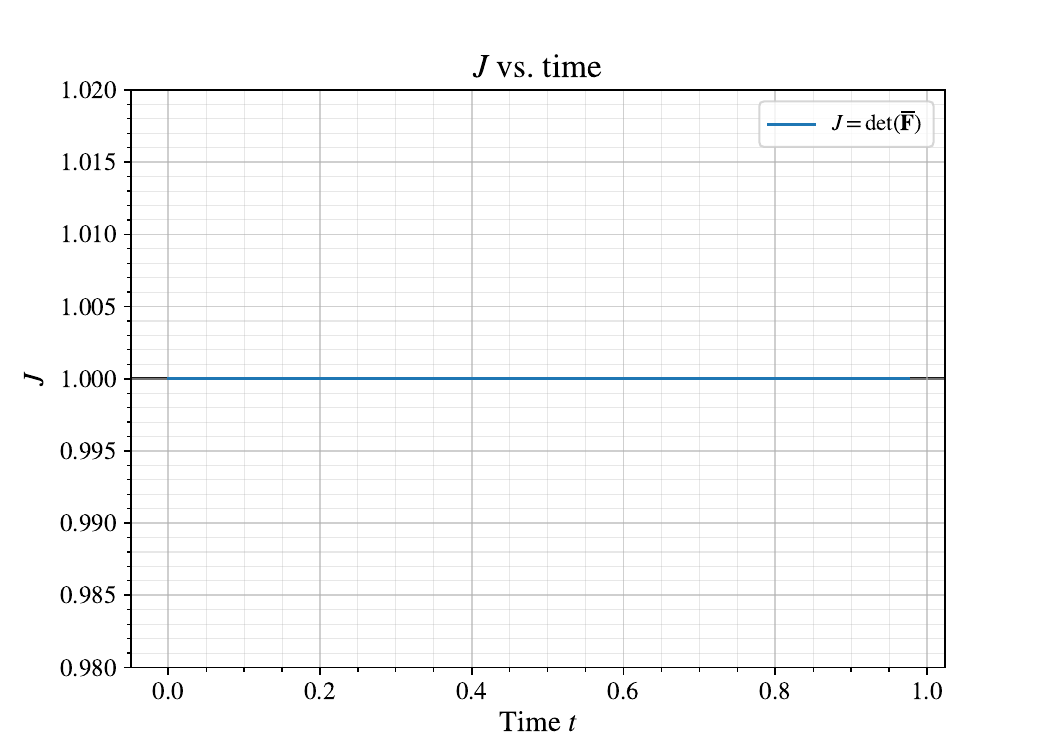}
    \caption{Homogenized Jacobian $J$ (volume change)}
    \label{fig:tet_sphere_js_plots_homo_J}
  \end{subfigure}

  \caption{Homogenized quantities for the single inclusion under strain loading.}
  \label{fig:tet_sphere_js_homogenized_quantities}
\end{figure}

\begin{figure}[htbp]
    \centering
    \begin{subfigure}[b]{0.49\linewidth}
        \centering
        \includegraphics[width=\linewidth]{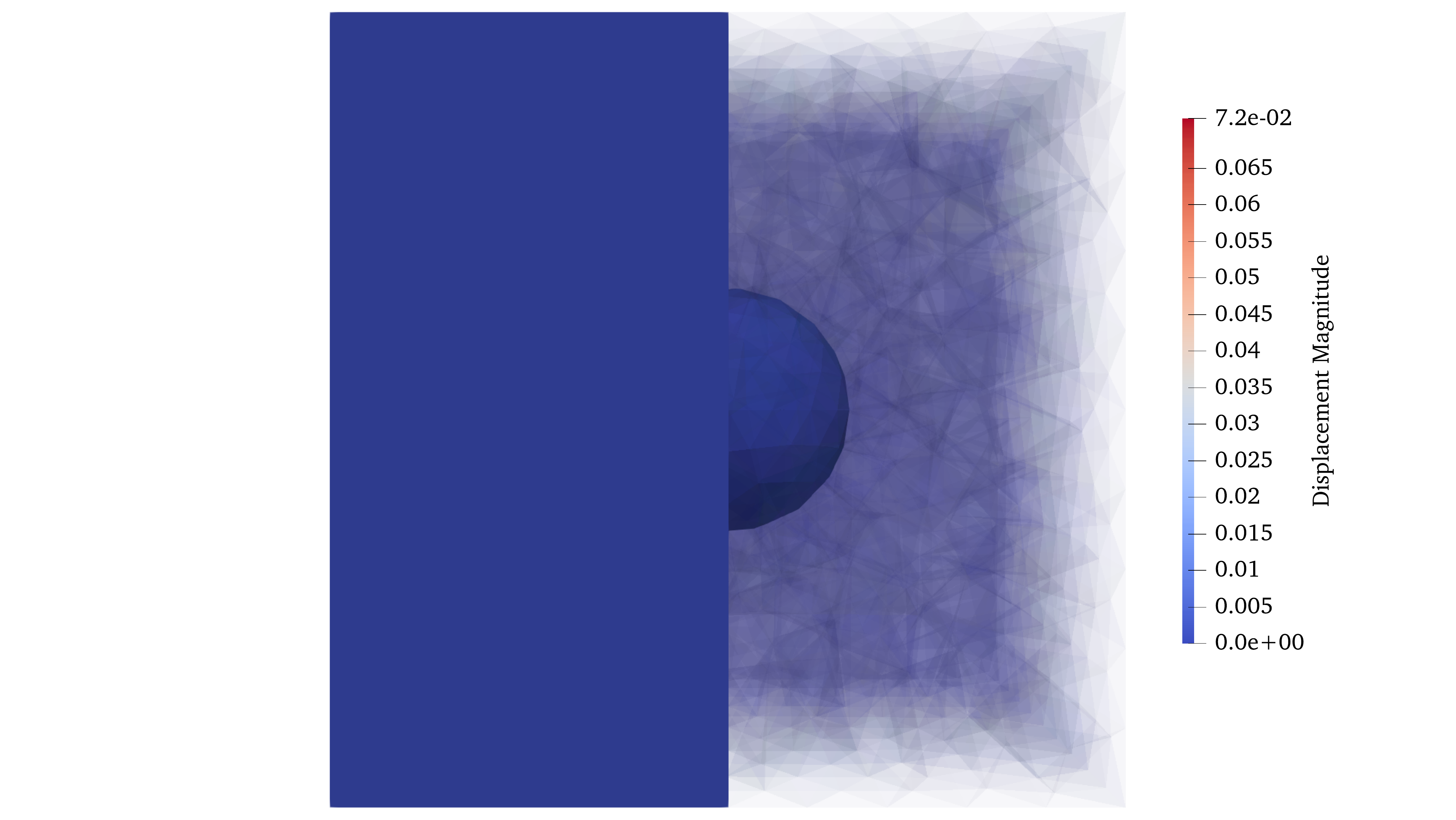}
        \caption{Initial state.}
        \label{fig:tet_sphere_js_initial}
    \end{subfigure}\hfill
    \begin{subfigure}[b]{0.49\linewidth}
        \centering
        \includegraphics[width=\linewidth]{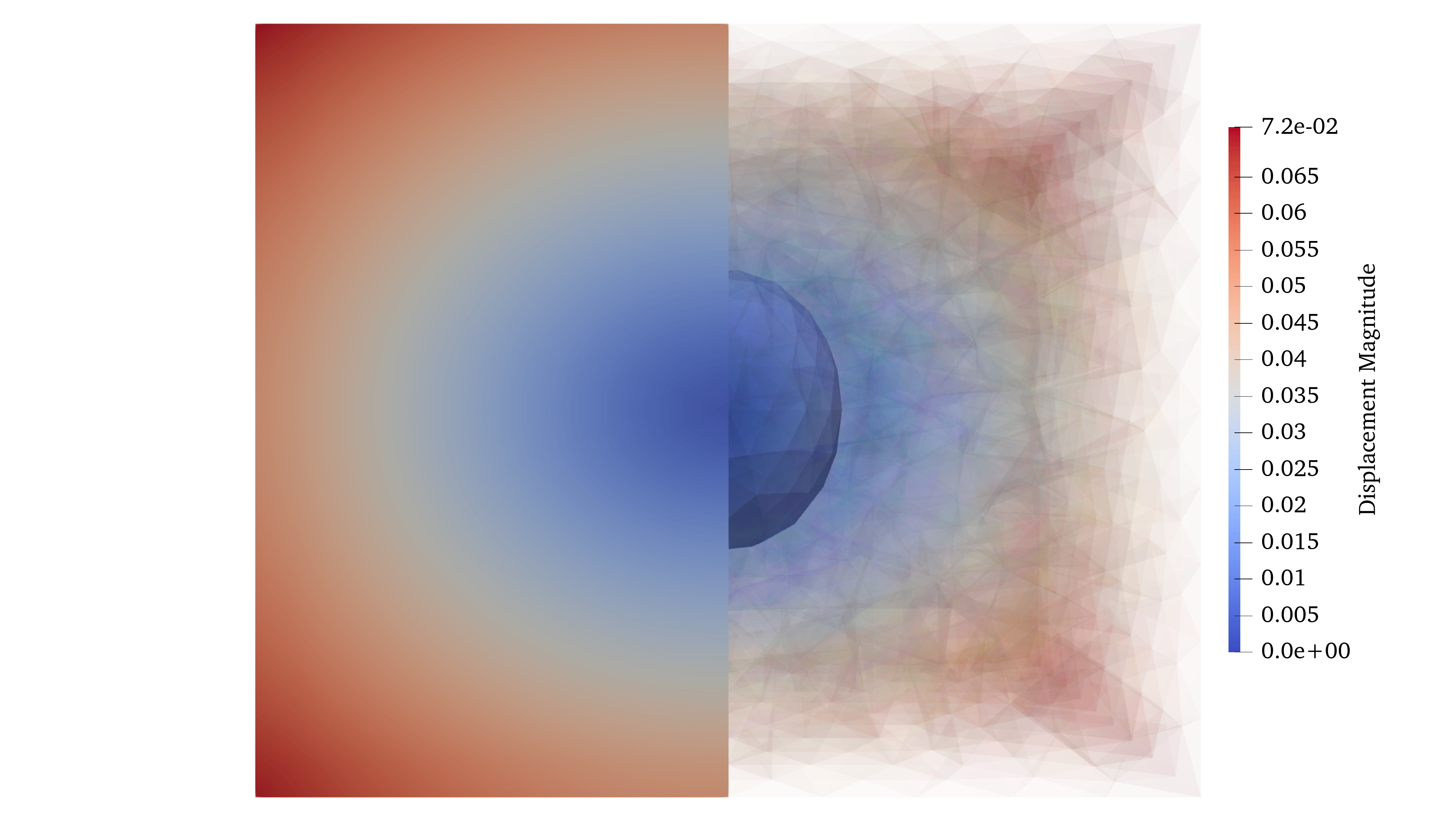}
        \caption{Final state.}
        \label{fig:tet_sphere_js_final}
    \end{subfigure}

    \caption{Spherical inclusion (tet mesh): initial and final configurations under the prescribed deformation.}
    \label{fig:tet_sphere_js_def}
\end{figure}

Figure~\ref{fig:tet_sphere_js_homogenized_quantities} shows that the volume-averaged deformation matches the prescribed macroscopic deformation to machine precision. Moreover, the Jacobian remains unity up to machine precision, confirming that the RVE maintains (near-)incompressibility throughout the loading path. The corresponding initial and final configurations of the spherical inclusion are shown in Figure~\ref{fig:tet_sphere_js_def}.

\paragraph{Purely magnetic loading: }
A purely magnetic loading path is prescribed through $\overline{\mathbf{B}}(t)$ while the macroscopic deformation is held fixed ($\overline{\mathbf{F}}=\mathbf{I}$). The homogenized magnetic response is reported as selected components of $\overline{\mathbf{H}}$ versus the applied induction.

\begin{figure}[htbp]
  \centering

  \begin{subfigure}[t]{0.32\linewidth}
    \centering
    \includegraphics[width=\linewidth]{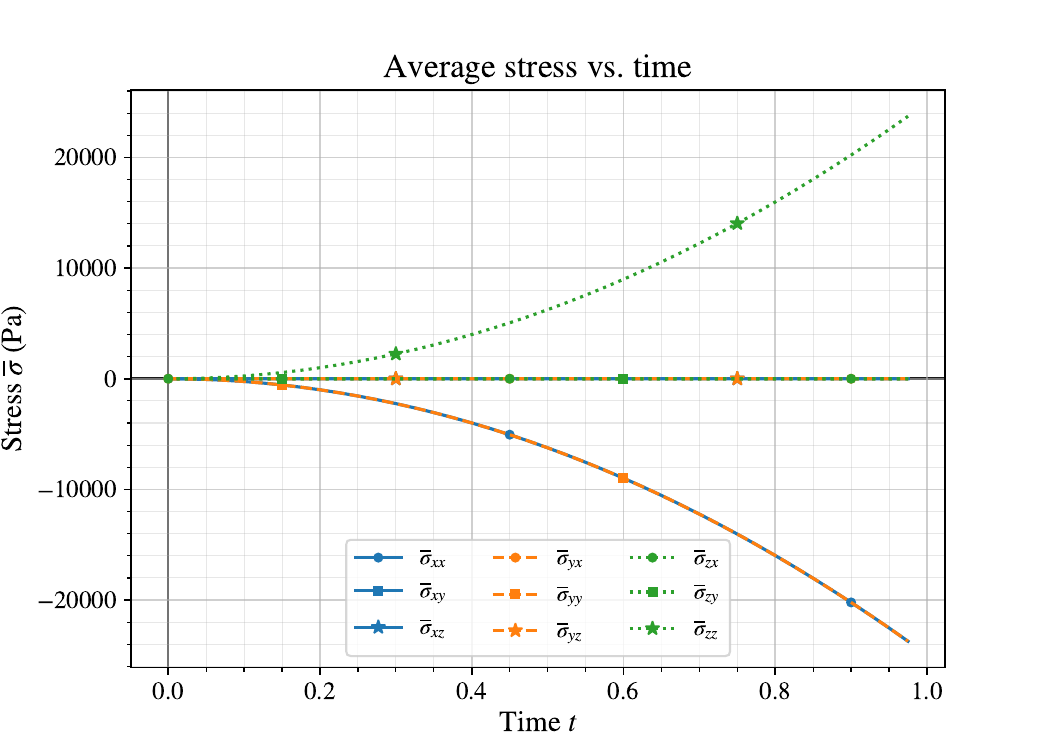}
    \caption{Homogenized stress response}
    \label{fig:tet_sphere_jm_plots_homo_stress}
  \end{subfigure}\hfill
  \begin{subfigure}[t]{0.32\linewidth}
    \centering
    \includegraphics[width=\linewidth]{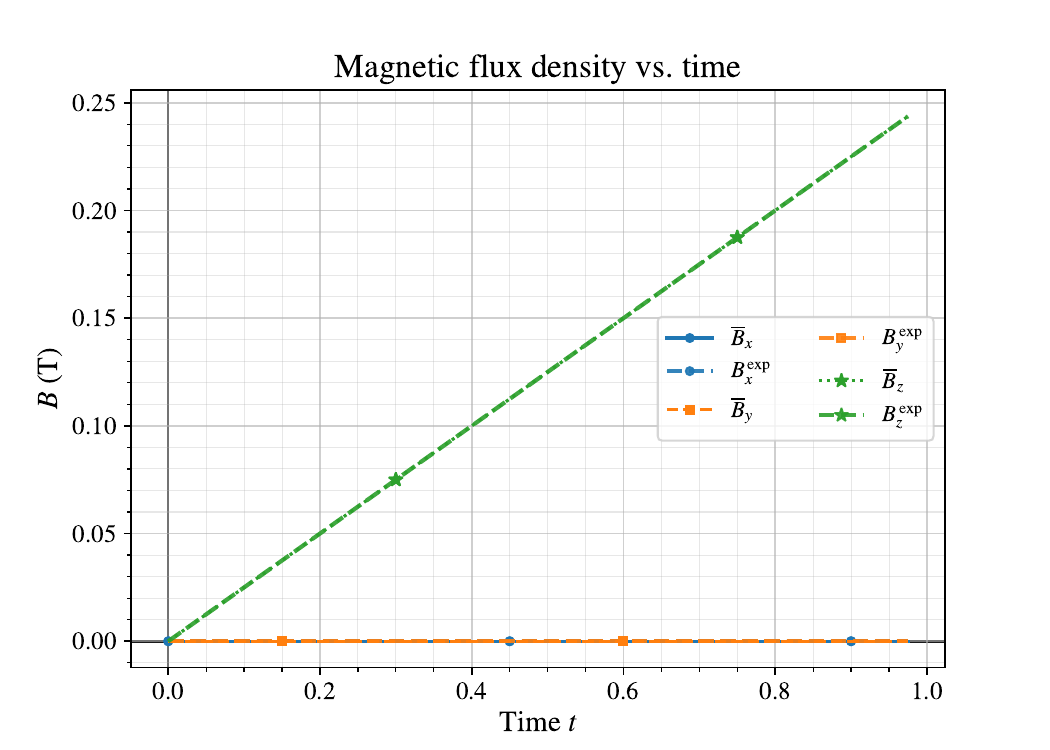}
    \caption{Homogenized magnetic induction $\bar{\mathbf{B}}$ vs.\ time}
    \label{fig:tet_sphere_jm_plots_homo_B}
  \end{subfigure}\hfill
  \begin{subfigure}[t]{0.32\linewidth}
    \centering
    \includegraphics[width=\linewidth]{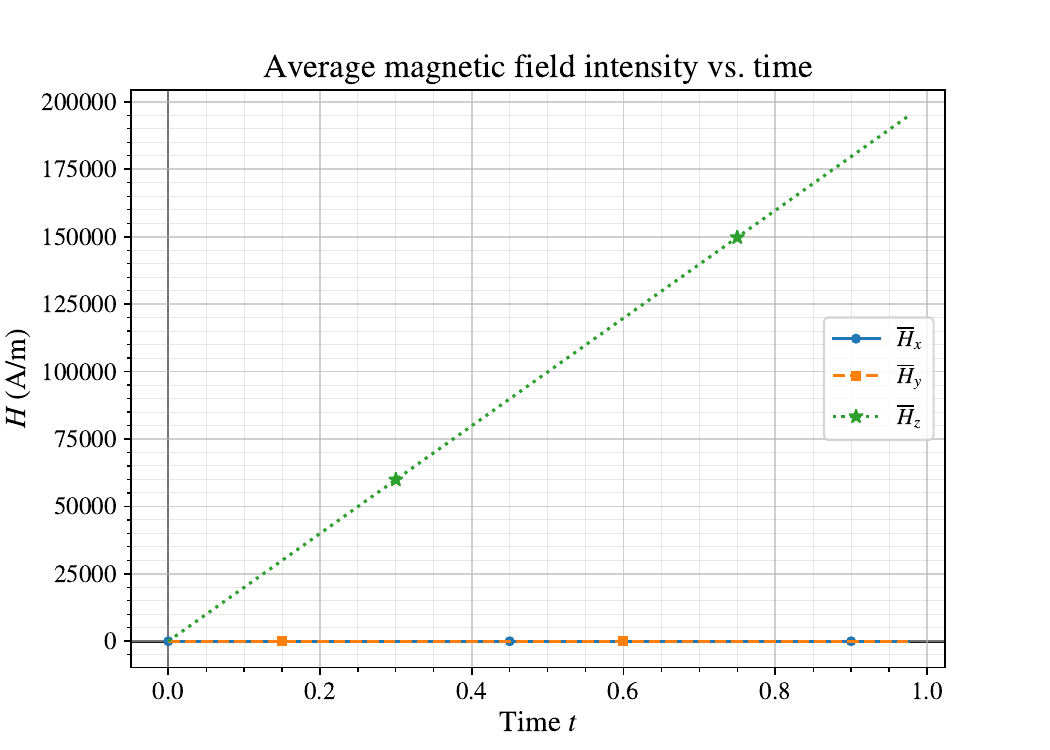}
    \caption{Homogenized magnetic field $\bar{\mathbf{H}}$ vs.\ time}
    \label{fig:tet_sphere_jm_plots_homo_H}
  \end{subfigure}

  \caption{Homogenized quantities for the spherical-inclusion RVE under magnetic loading.}
  \label{fig:tet_sphere_jm_homogenized_quantities}
\end{figure}

\begin{figure}[htbp]
    \centering
    \begin{subfigure}[b]{0.49\linewidth}
        \centering
        \includegraphics[width=\linewidth]{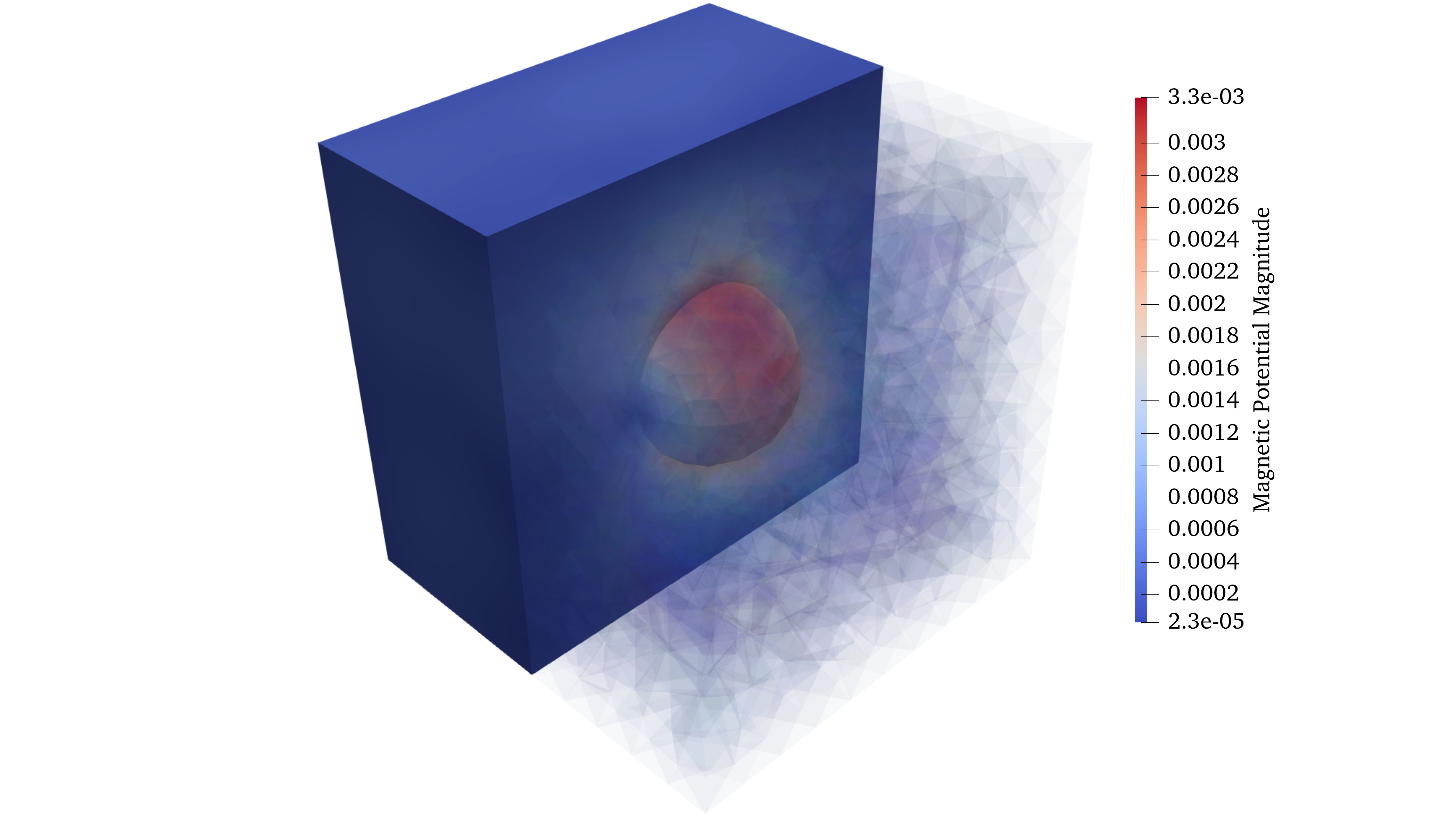}
        \caption{Magnetic Potential Magnitude.}
        \label{fig:tet_sphere_jm_mag_pot}
    \end{subfigure}\hfill
    \begin{subfigure}[b]{0.49\linewidth}
        \centering
        \includegraphics[width=\linewidth]{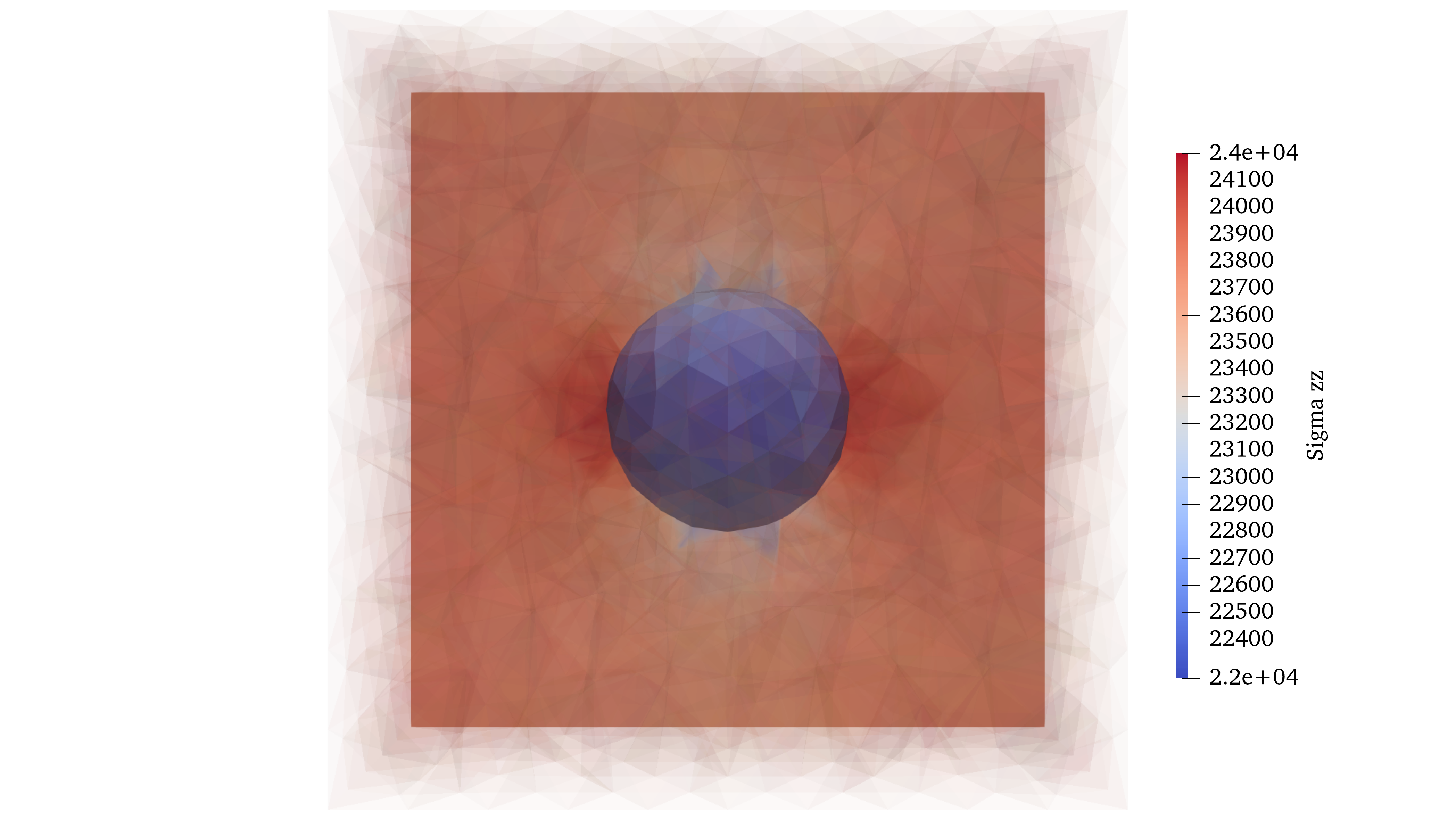}
        \caption{$\sigma_{zz}$}
        \label{fig:tet_sphere_jm_sigma_zz}
    \end{subfigure}

    \caption{Spherical inclusion (tet mesh): Magnetic Potential Magnitude and $\sigma_{zz}$.}
    \label{fig:tet_sphere_jm_figs}
\end{figure}

Figure ~\ref{fig:tet_sphere_jm_homogenized_quantities} shows that under magnetic loading, the RVE outputs the expected average magnetic induction as it is the same as the input quantity. It also shows how it maintains a constrained average strain. The stress caused by the magnetic loading is of similar magnitude to that caused by strain loading. The magnitude of the magnetic potential and the stress in the direction of magnetic induction ($\sigma_{zz}$) is shown in Figure ~\ref{fig:tet_sphere_jm_figs}.

\begin{figure}[htbp]
    \centering
    \begin{subfigure}[b]{0.49\linewidth}
        \centering
        \includegraphics[width=\linewidth]{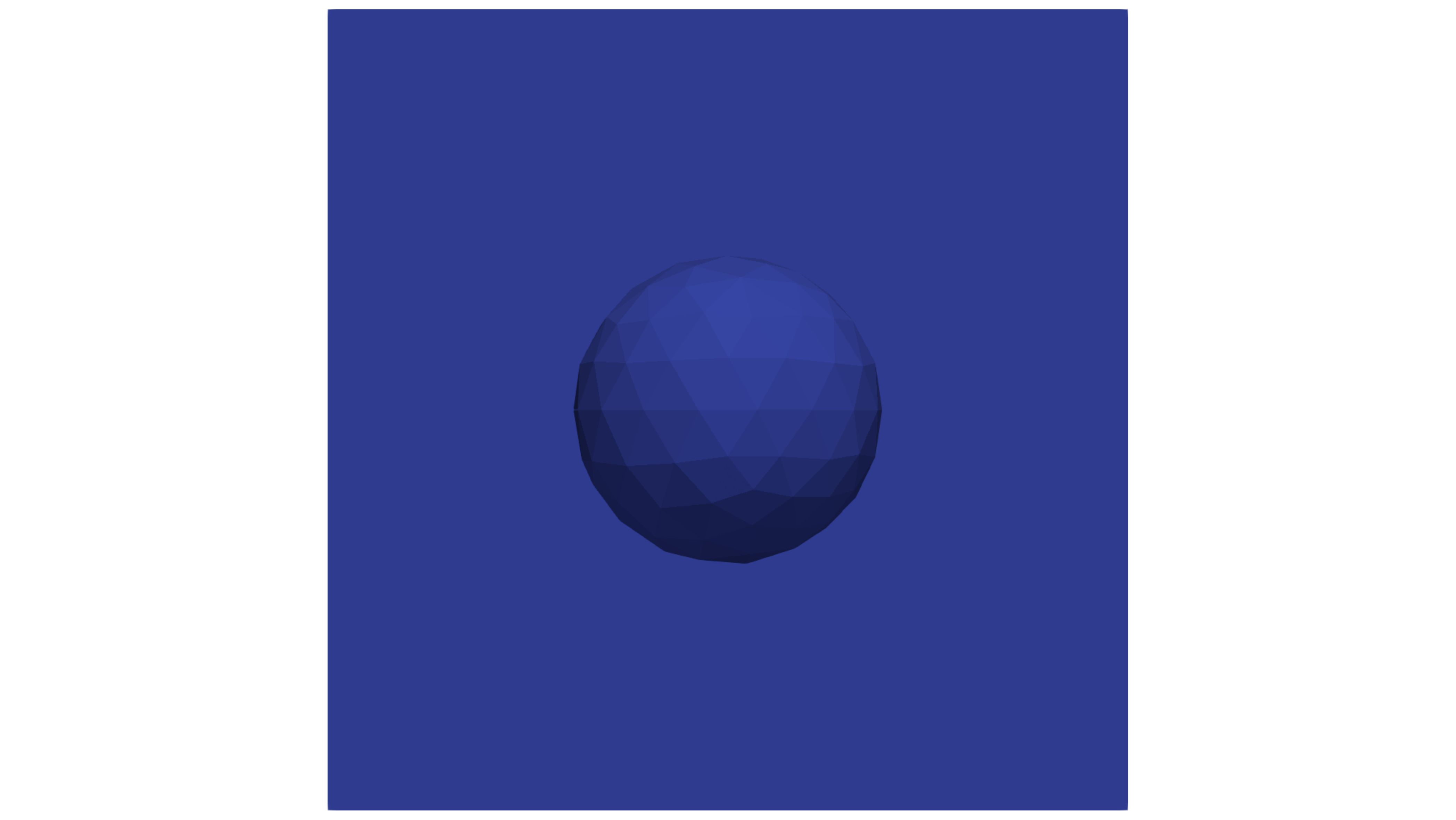}
        \caption{Initial state.}
        \label{fig:tet_sphere_jm_dis_init}
    \end{subfigure}\hfill
    \begin{subfigure}[b]{0.49\linewidth}
        \centering
        \includegraphics[width=\linewidth]{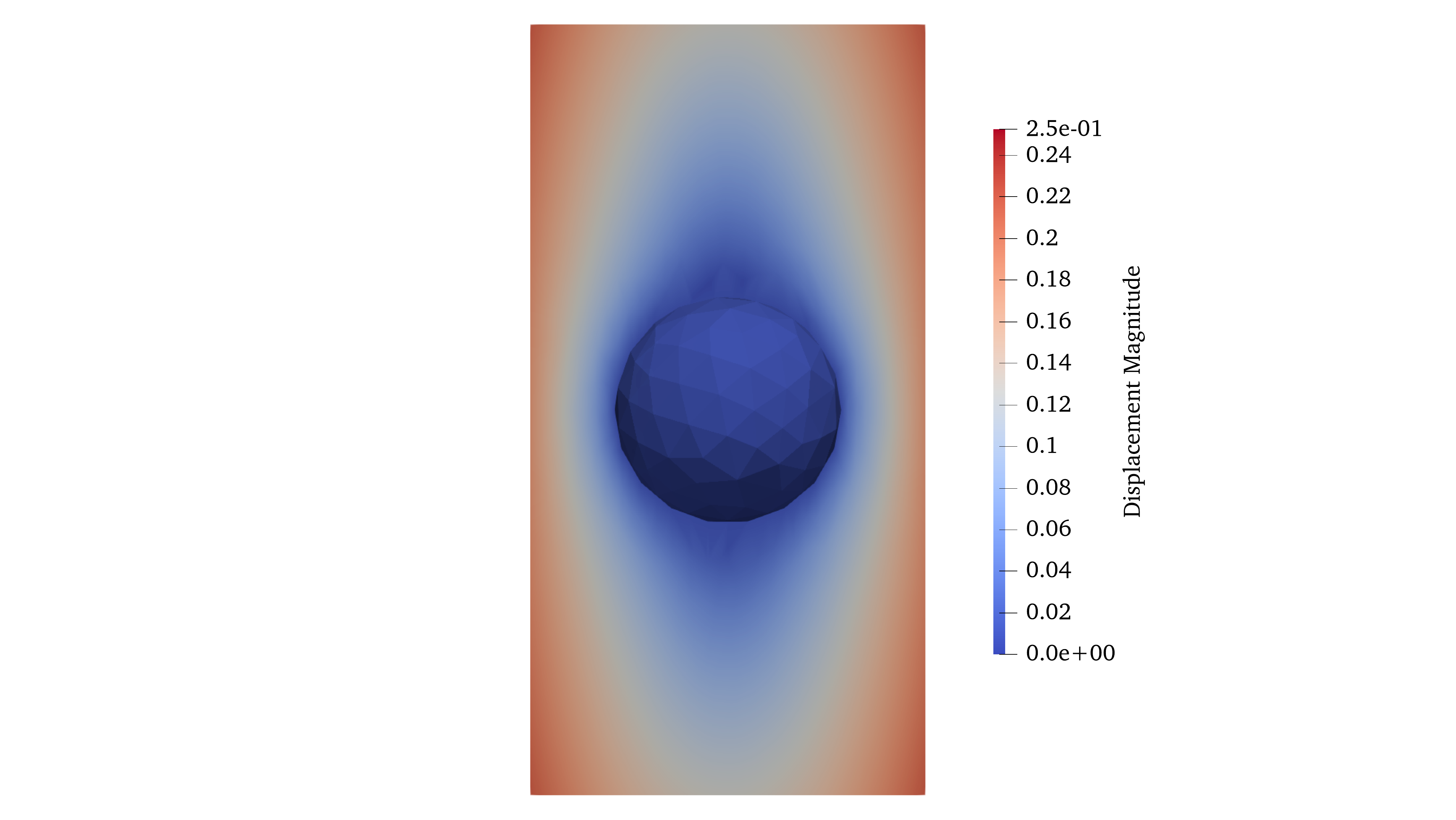}
        \caption{Final state.}
        \label{fig:tet_sphere_jm_dis_final}
    \end{subfigure}

    \caption{Spherical inclusion with a tetrahedral mesh in the initial and final configurations under the prescribed magnetic induction.}
    \label{fig:tet_sphere_jm_def}
\end{figure}

\begin{figure}[htbp]
    \centering
    \begin{subfigure}[b]{0.49\linewidth}
        \centering
        \includegraphics[width=\linewidth]{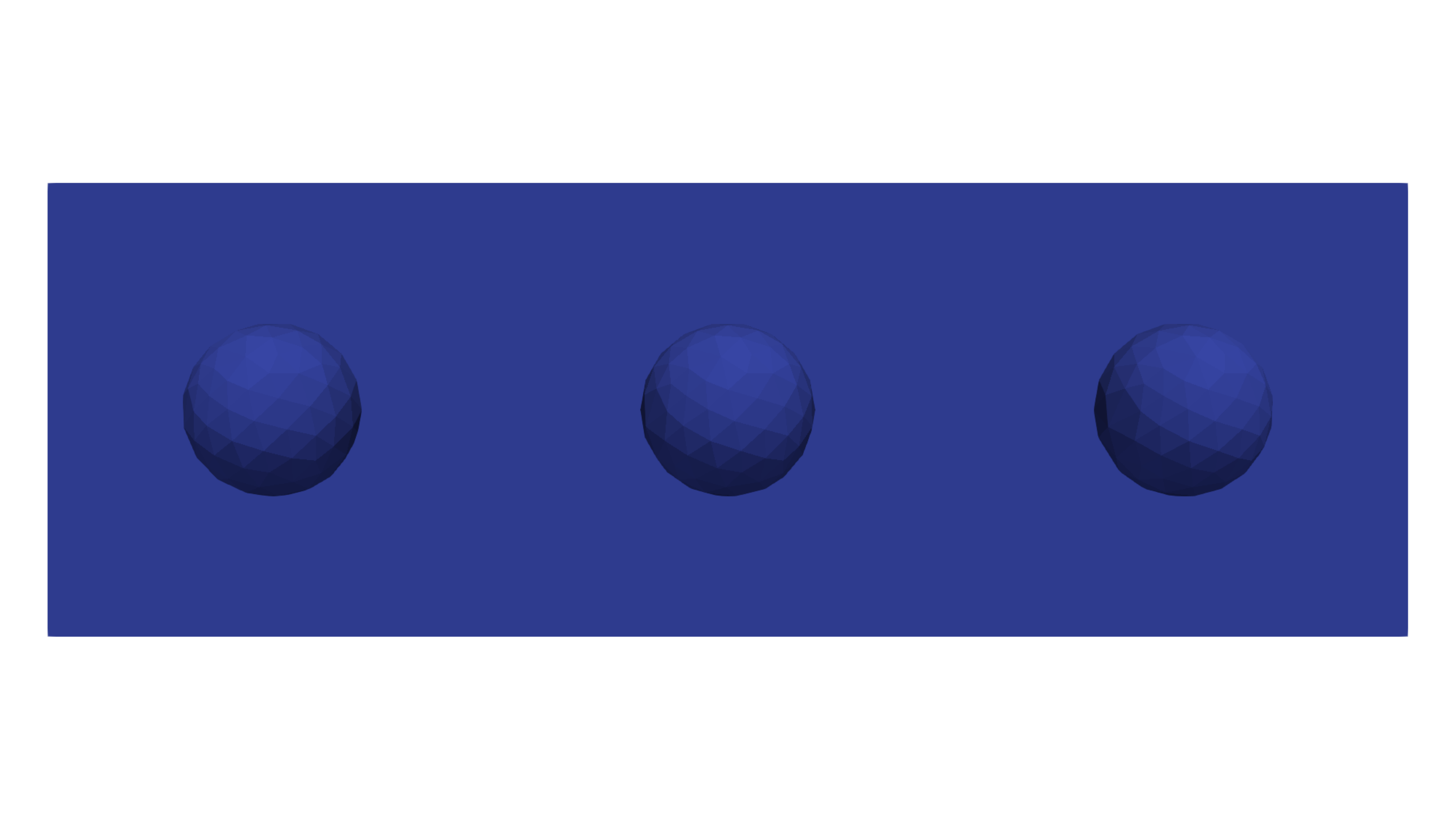}
        \caption{Initial periodic configuration.}
        \label{fig:sphere_full_movement_init}
    \end{subfigure}\hfill
    \begin{subfigure}[b]{0.49\linewidth}
        \centering
        \includegraphics[width=\linewidth]{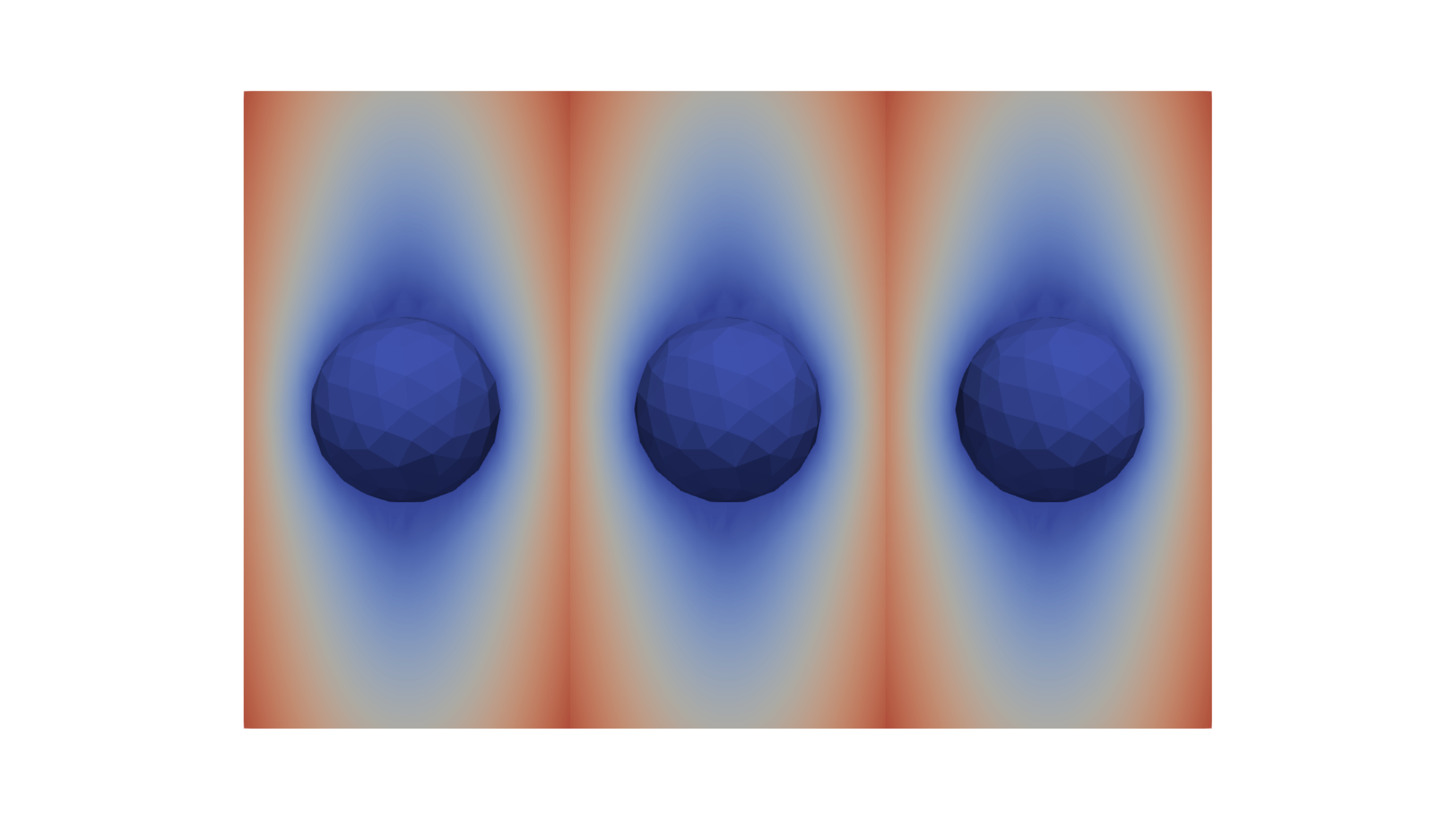}
        \caption{Final periodic configuration.}
        \label{fig:sphere_full_movement_final}
    \end{subfigure}

    \caption{Initial and final periodic configurations for the spherical-inclusion RVE under stress-relaxed boundary conditions.}
    \label{fig:sphere_full_movement}
\end{figure}

\begin{figure}[htbp]
    \centering
    \includegraphics[width=0.95\linewidth]{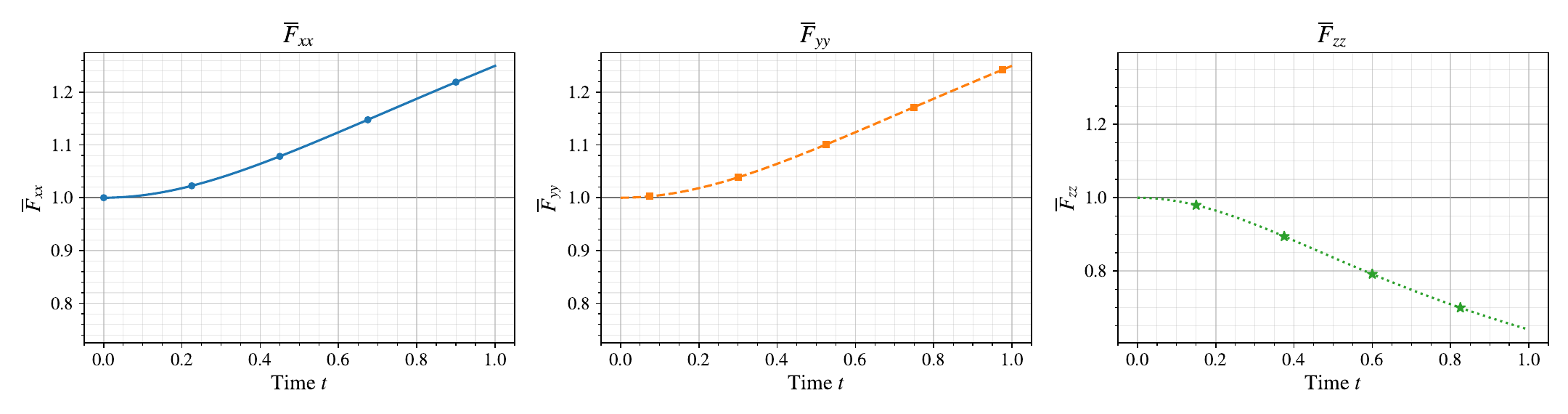}
    \caption{Homogenized deformation response under stress-relaxed boundary conditions, with the residual stress tolerance set to $10^{-3}$.}
    \label{fig:tet_sphere_ns_def}
\end{figure}

Additionally, using the stress-relaxed boundary conditions described in Section~\ref{sec:stress_relaxed_bc}, we can observe the deformation of the inclusion and surrounding matrix induced by the applied magnetic loading, as shown in Figure~\ref{fig:tet_sphere_jm_def}. The corresponding homogenized deformation response is shown in Figure~\ref{fig:tet_sphere_ns_def}. In the deformed configuration, the inclusion exhibits a predominantly symmetric motion, with the upper and lower portions moving toward one another along the direction of the applied magnetic induction. This indicates a contraction in the \(z\)-direction, while the surrounding matrix deforms compatibly to accommodate the inclusion response under the stress-relaxed setting. The resulting motion is therefore characterized primarily by field-direction compression rather than rigid translation, with only minor shear-driven distortion visible in the final configuration. This interpretation is also consistent with the effective magnetostrictive response discussed in Section~\ref{sec:magnetostriction_compare}, in the sense that the local inclusion kinematics and the homogenized deformation both reflect a field-induced contraction along the loading direction. Owing to the periodic boundary conditions, however, this motion should be understood as the local response of a repeating microstructure rather than the behavior of an isolated particle in free space. In that sense, if additional inclusions were present in neighboring periodic images of the cell, the same deformation pattern could also be interpreted on a larger scale as relative approach or rigid-body-like translation between particles along the field direction, as illustrated in Figure~\ref{fig:sphere_full_movement}.

\paragraph{Coupled magneto--mechanical loading:}
Finally, a coupled loading path is applied by prescribing $(\overline{\mathbf{F}}(t),\overline{\mathbf{B}}(t))$ simultaneously. The coupled response is summarized by reporting both $\overline{\mathbf{\sigma}}$ and $\overline{\mathbf{H}}$ along the loading history, and comparing against the corresponding purely mechanical and purely magnetic responses to highlight the magneto--mechanical interaction.

\begin{figure}[htbp]
  \centering

  \begin{subfigure}[t]{0.32\linewidth}
    \centering
    \includegraphics[width=\linewidth]{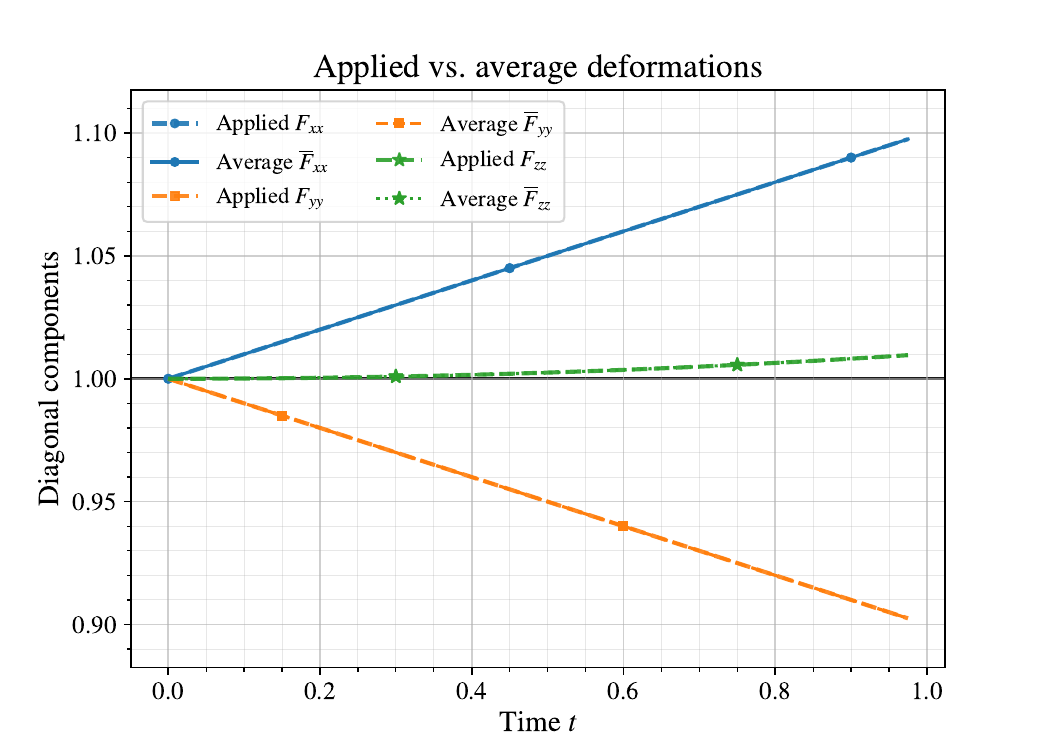}
    \caption{Homogenized deformation (average vs.\ applied)}
    \label{fig:tet_sphere_both_plots_homo_def}
  \end{subfigure}\hfill
  \begin{subfigure}[t]{0.32\linewidth}
    \centering
    \includegraphics[width=\linewidth]{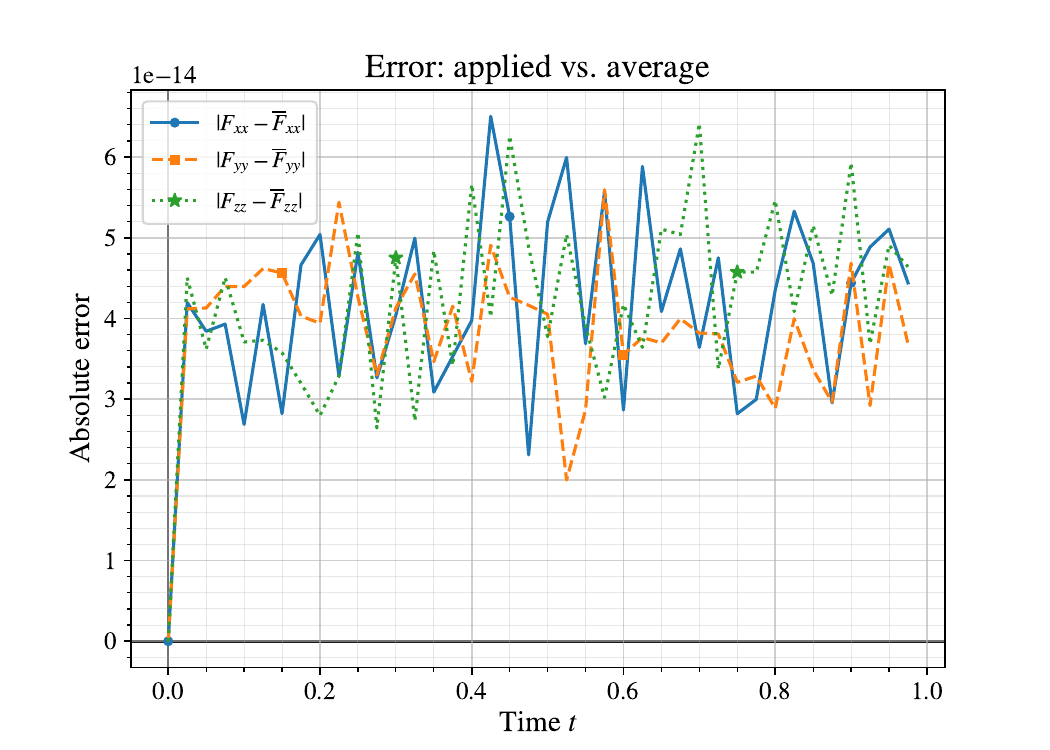}
    \caption{Deformation mismatch (average $-$ applied)}
    \label{fig:tet_sphere_both_plots_homo_def_error}
  \end{subfigure}\hfill
  \begin{subfigure}[t]{0.32\linewidth}
    \centering
    \includegraphics[width=\linewidth]{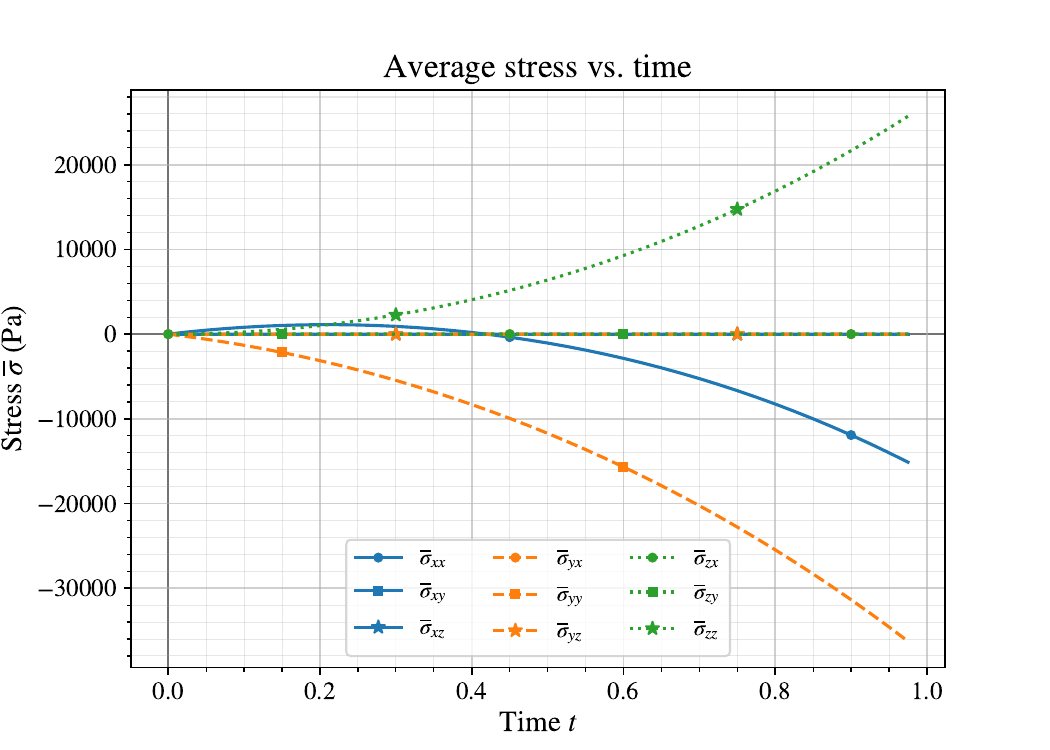}
    \caption{Homogenized stress response}
    \label{fig:tet_sphere_both_plots_homo_stress}
  \end{subfigure}

  \vspace{0.6em}

  \begin{subfigure}[t]{0.32\linewidth}
    \centering
    \includegraphics[width=\linewidth]{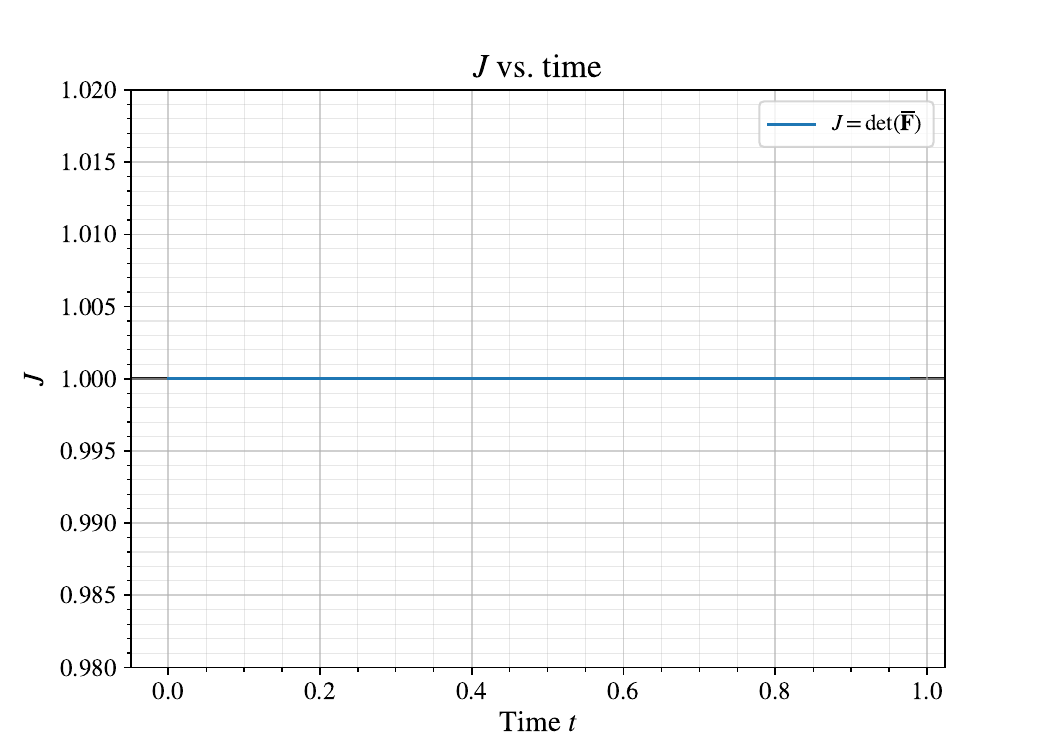}
    \caption{Homogenized Jacobian $J$ (volume change)}
    \label{fig:tet_sphere_both_plots_homo_J}
  \end{subfigure}\hfill
  \begin{subfigure}[t]{0.32\linewidth}
    \centering
    \includegraphics[width=\linewidth]{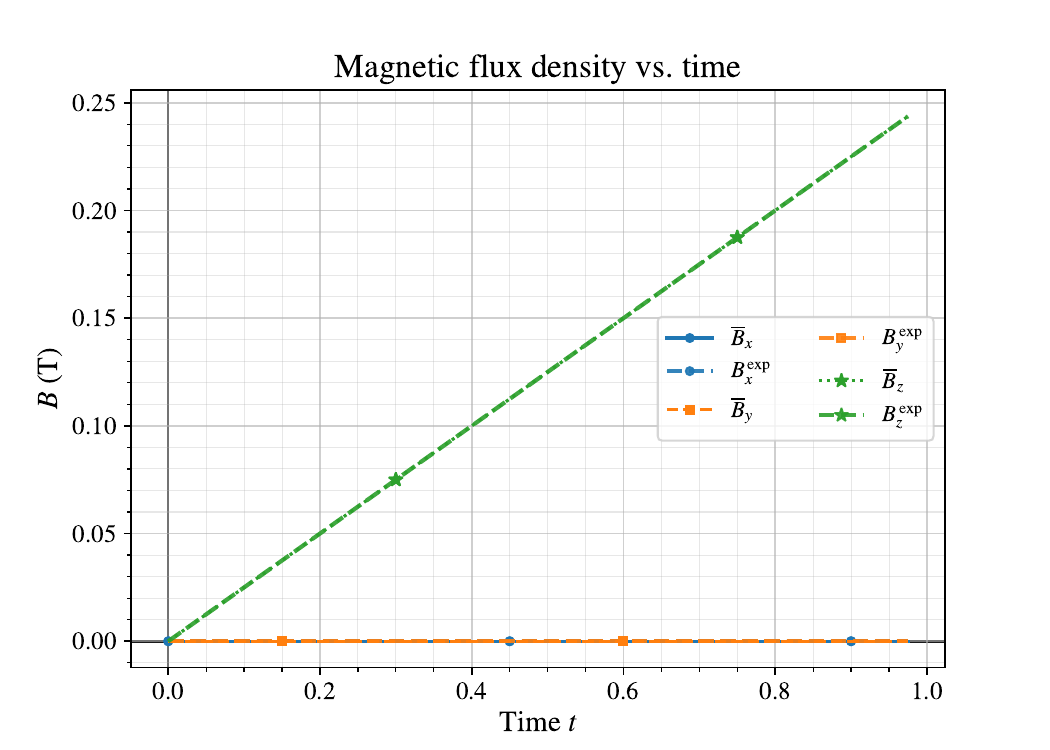}
    \caption{Homogenized magnetic induction $\bar{\mathbf{B}}$ vs.\ time}
    \label{fig:tet_sphere_both_plots_homo_B}
  \end{subfigure}\hfill
  \begin{subfigure}[t]{0.32\linewidth}
    \centering
    \includegraphics[width=\linewidth]{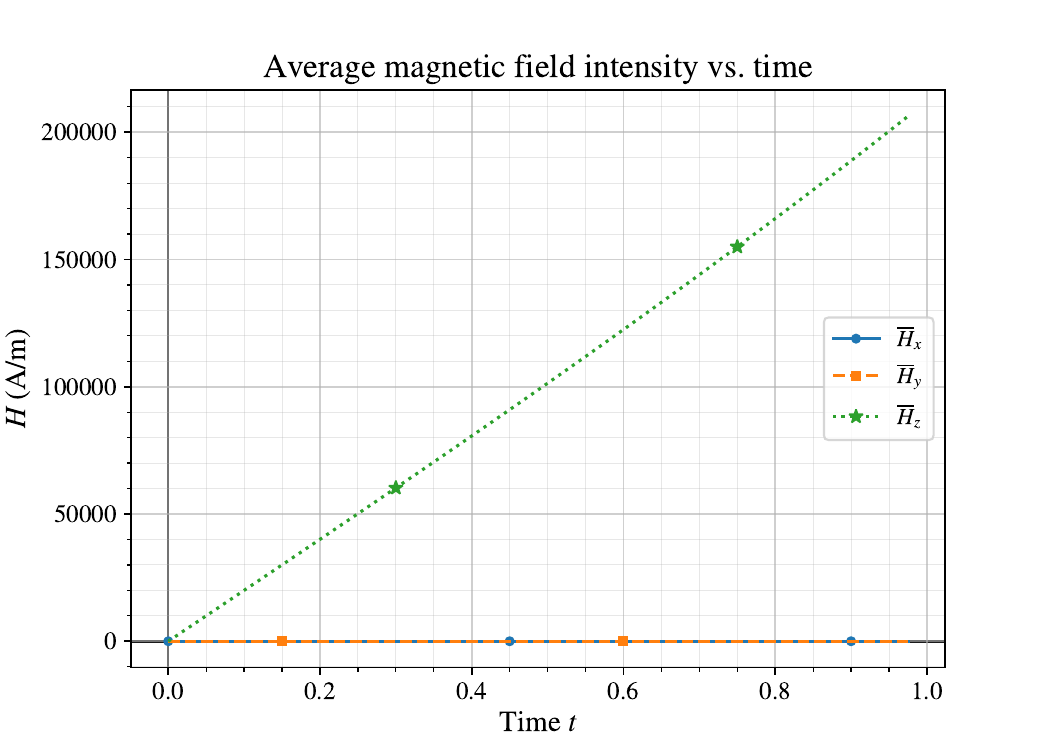}
    \caption{Homogenized magnetic field $\bar{\mathbf{H}}$ vs.\ time}
    \label{fig:tet_sphere_both_plots_homo_H}
  \end{subfigure}

  \caption{Homogenized quantities for the tet spherical inclusion RVE under the prescribed magneto-mechanical loading.}
  \label{fig:tet_sphere_both_homogenized_quantities}
\end{figure}

The combined influence of the strain loading and magnetic loading is most clearly observed in Figure~\ref{fig:tet_sphere_both_plots_homo_stress}. For $\sigma_{xx}$, the response is initially dominated by the mechanically induced stress, as evidenced by the positive stress contribution up to approximately $t \approx 0.2$. As loading progresses, the magnetic contribution becomes increasingly significant and eventually overtakes the mechanical contribution. This transition is reflected in the reversal of the stress direction at approximately $t \approx 0.45$, indicating that the magnetic loading not only counteracts the mechanically induced stress but ultimately governs the overall response. This behavior highlights the strongly nonlinear competition between the mechanical and magnetic effects in the coupled loading case.

\begin{figure}[htbp]
    \centering
    \begin{subfigure}[b]{0.32\linewidth}
        \centering
        \includegraphics[width=\linewidth]{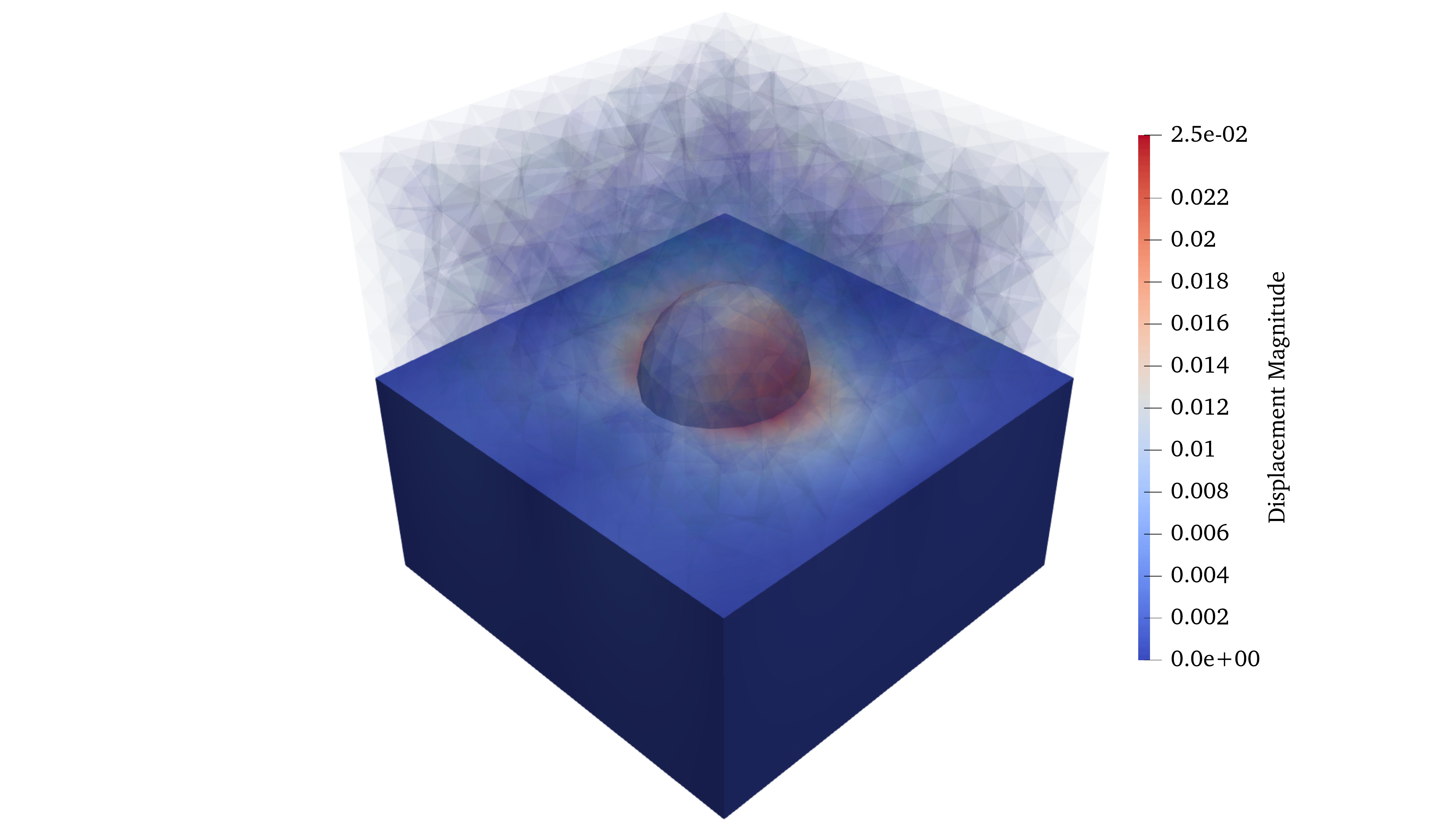}
        \caption{Combined Displacement Magnitude}
        \label{fig:tet_sphere_both_disp_final}
    \end{subfigure}\hfill
    \begin{subfigure}[b]{0.32\linewidth}
        \centering
        \includegraphics[width=\linewidth]{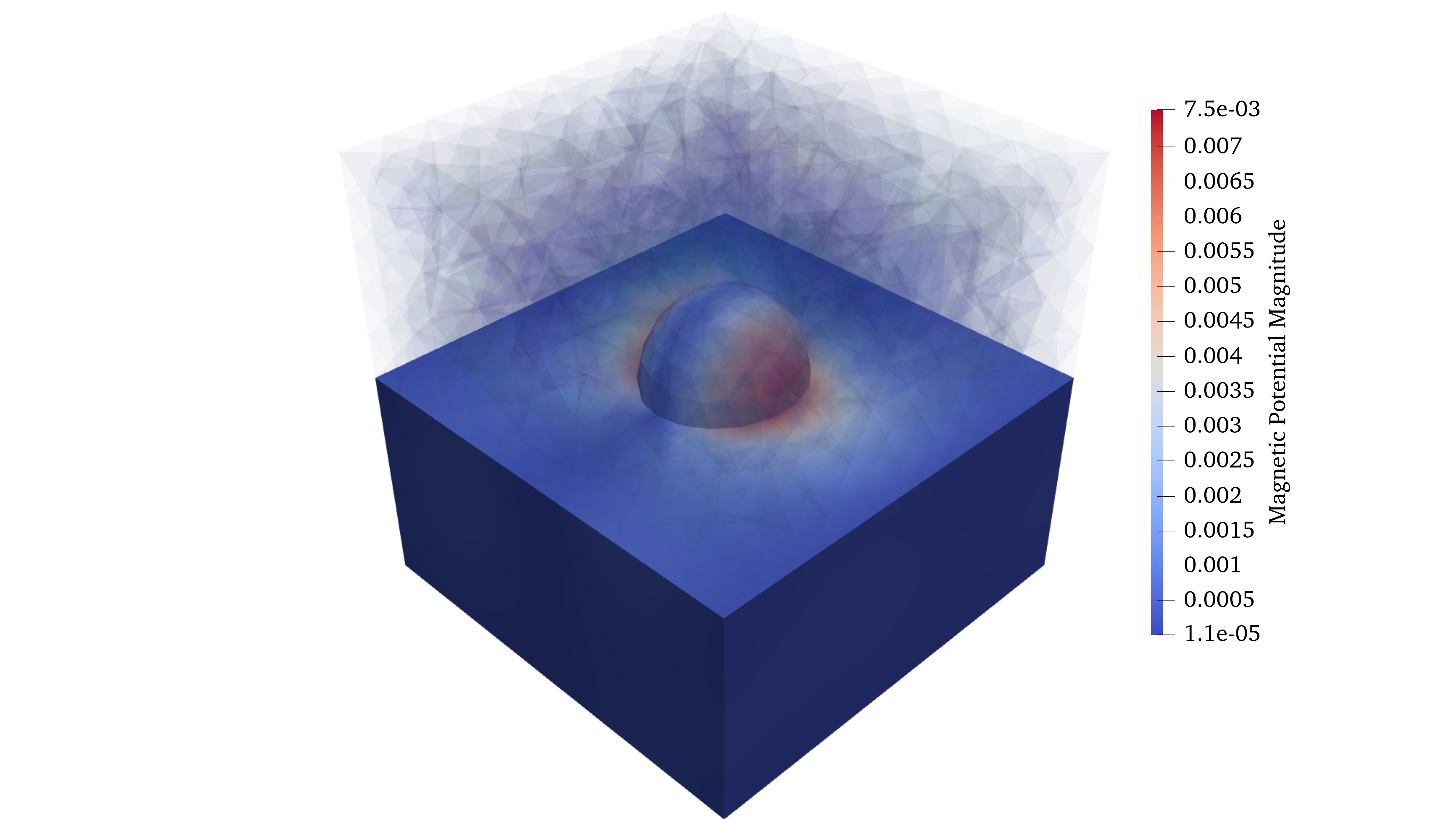}
        \caption{Combined Magnetic Potential}
        \label{fig:tet_sphere_both_magpot_final}
    \end{subfigure}\hfill
    \begin{subfigure}[b]{0.32\linewidth}
        \centering
        \includegraphics[width=\linewidth]{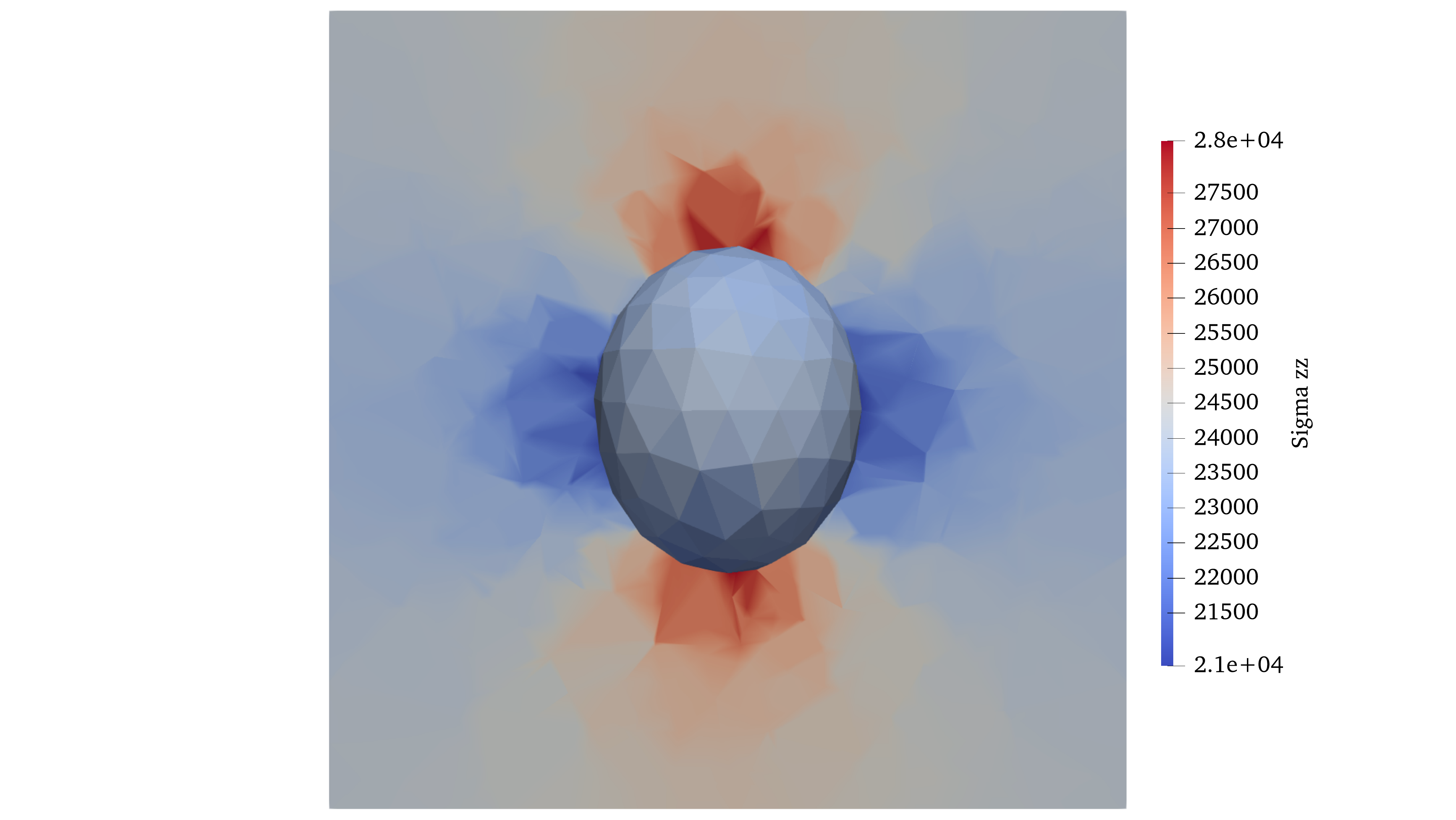}
        \caption{Final $\sigma_{zz}$}
        \label{fig:tet_sphere_both_sigmazz_final}
    \end{subfigure}

    \caption{Spherical inclusion (tet mesh): displacement magnitude, magnetic potential, and $\sigma_{zz}$ under the prescribed magneto-mechanical loading.}
    \label{fig:tet_sphere_both_para_shots}
\end{figure}

\subsubsection{Hexahedral Response}
\label{sec:hex_sphere}

\begin{figure}[htbp]
    \centering

    \begin{subfigure}[t]{0.32\linewidth}
        \centering
        \vspace{0pt}
        \includegraphics[width=\linewidth,height=4.2cm,keepaspectratio]{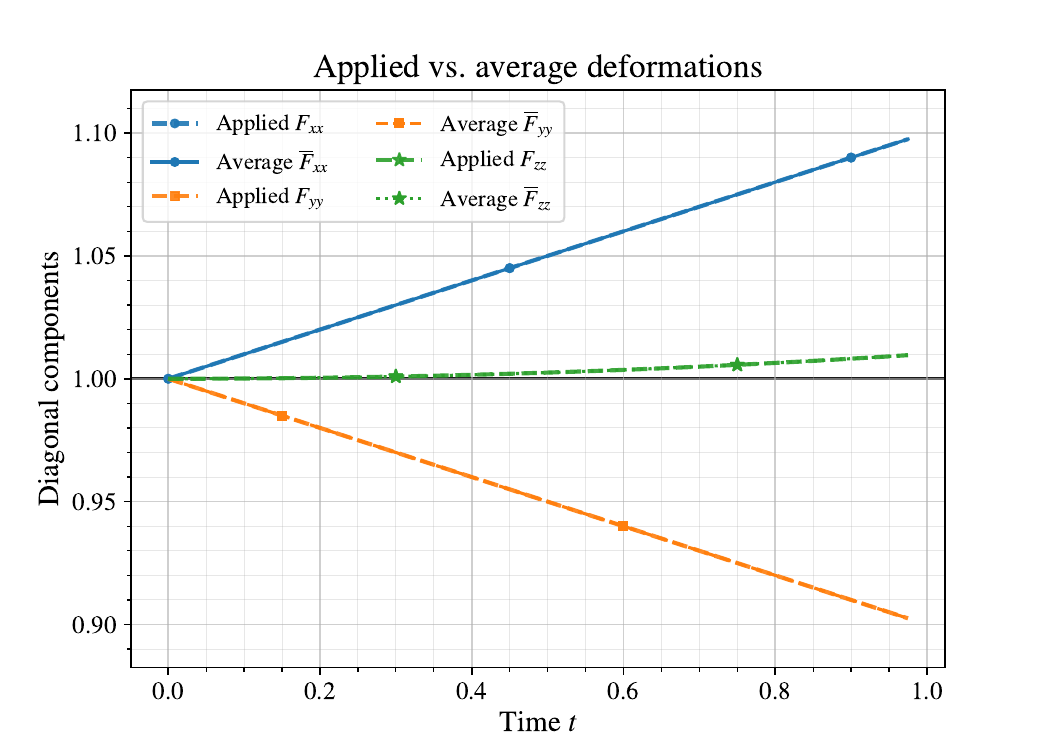}
        \caption{Homogenized deformation (average vs.\ applied)}
        \label{fig:hex_sphere_homo_def}
    \end{subfigure}\hfill
    \begin{subfigure}[t]{0.32\linewidth}
        \centering
        \vspace{0pt}
        \includegraphics[width=\linewidth,height=4.2cm,keepaspectratio]{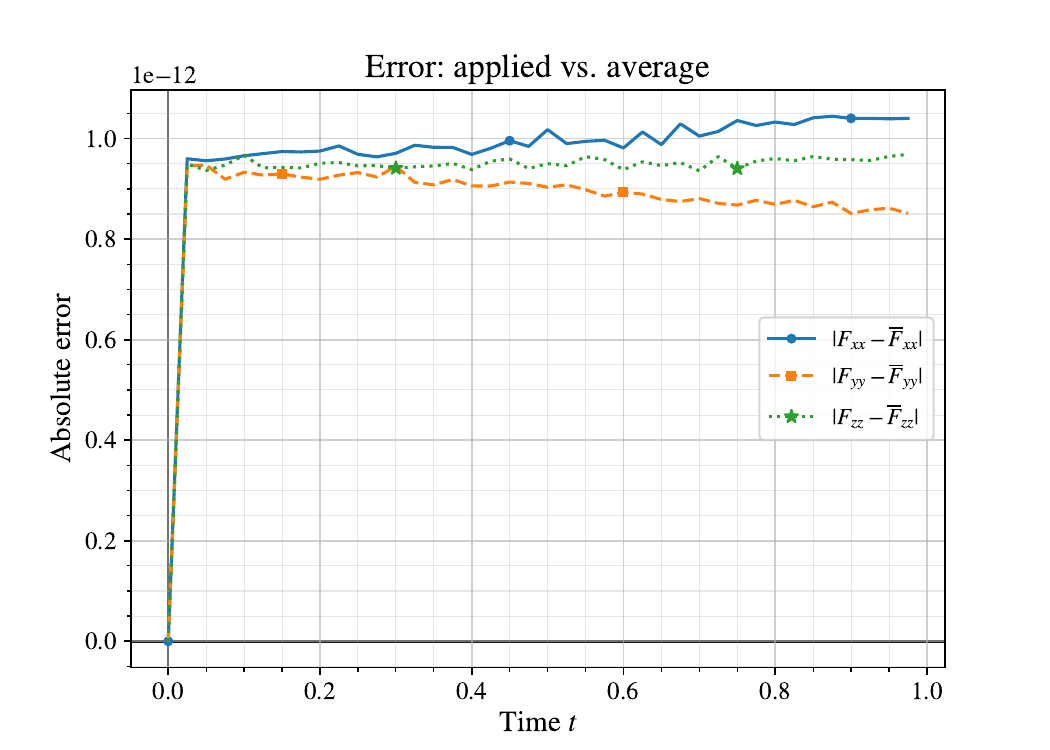}
        \caption{Deformation mismatch (average $-$ applied)}
        \label{fig:hex_sphere_homo_def_error}
    \end{subfigure}\hfill
    \begin{subfigure}[t]{0.32\linewidth}
        \centering
        \vspace{0pt}
        \includegraphics[width=\linewidth,height=4.2cm,keepaspectratio]{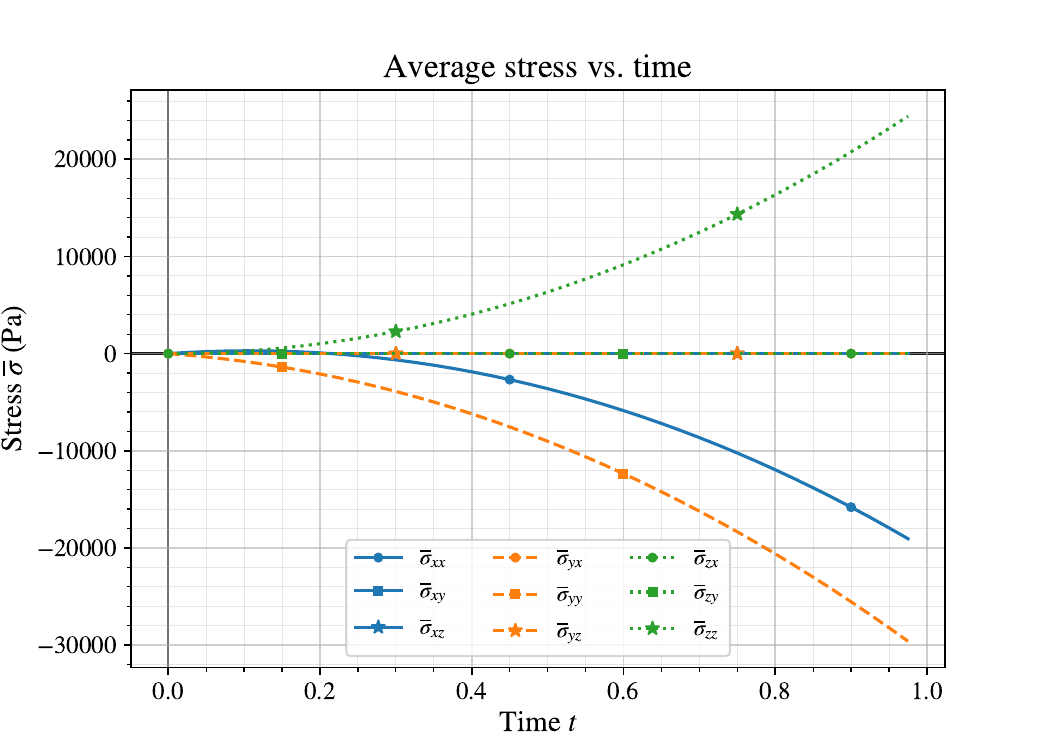}
        \caption{Homogenized stress response}
        \label{fig:hex_sphere_homo_stress}
    \end{subfigure}

    \vspace{0.6em}

    \begin{subfigure}[t]{0.32\linewidth}
        \centering
        \vspace{0pt}
        \includegraphics[width=\linewidth,height=4.2cm,keepaspectratio]{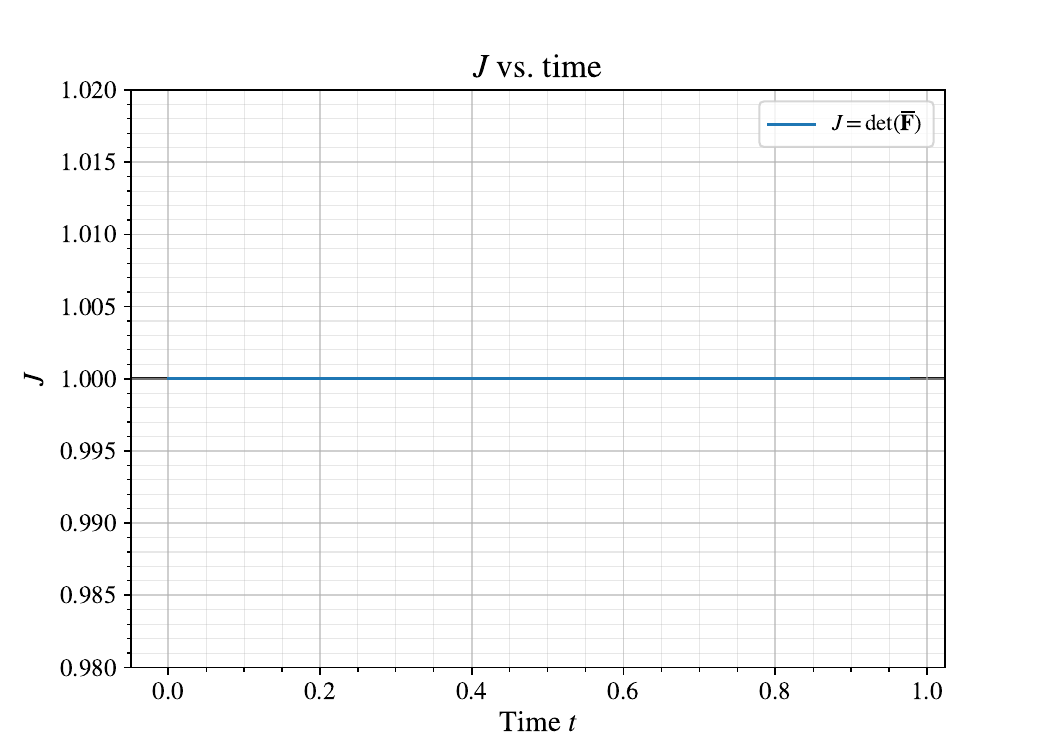}
        \caption{Homogenized Jacobian $J$}
        \label{fig:hex_sphere_homo_J}
    \end{subfigure}\hfill
    \begin{subfigure}[t]{0.32\linewidth}
        \centering
        \vspace{0pt}
        \includegraphics[width=\linewidth,height=4.2cm,keepaspectratio]{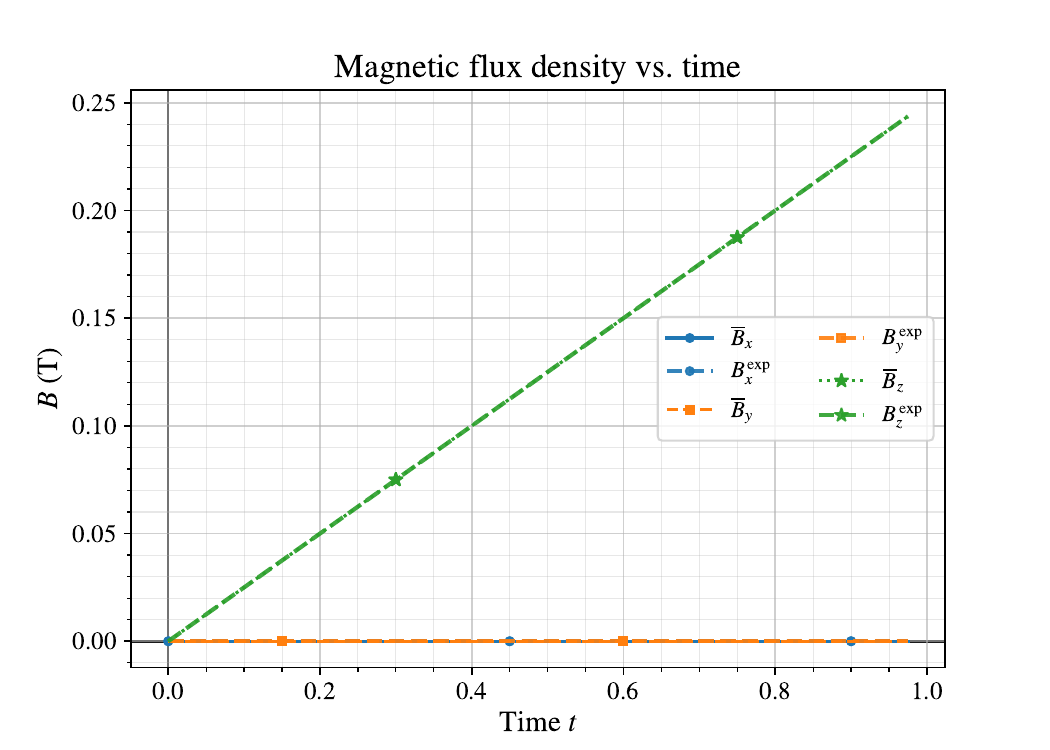}
        \caption{Homogenized magnetic induction $\bar{\mathbf{B}}$}
        \label{fig:hex_sphere_homo_B}
    \end{subfigure}\hfill
    \begin{subfigure}[t]{0.32\linewidth}
        \centering
        \vspace{0pt}
        \includegraphics[width=\linewidth,height=4.2cm,keepaspectratio]{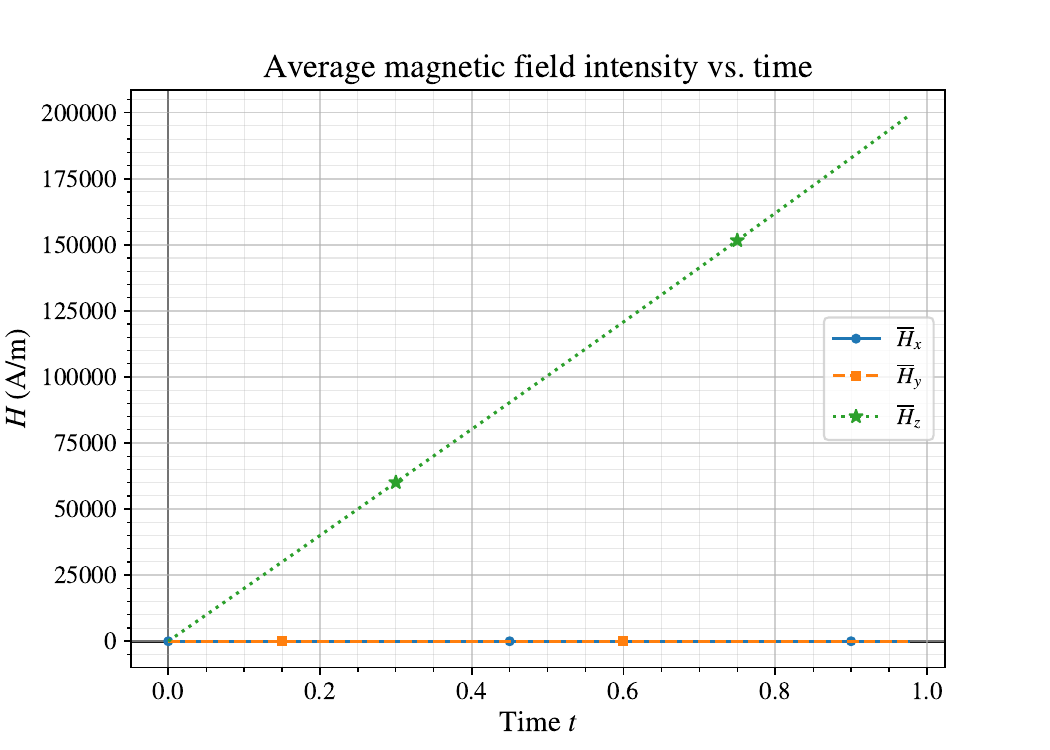}
        \caption{Homogenized magnetic field $\bar{\mathbf{H}}$}
        \label{fig:hex_sphere_homo_H}
    \end{subfigure}

    \caption{Homogenized deformation, stress, volumetric, and magnetic responses for the hex single-inclusion RVE under the prescribed magneto-mechanical loading.}
    \label{fig:hex_sphere_homos}
\end{figure}

To demonstrate the behavior of the hexahedral discretization, the hexahedral spherical-inclusion mesh was subjected to the same coupled magneto--mechanical loading conditions described above. The resulting homogenized response is shown in Figure~\ref{fig:hex_sphere_homos}.

\begin{figure}[htbp]
    \centering
    \begin{subfigure}[t]{0.32\linewidth}
        \centering
        \includegraphics[height=4.2cm]{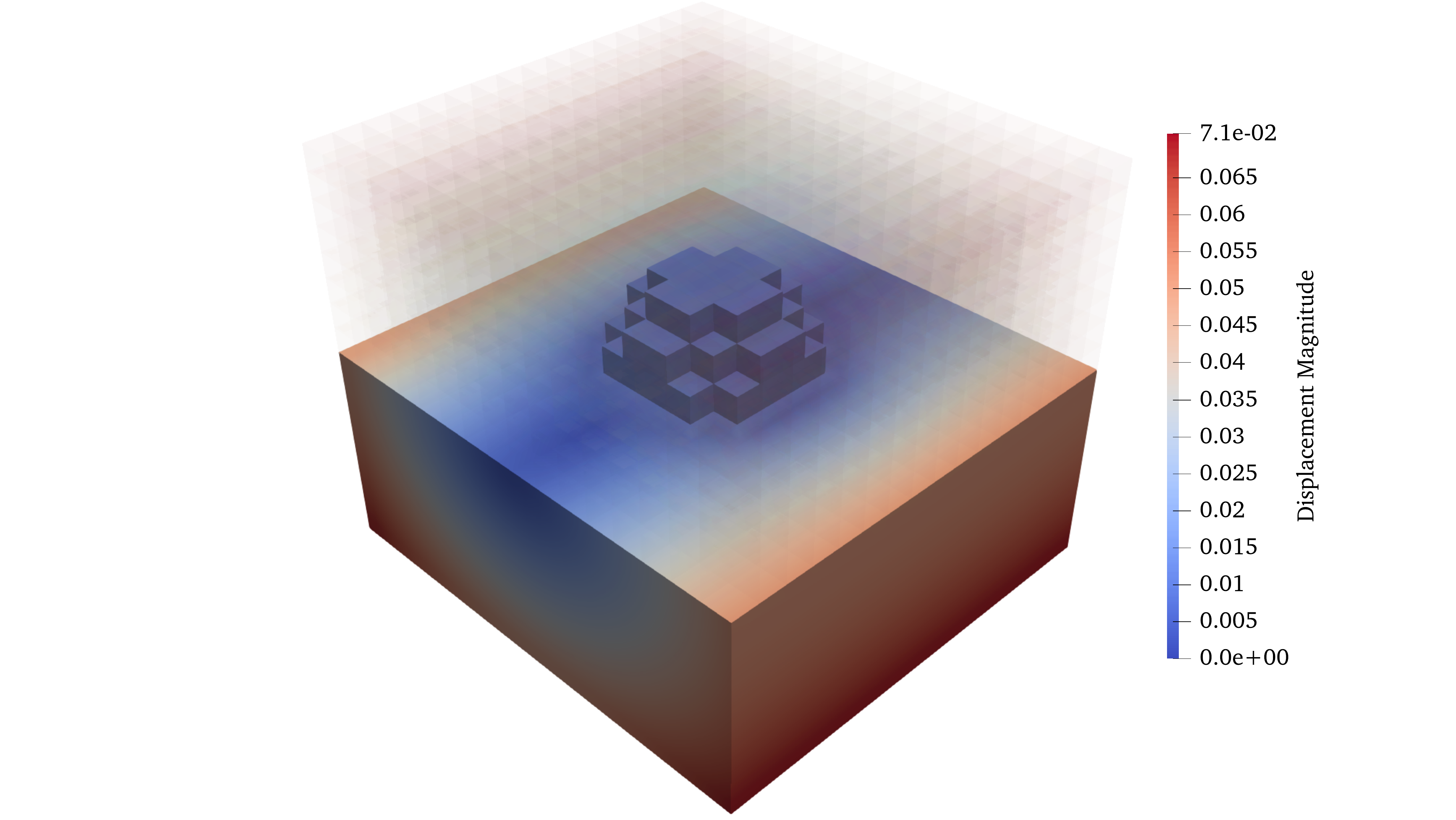}
        \caption{Displacement magnitude}
        \label{fig:hex_sphere_disp_final}
    \end{subfigure}\hfill
    \begin{subfigure}[t]{0.32\linewidth}
        \centering
        \includegraphics[height=4.2cm]{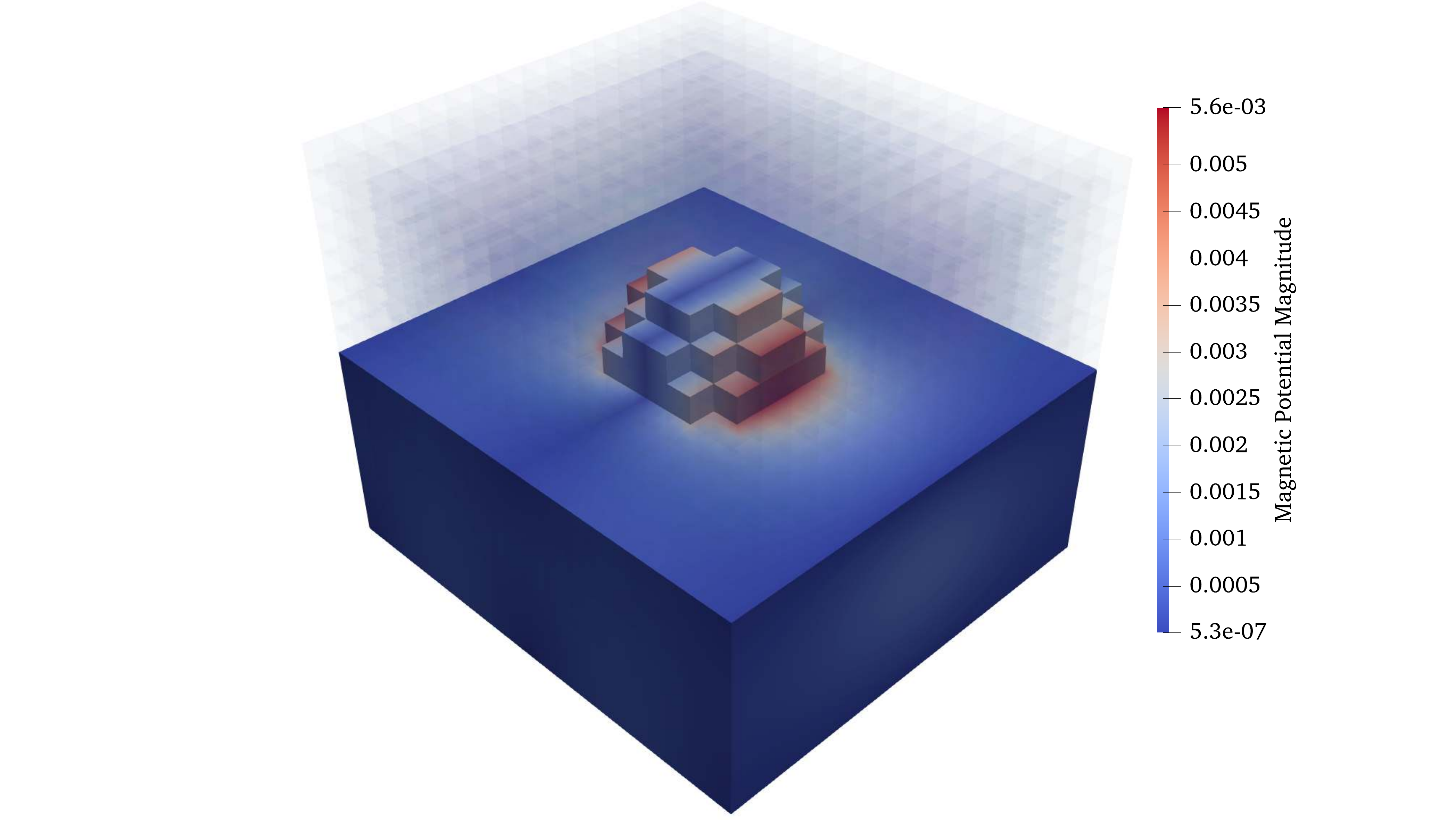}
        \caption{Magnetic potential}
        \label{fig:hex_sphere_magpot_final}
    \end{subfigure}\hfill
    \begin{subfigure}[t]{0.32\linewidth}
        \centering
        \includegraphics[height=4.2cm]{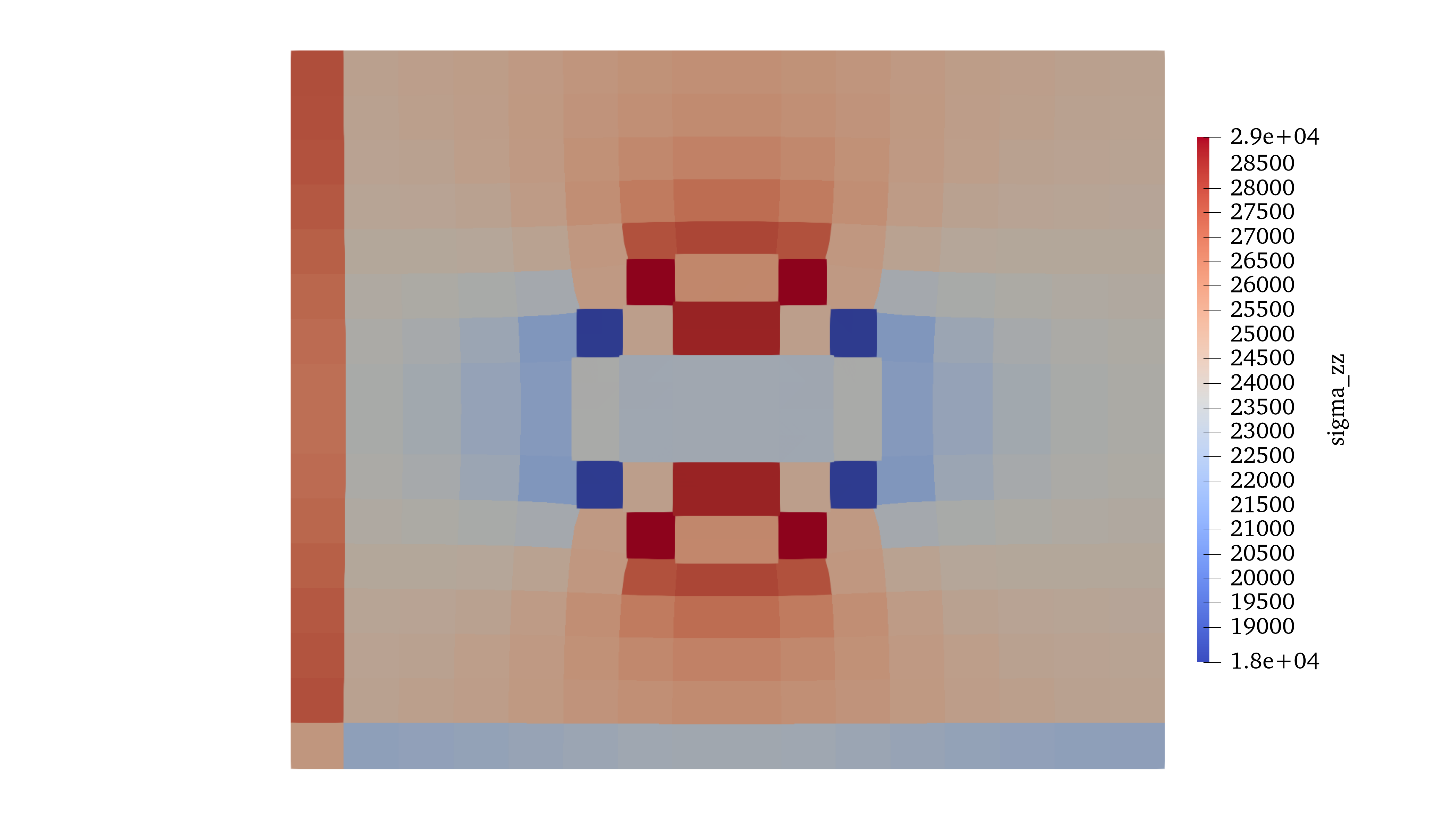}
        \caption{Final $\sigma_{zz}$}
        \label{fig:hex_sphere_sigmazz_final}
    \end{subfigure}

    \caption{Spherical inclusion (hex mesh): displacement magnitude, magnetic potential, and final $\sigma_{zz}$ under the prescribed magneto-mechanical loading.}
    \label{fig:hex_sphere_images}
\end{figure}

Additionally, Figure~\ref{fig:hex_sphere_images} shows the displacement magnitude, magnetic potential, and final $\sigma_{zz}$ field for the hexahedral spherical-inclusion simulation.

This example is included to illustrate the response obtained with the hexahedral discretization rather than to provide a direct comparison with the tetrahedral results. In the present three-dimensional setting, achieving comparable convergence with the hexahedral formulation would require substantially greater mesh refinement, which is computationally very expensive. Accordingly, the hexahedral results are presented here as a qualitative demonstration of the formulation and its behavior under coupled loading.

\subsection{Particle Chains}
\label{sec:particle_chains}

To highlight microstructure-induced interactions beyond a single inclusion, we next consider RVEs containing short chains of
spherical particles embedded in the elastomeric matrix. This configuration introduces strong particle--particle coupling and field
localization in the inter-particle ligaments, providing a sensitive test of the fully coupled formulation under heterogeneous
loading. In contrast to the simple inclusion benchmark, all particle-chain examples are discretized exclusively with unstructured
tetrahedral meshes to accommodate the narrow gaps between neighboring particles and to enable targeted local refinement in regions
of high deformation and magnetic-field gradients.

Periodic boundary conditions are enforced on all faces for both the displacement fluctuation and the magnetic vector potential, and
a minimal anchoring/gauge constraint is applied to remove the rigid-body and magnetic null modes. Homogenized quantities
$(\overline{\mathbf{\sigma}},\overline{\mathbf{H}})$ are extracted using the same volume-averaging procedure described in
Section~\ref{sec:homogenization}. The particle-chain cases are used to (i) demonstrate the emergence of anisotropic effective
behavior due to chain alignment, (ii) visualize localization of mechanical and magnetic fields near particle contacts, and (iii) assess the robustness of the tetrahedral simplex implementation for geometries that are difficult to represent with structured
hexahedral meshes.

\subsubsection{Mesh}
\label{sec:particle_chains_mesh}

\begin{figure}[htbp]
  \centering
  \begin{subfigure}[b]{0.48\linewidth}
    \centering
    \includegraphics[width=\linewidth]{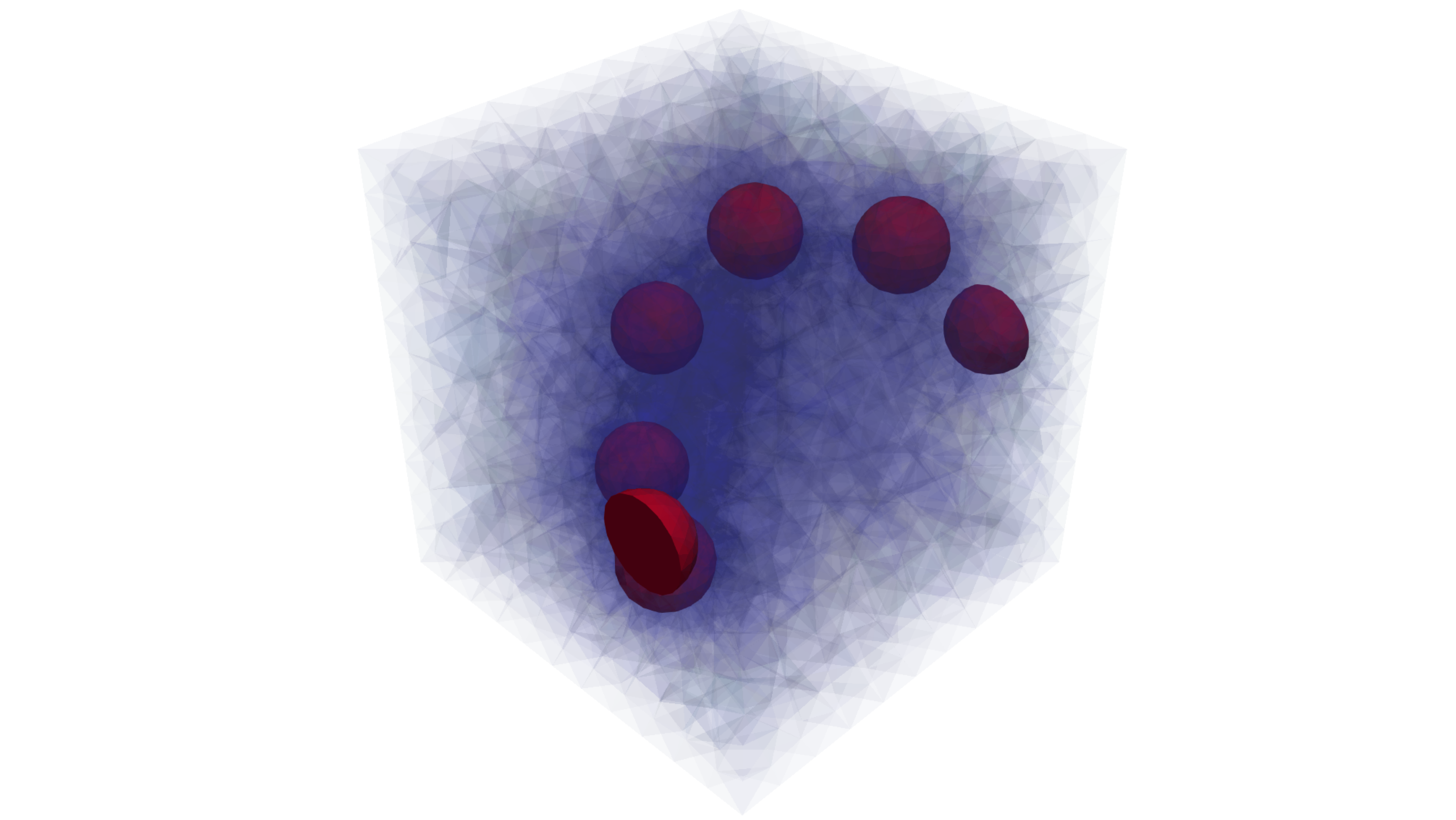}
    \caption{Particle Chain Corner View}
    \label{fig:PC_mesh}
  \end{subfigure}
  \hfill
  \begin{subfigure}[b]{0.48\linewidth}
    \centering
    \includegraphics[width=\linewidth]{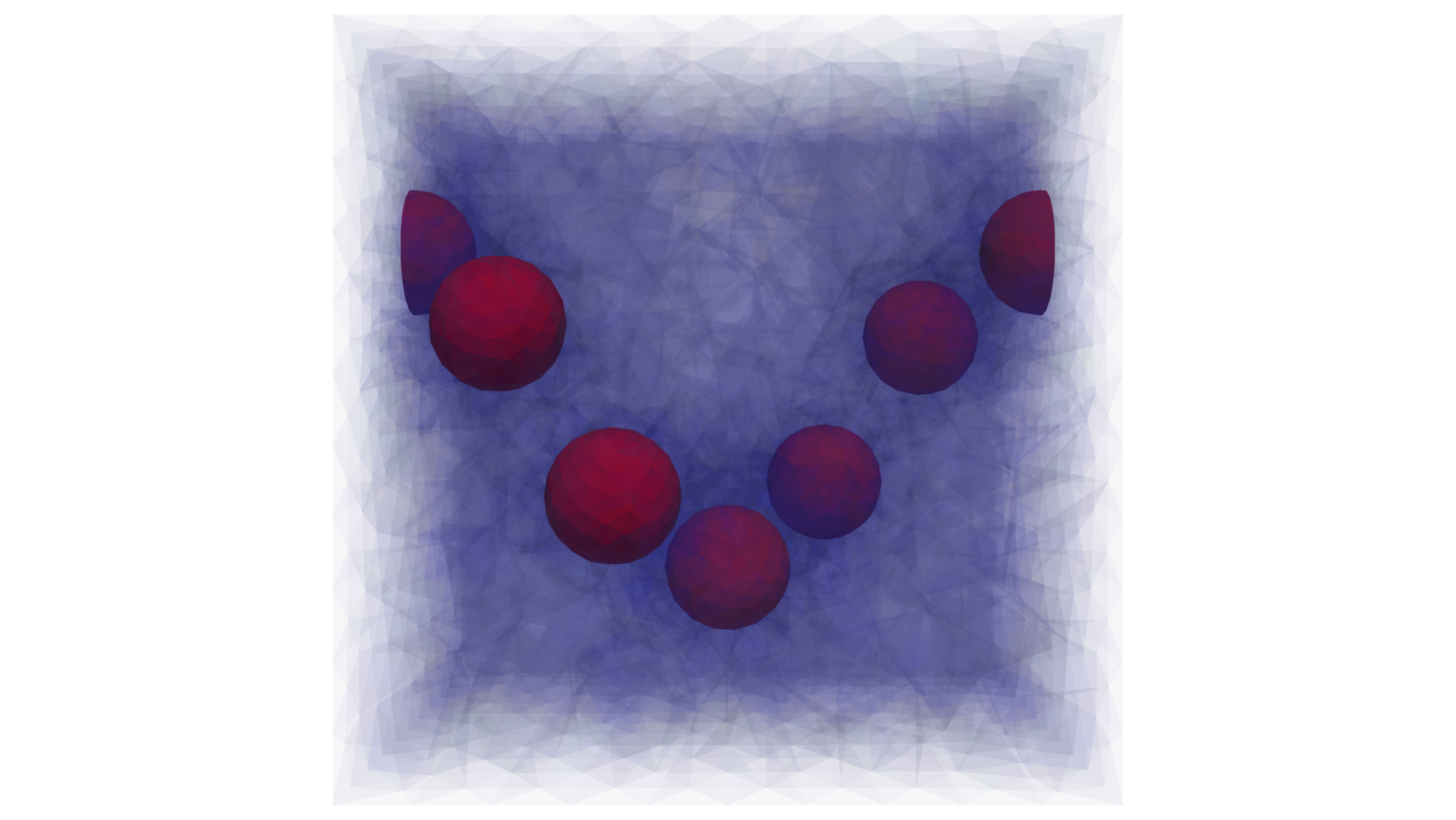}
    \caption{Particle Chain Side View}
    \label{fig:PC_mesh_side}
  \end{subfigure}
  \caption{Particle Chain Mesh}
  \label{fig:PC_mesh_full}
\end{figure}

The particle chain consists of six particles, each with radius $r=0.1$. The chain is generated centered about the $Z$-axis, with each
successive particle positioned such that the line connecting neighboring particle centers rotates by $60^\circ$ relative to its predecessor, producing a helical chain morphology. The tetrahedral mesh used for this example is shown in Figure~\ref{fig:PC_mesh_full}.

\subsubsection{Homogenized Responses}
\label{sec:particle_chains_homo_responses}

\paragraph{Mechanical loading:}
The particle-chain RVE is first subjected to the same purely mechanical loading path introduced in Paragraph~\ref{par:sphere_mech_loading}. In this case, the macroscopic deformation gradient $\overline{\mathbf{F}}(t)$ is prescribed while the macroscopic magnetic induction is set to $\overline{\mathbf{B}}=\mathbf{0}$. The resulting response is therefore driven entirely by the imposed mechanical deformation, and the homogenized behavior is characterized primarily through the macroscopic Cauchy stress $\overline{\boldsymbol{\sigma}}$. Relative to the single-inclusion case, the chain morphology introduces stronger microstructural anisotropy and particle--particle interactions, which lead to more localized deformation within the ligaments between neighboring particles and alter the effective stress response.

\begin{figure}[htbp]
  \centering

  \begin{subfigure}[t]{0.32\linewidth}
    \centering
    \includegraphics[width=\linewidth]{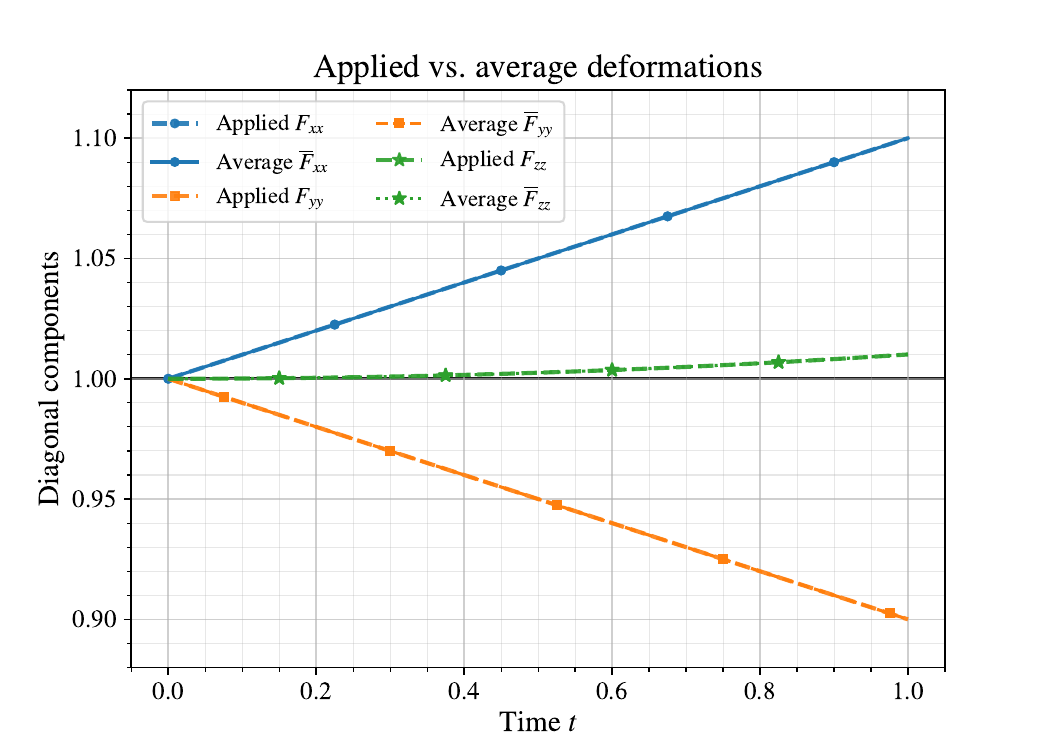}
    \caption{Homogenized deformation (average vs.\ applied)}
    \label{fig:pc_js_plots_homo_def}
  \end{subfigure}\hfill
  \begin{subfigure}[t]{0.32\linewidth}
    \centering
    \includegraphics[width=\linewidth]{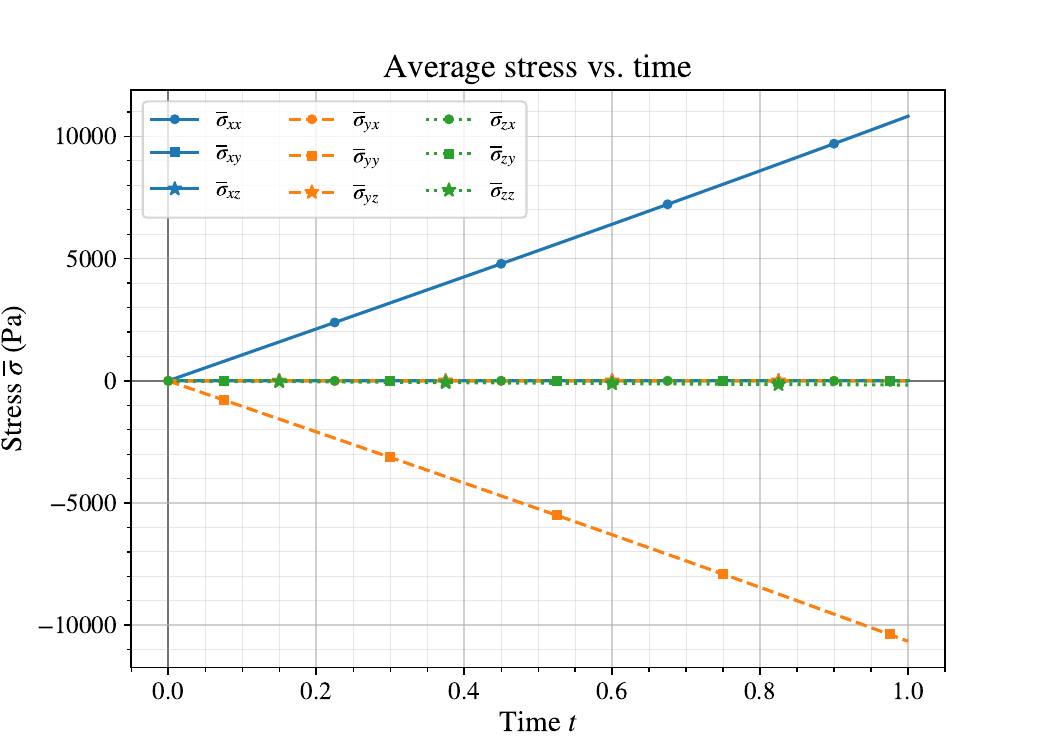}
    \caption{Homogenized stress response}
    \label{fig:pc_js_plots_homo_stress}
  \end{subfigure}\hfill
  \begin{subfigure}[t]{0.32\linewidth}
    \centering
    \includegraphics[width=\linewidth]{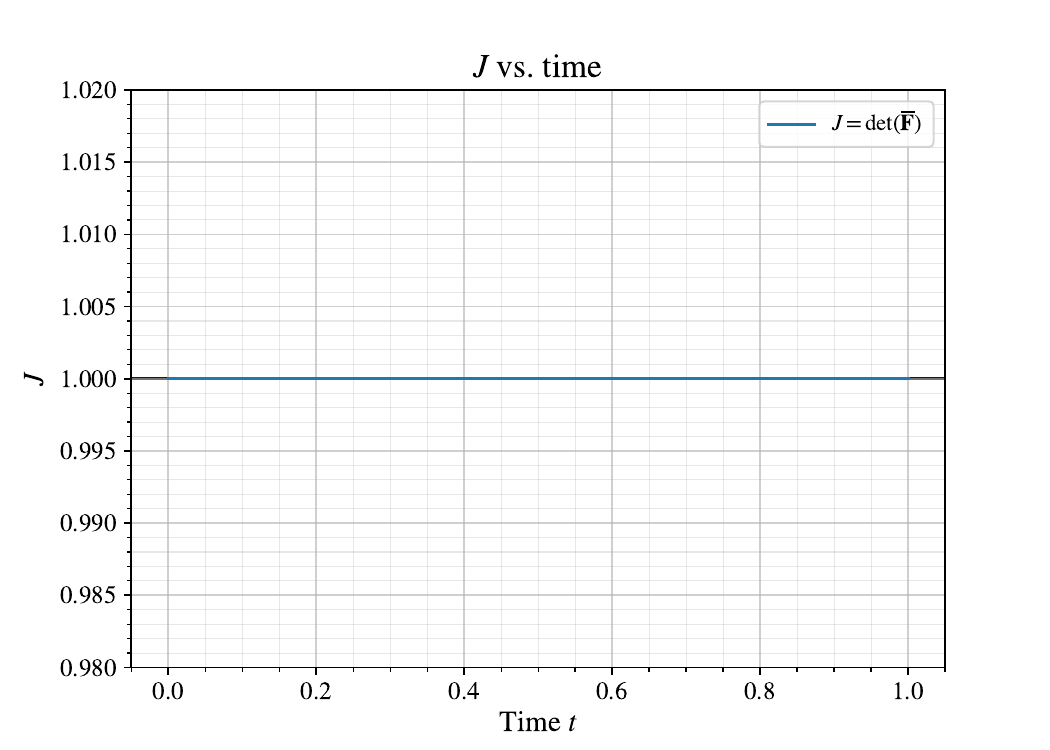}
    \caption{Homogenized Jacobian $J$ (volume change)}
    \label{fig:pc_js_plots_homo_J}
  \end{subfigure}

  \caption{Homogenized quantities for the particle chain under strain loading.}
  \label{fig:pc_js_homogenized_quantities}
\end{figure}

\begin{figure}[htbp]
    \centering
    \begin{subfigure}[b]{0.49\linewidth}
        \centering
        \includegraphics[width=\linewidth]{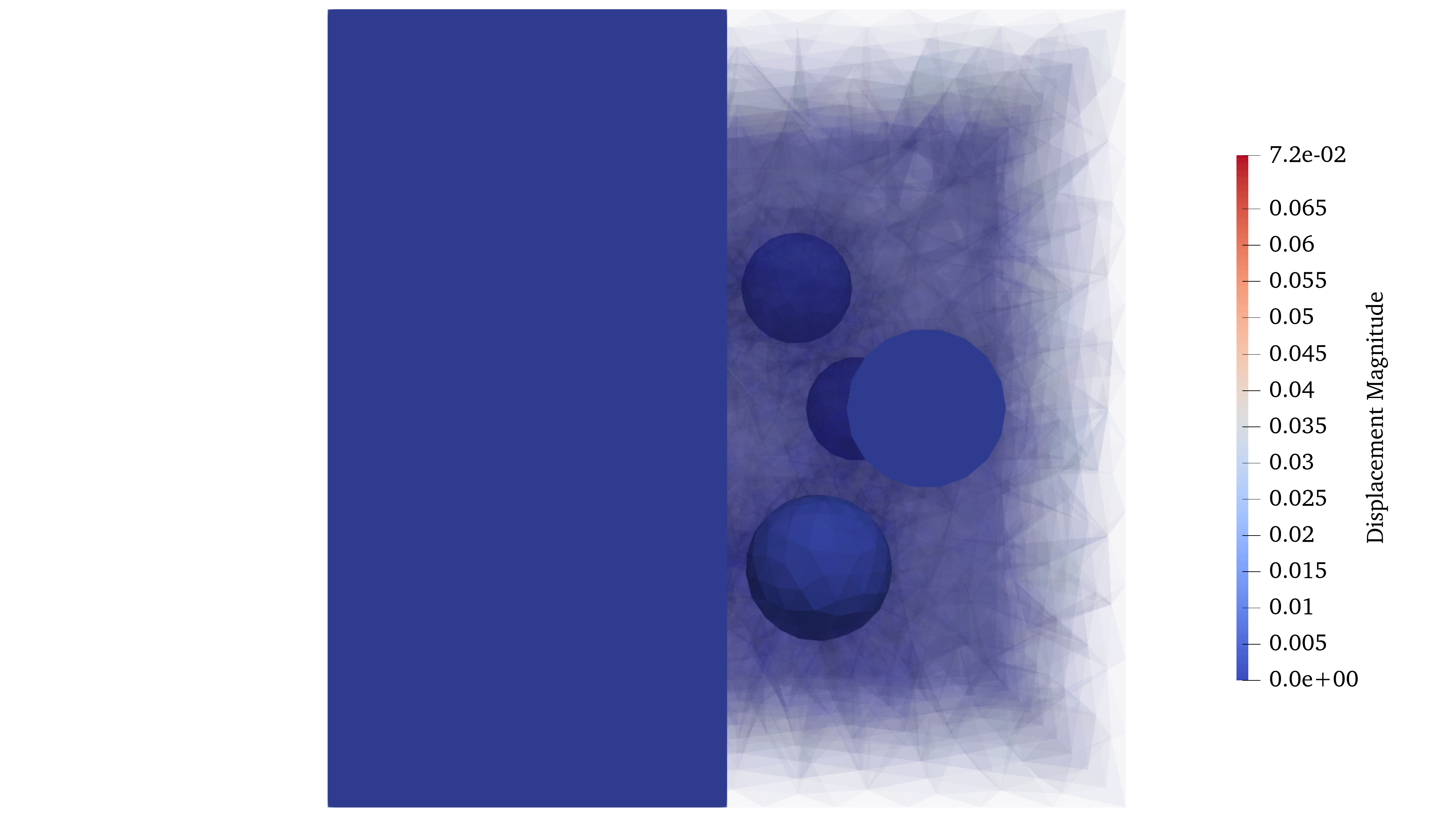}
        \caption{Initial state.}
        \label{fig:pc_js_initial}
    \end{subfigure}\hfill
    \begin{subfigure}[b]{0.49\linewidth}
        \centering
        \includegraphics[width=\linewidth]{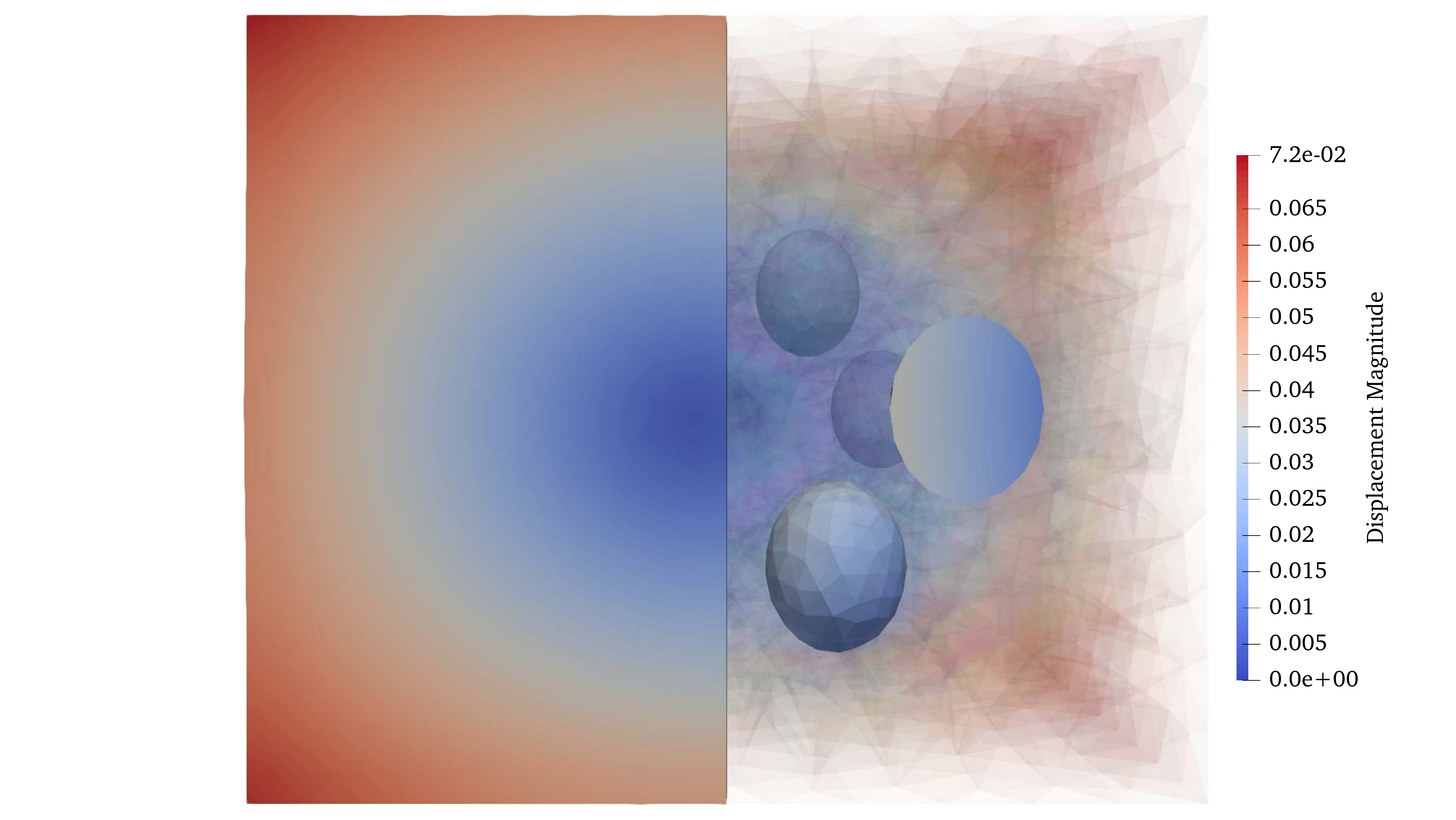}
        \caption{Final state.}
        \label{fig:pc_js_final}
    \end{subfigure}

    \caption{Particle Chain: initial and final configurations under the prescribed deformation.}
    \label{fig:pc_js_def}
\end{figure}

The particle-chain RVE exhibits a macroscopic strain response that is very similar to that of the single spherical-inclusion case. This similarity is consistent with the fact that both configurations have the same constituent material properties and approximately the same particle volume fraction, which largely control the overall effective stiffness. However, while the averaged response remains similar, the chain morphology is still expected to modify the local deformation patterns and stress localization, particularly in the narrow ligaments between neighboring particles. This effect is clearly observed in Figure~\ref{fig:pc_js_def}.

\paragraph{Magnetic loading:}
The same purely magnetic loading path used for the single-inclusion case is applied to the particle-chain RVE. Specifically, the macroscopic magnetic induction $\overline{\mathbf{B}}(t)$ is prescribed while the macroscopic deformation is held fixed at $\overline{\mathbf{F}}=\mathbf{I}$. The homogenized response is reported in terms of the macroscopic magnetic field $\overline{\mathbf{H}}$ and the induced macroscopic stress $\overline{\boldsymbol{\sigma}}$. Compared with the single-inclusion benchmark, the chain geometry is expected to enhance particle--particle interactions and to produce more localized magnetic and mechanical fields within the RVE.

\begin{figure}[htbp]
  \centering

  \begin{subfigure}[t]{0.32\linewidth}
    \centering
    \includegraphics[width=\linewidth]{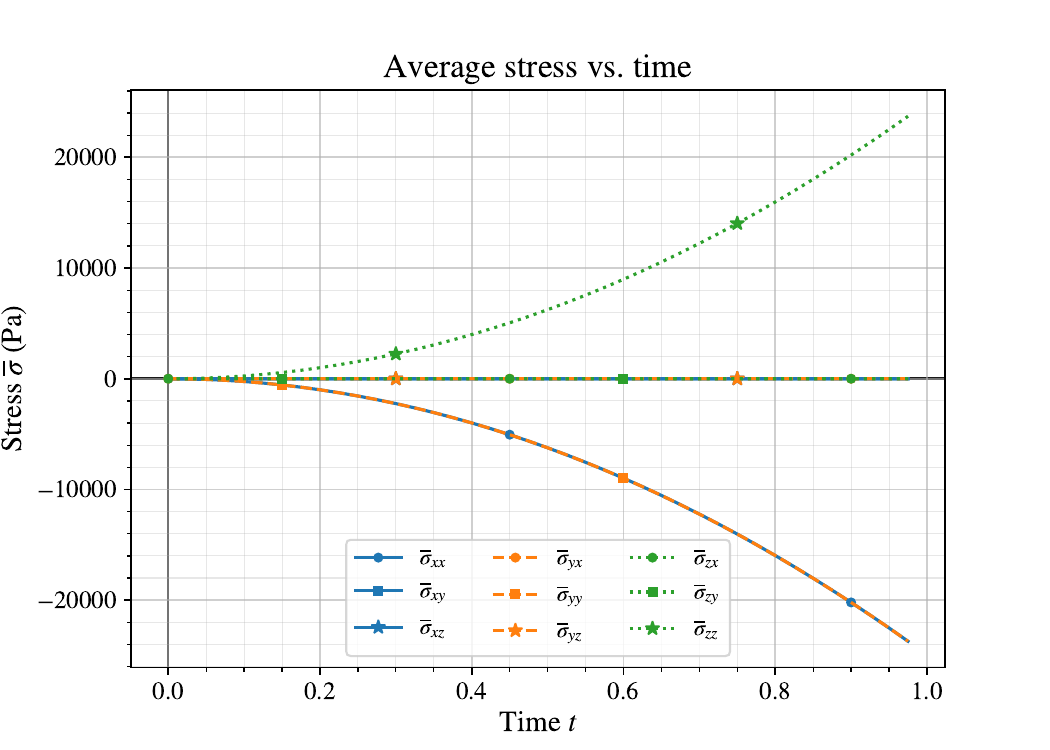}
    \caption{Homogenized stress response}
    \label{fig:pc_jm_plots_homo_stress}
  \end{subfigure}\hfill
  \begin{subfigure}[t]{0.32\linewidth}
    \centering
    \includegraphics[width=\linewidth]{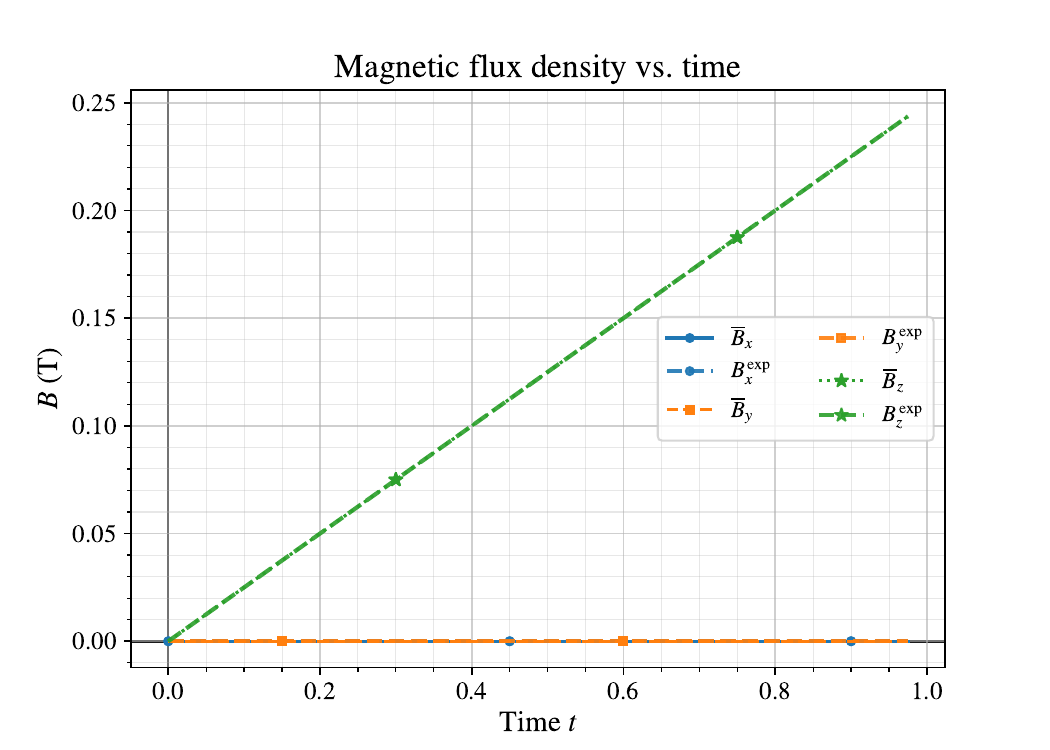}
    \caption{Homogenized magnetic induction $\bar{\mathbf{B}}$ vs.\ time}
    \label{fig:pc_jm_plots_homo_B}
  \end{subfigure}\hfill
  \begin{subfigure}[t]{0.32\linewidth}
    \centering
    \includegraphics[width=\linewidth]{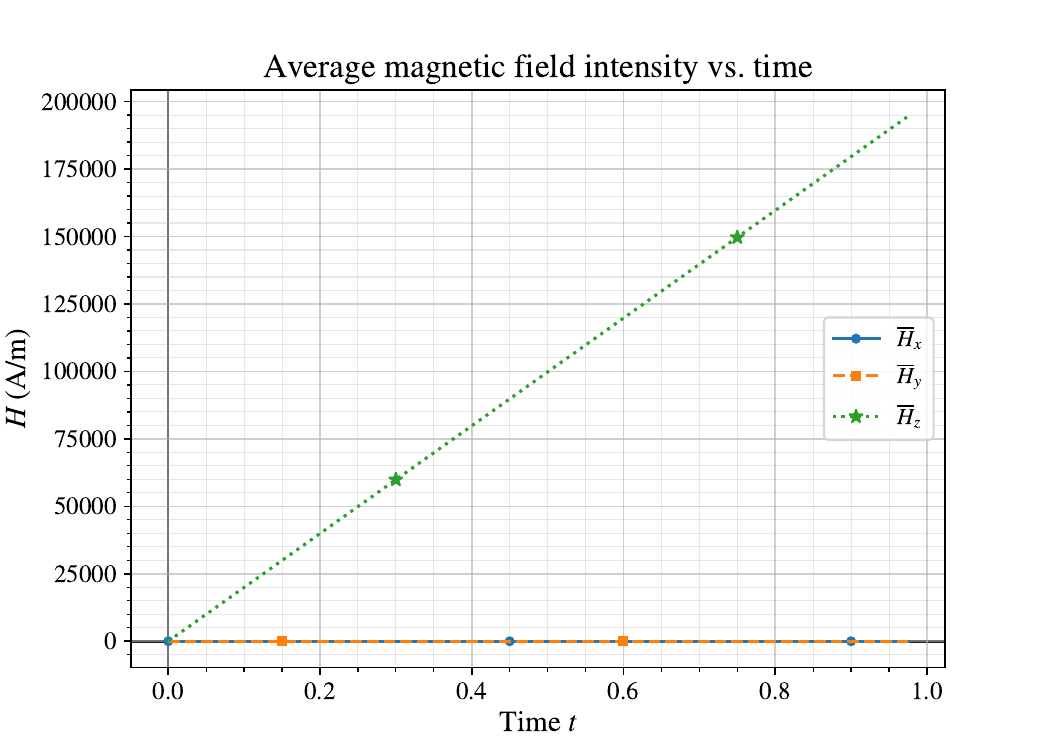}
    \caption{Homogenized magnetic field $\bar{\mathbf{H}}$ vs.\ time}
    \label{fig:pc_jm_plots_homo_H}
  \end{subfigure}

  \caption{Homogenized quantities for the particle chain under magnetic loading.}
  \label{fig:pc_jm_homogenized_quantities}
\end{figure}

\begin{figure}[htbp]
    \centering
    \begin{subfigure}[b]{0.49\linewidth}
        \centering
        \includegraphics[width=\linewidth]{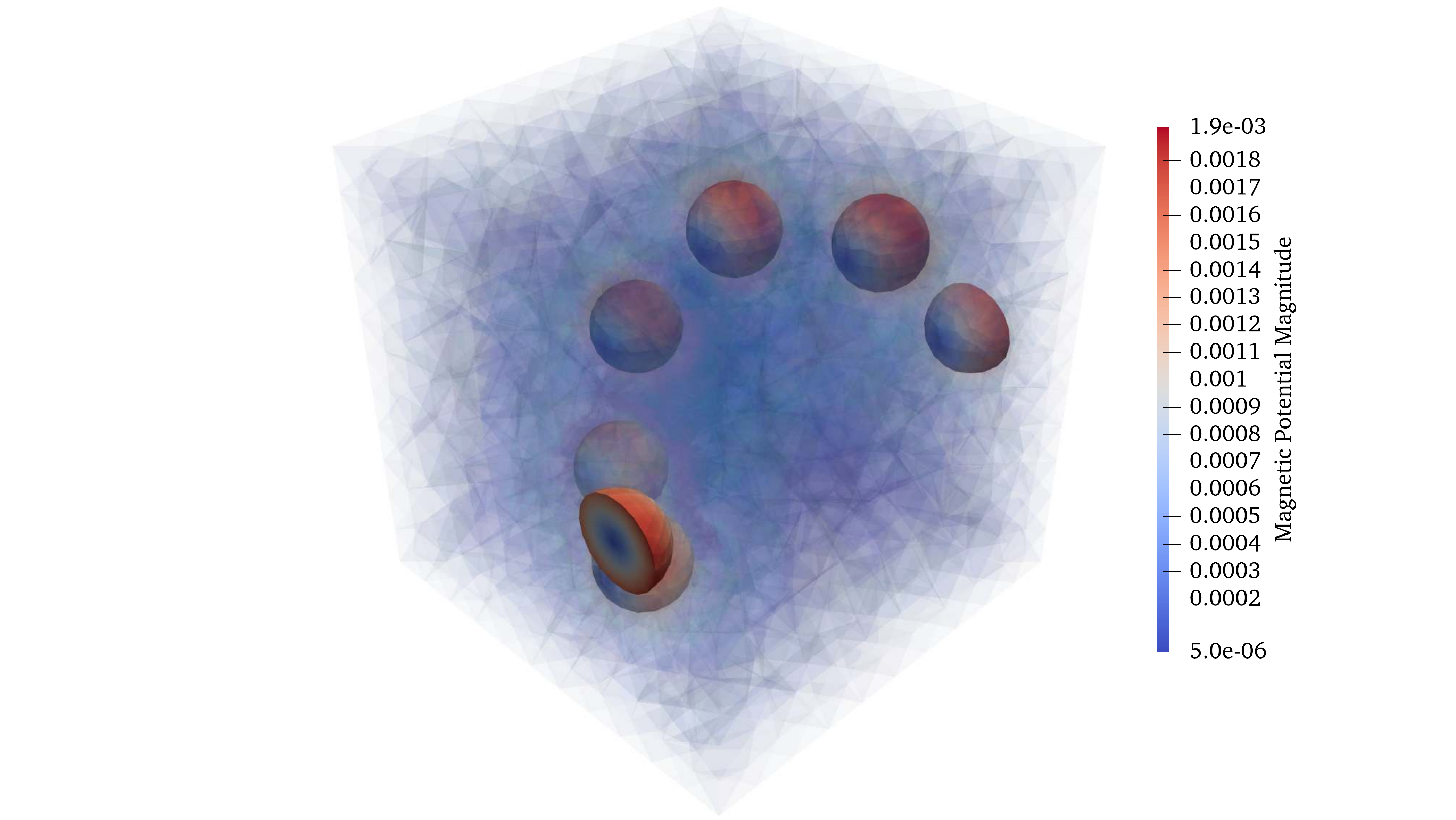}
        \caption{Magnetic Potential}
        \label{fig:pc_jm_mag_pot}
    \end{subfigure}\hfill
    \begin{subfigure}[b]{0.49\linewidth}
        \centering
        \includegraphics[width=\linewidth]{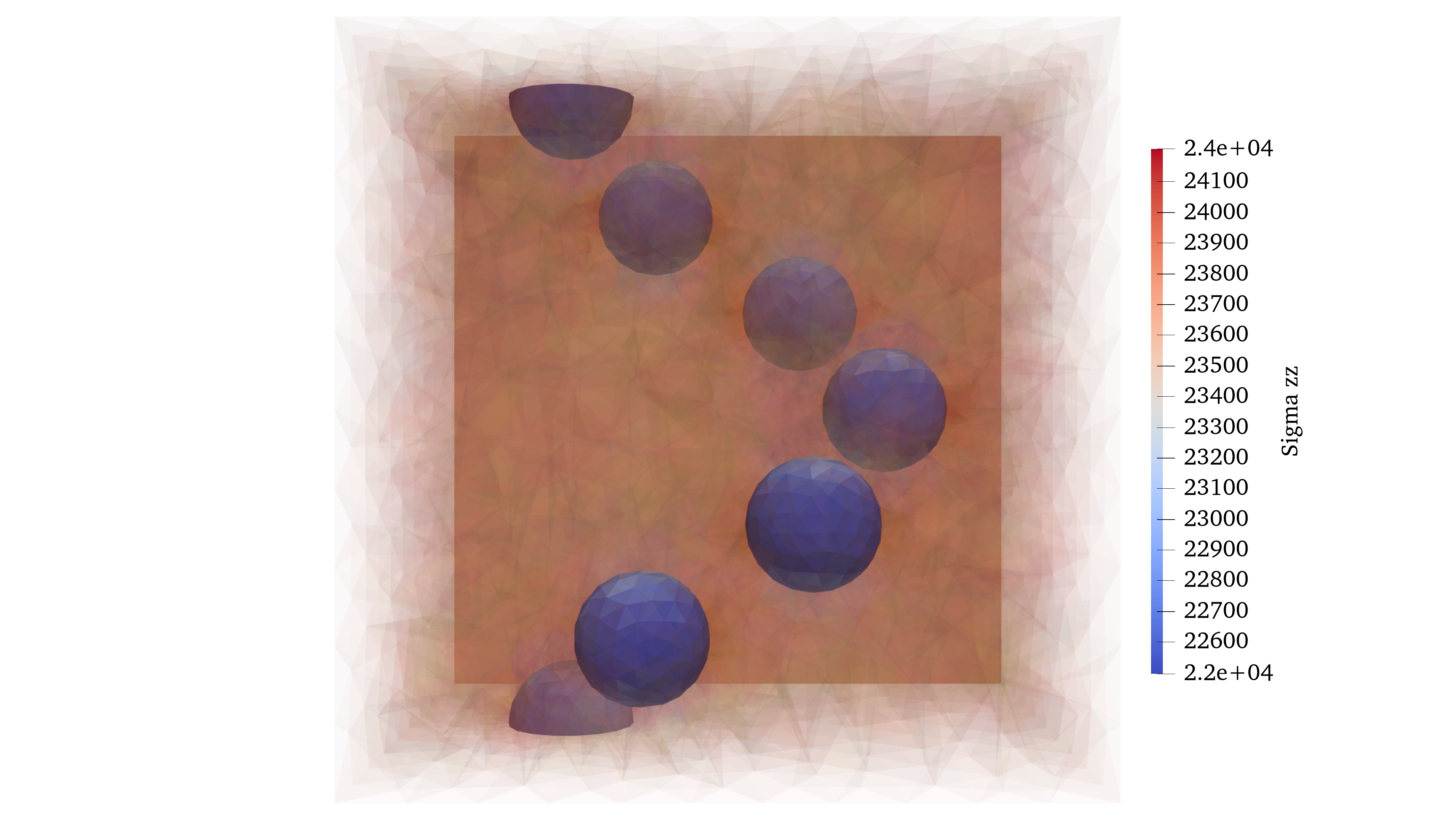}
        \caption{$\sigma_{zz}$}
        \label{fig:pc_js_sig_zz}
    \end{subfigure}

    \caption{Particle Chain: Magnetic potential and $\sigma_{zz}$ under magnetic loading.}
    \label{fig:pc_jm_mag_and_sig}
\end{figure}

Like the strain response, the macroscopic magnetic response of the particle-chain RVE is very similar to that of the spherical-inclusion case. This similarity is expected because both configurations share the same constituent material properties and nearly the same particle volume fraction. However, the microscopic response differs substantially. In particular, the chain geometry produces a markedly different spatial distribution of magnetic potential and stress, as shown in Figure~\ref{fig:pc_jm_mag_and_sig}.

\begin{figure}[htbp]
  \centering

  \begin{subfigure}[b]{0.40\linewidth}
    \centering
    \includegraphics[width=\linewidth,height=4.4cm,keepaspectratio]{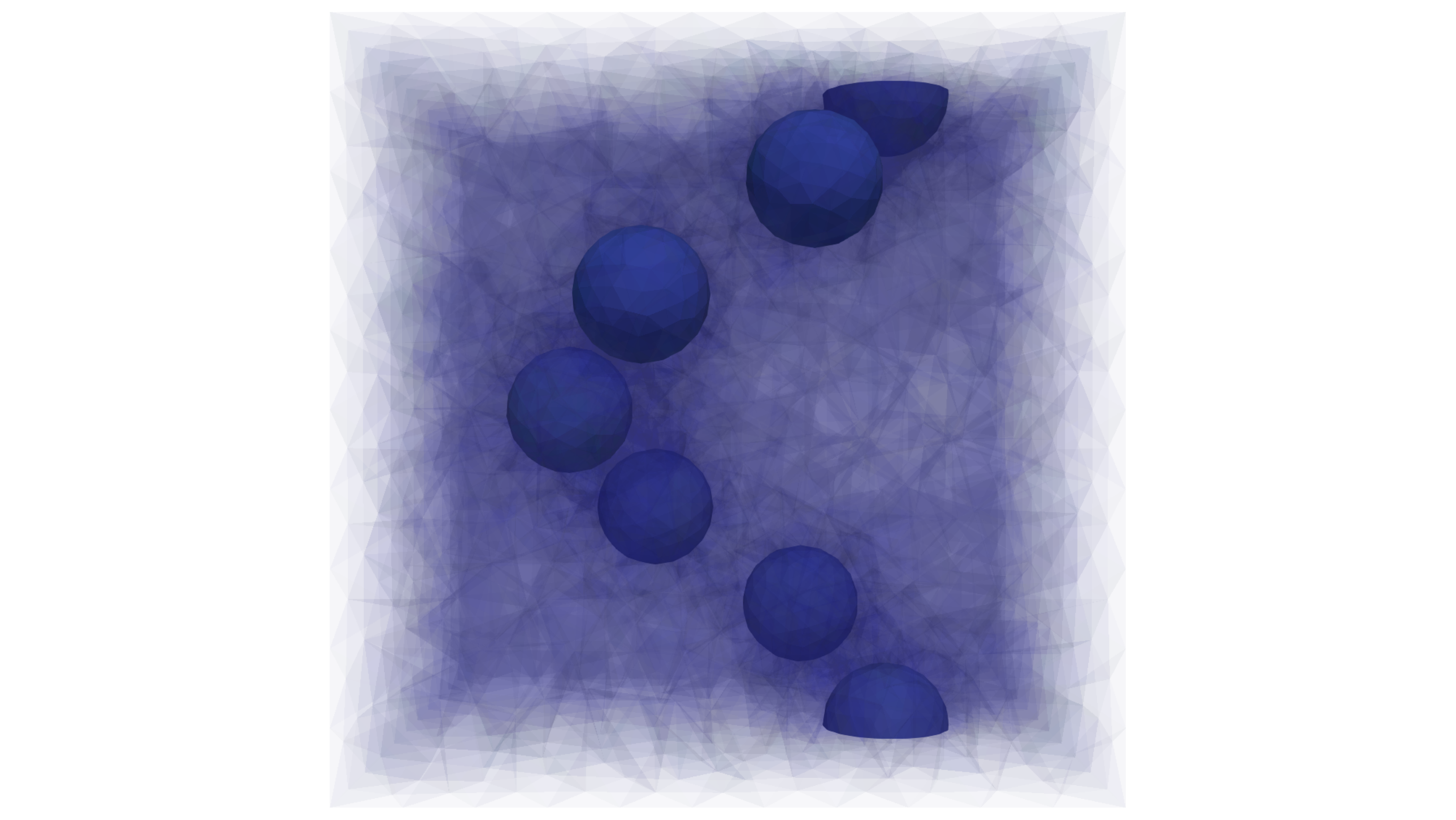}
    \caption{Initial configuration.}
    \label{fig:PC_move_t0}
  \end{subfigure}
  \hspace{0.01\linewidth}
  \begin{subfigure}[b]{0.40\linewidth}
    \centering
    \includegraphics[width=\linewidth,height=4.4cm,keepaspectratio]{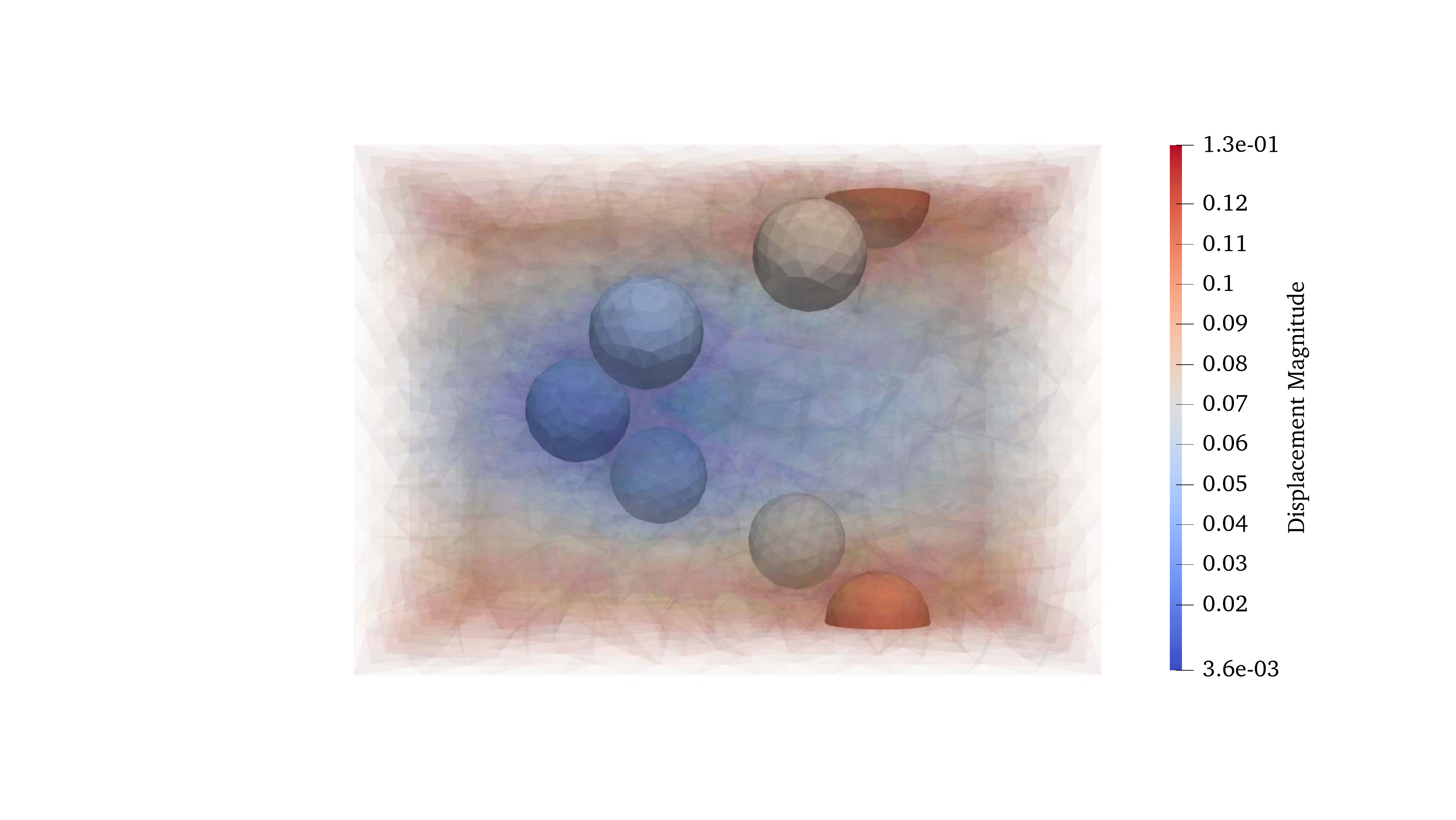}
    \caption{Final configuration.}
    \label{fig:PC_move_tfinal}
  \end{subfigure}

  \caption{Initial and final particle-chain configurations under stress-relaxed boundary conditions.}
  \label{fig:PC_move_displacement}
\end{figure}

The microscopic response remains more complex when particle motion is permitted through the stress-relaxed boundary conditions, although the corrected physical displacement field indicates a more structured and cooperative response than initially suggested. In contrast to the earlier interpretation based on the uncorrected displacement field, the particles do not primarily exhibit pronounced elongation, large rigid-body rotation, or substantial translation toward one another. Instead, the periodic particle assembly undergoes a predominantly symmetric rearrangement along the direction of the applied magnetic flux. In particular, the particle centroids move toward the central plane of the chain in the \(z\)-direction, with the magnitude of this motion increasing with distance from the chain midpoint. Thus, the dominant effect is a cooperative field-aligned approach of the periodic particle structure, while the surrounding matrix deforms compatibly to accommodate this motion. Small accompanying \(x\)- and \(y\)-components are also present, but these remain secondary and are best interpreted as local compatibility-driven adjustments of the periodic microstructure rather than the primary mode of particle rearrangement. This behavior is illustrated in Figure~\ref{fig:PC_move_displacement}.

\begin{figure}[htbp]
    \centering
    \includegraphics[width=0.5\linewidth]{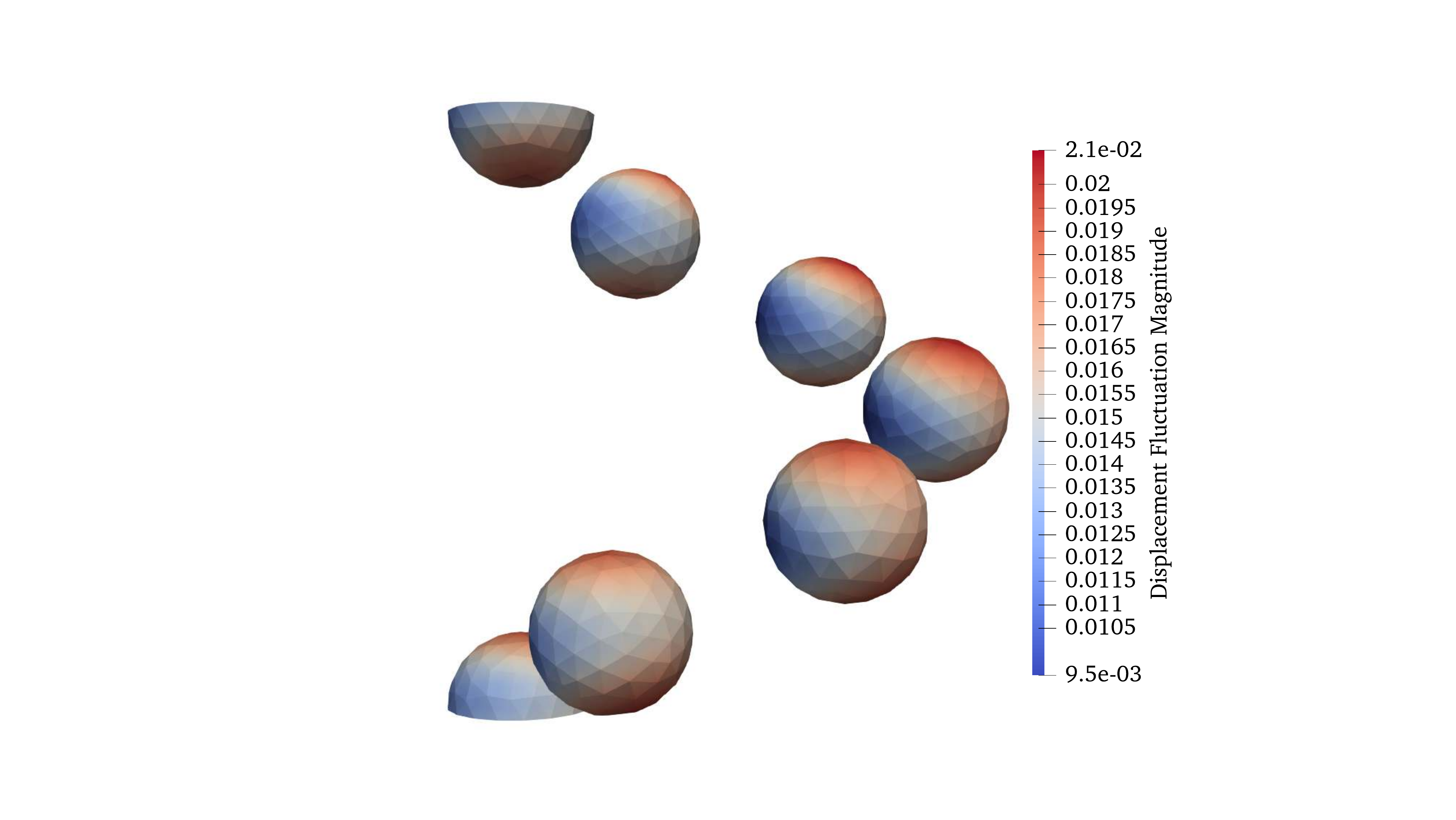}
    \caption{Displacement fluctuation field for the particle-chain RVE under stress-relaxed boundary conditions.}
    \label{fig:pc_move_fluctuation}
\end{figure}

\begin{figure}[htbp]
    \centering
    \includegraphics[width=0.95\linewidth]{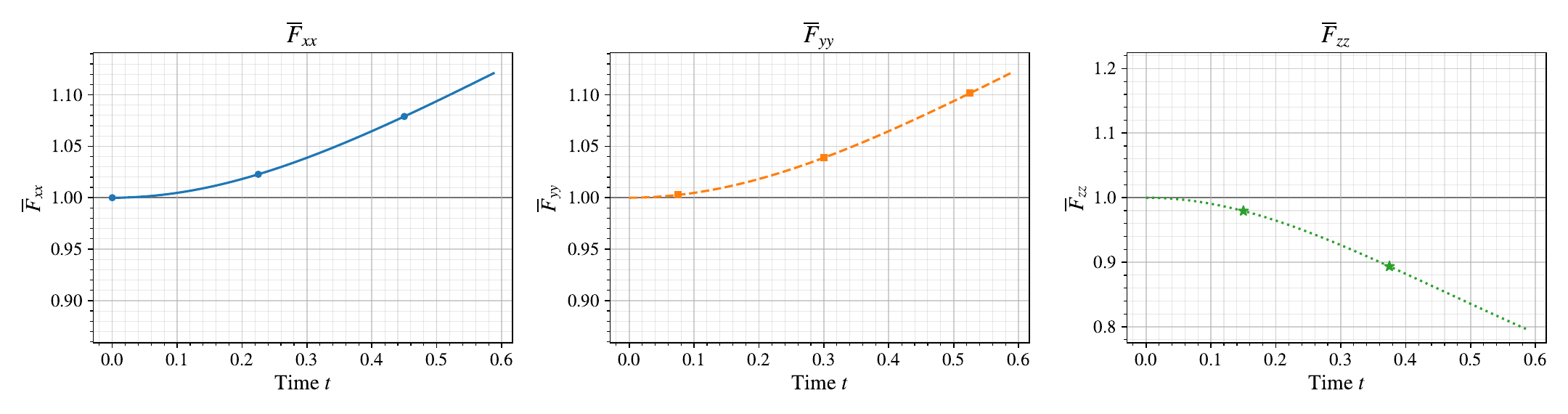}
    \caption{Homogenized deformation response under stress-relaxed boundary conditions, with the residual stress tolerance set to $10^{-3}$ and the loading applied up to approximately 60\% of the full prescribed value.}
    \label{fig:pc_ns_def}
\end{figure}

Because the RVE is periodic, this motion should not be interpreted as the behavior of isolated particles in free space. Rather, it represents the local kinematics of a repeating chain-like microstructure, in which the boundary particles are continuations of neighboring periodic images. In that sense, the observed response may be interpreted on a larger scale as a relative approach of neighboring particles along the loading direction, rather than as a simple local shape change confined to a single cell. To better highlight the role of the surrounding inclusions, Figure~\ref{fig:pc_move_fluctuation} shows the displacement fluctuation field. Since this quantity removes the affine macroscopic contribution, it more clearly exposes the spatial variation induced by particle--particle interaction and periodic compatibility. In particular, it reveals how the local motion departs from the average homogenized deformation and how neighboring inclusions perturb one another within the chain. This interpretation is also consistent with the homogenized deformation response shown later in Figure~\ref{fig:pc_ns_def}, which reflects the same field-direction contraction at the macroscale.

Additionally, the homogenized deformation gradient obtained from the stress-relaxed simulations is shown in Figure~\ref{fig:pc_ns_def}. Owing to the computational cost of these simulations, the loading was applied only up to approximately 60\% of the full prescribed value. The homogenized response is consistent with the local kinematics observed in the RVE, namely a predominantly contractive deformation with the strongest compression aligned with the magnetic loading direction.

\paragraph{Coupled magneto--mechanical loading:}
The particle-chain RVE is finally subjected to a coupled loading path in which the macroscopic deformation gradient $\overline{\mathbf{F}}(t)$ and the macroscopic magnetic induction $\overline{\mathbf{B}}(t)$ are prescribed simultaneously. Under these conditions, the homogenized response is reported in terms of both the macroscopic Cauchy stress $\overline{\boldsymbol{\sigma}}$ and the macroscopic magnetic field $\overline{\mathbf{H}}$. Relative to the purely mechanical and purely magnetic cases, this loading path highlights the nonlinear interaction between deformation and magnetic effects. In particular, the chain morphology is expected to promote stronger anisotropy and more localized coupled fields due to the combined influence of particle alignment, inter-particle interactions, and magnetic-field-induced stress development.

\begin{figure}[htbp]
  \centering

  \begin{subfigure}[t]{0.32\linewidth}
    \centering
    \includegraphics[width=\linewidth]{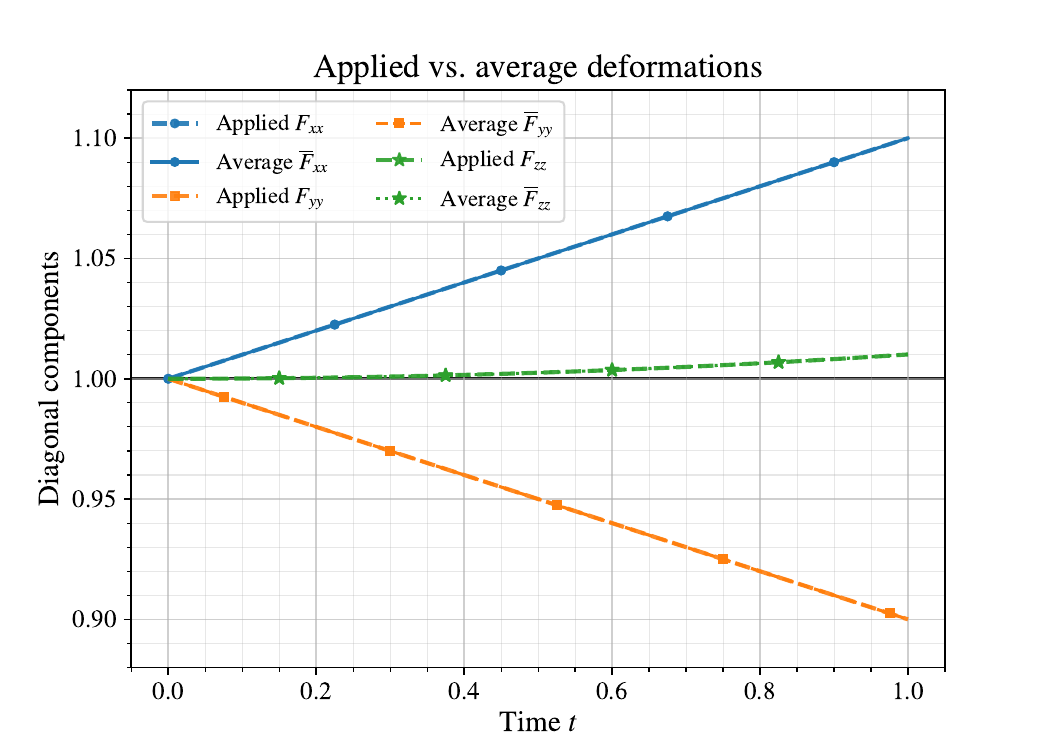}
    \caption{Homogenized deformation (average vs.\ applied)}
    \label{fig:PC_plots_homo_def}
  \end{subfigure}\hfill
  \begin{subfigure}[t]{0.32\linewidth}
    \centering
    \includegraphics[width=\linewidth]{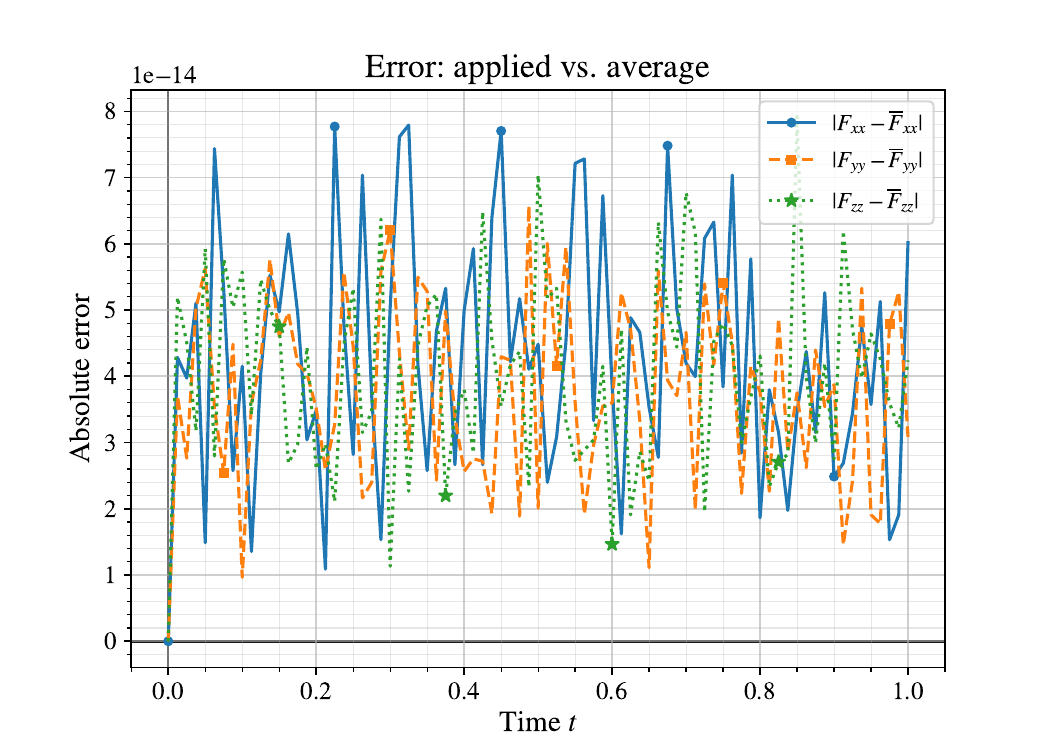}
    \caption{Deformation mismatch (average $-$ applied)}
    \label{fig:PC_plots_homo_def_error}
  \end{subfigure}\hfill
  \begin{subfigure}[t]{0.32\linewidth}
    \centering
    \includegraphics[width=\linewidth]{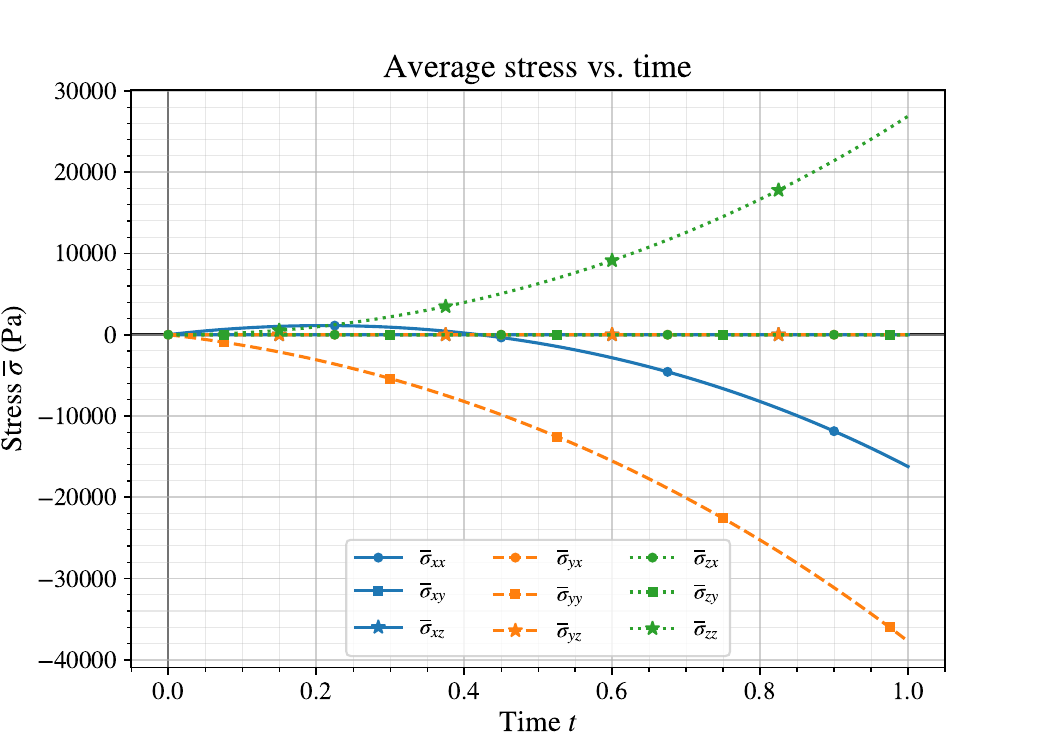}
    \caption{Homogenized stress response}
    \label{fig:PC_plots_homo_stress}
  \end{subfigure}

  \vspace{0.6em}

  \begin{subfigure}[t]{0.32\linewidth}
    \centering
    \includegraphics[width=\linewidth]{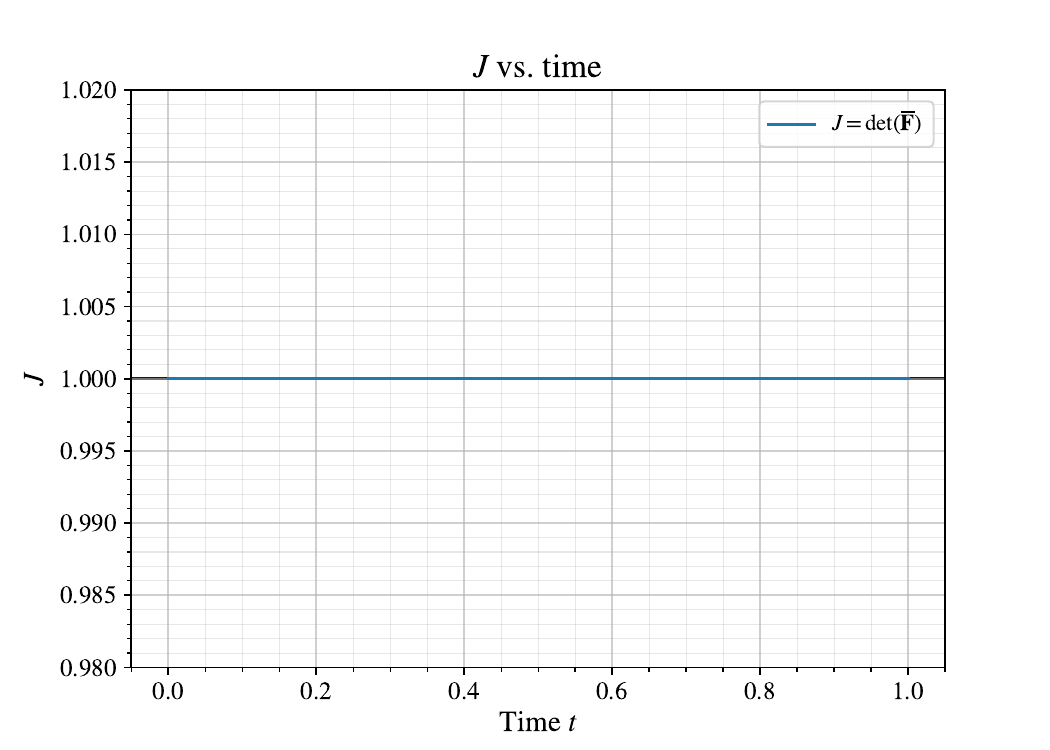}
    \caption{Homogenized Jacobian $J$ (volume change)}
    \label{fig:PC_plots_homo_J}
  \end{subfigure}\hfill
  \begin{subfigure}[t]{0.32\linewidth}
    \centering
    \includegraphics[width=\linewidth]{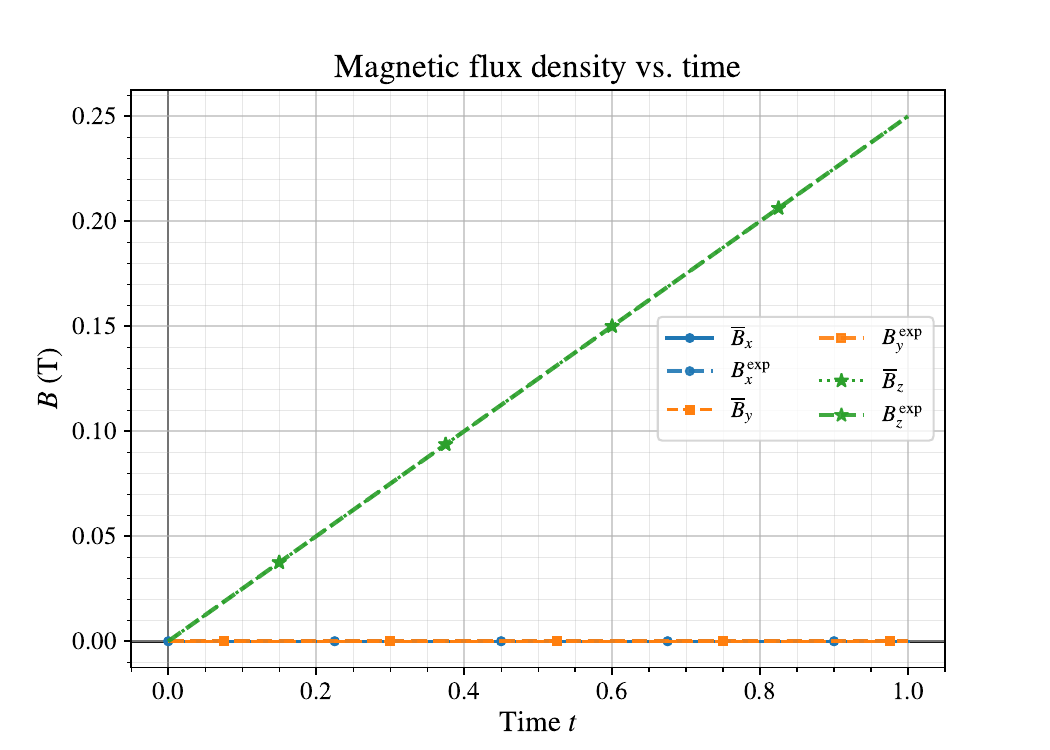}
    \caption{Homogenized magnetic induction $\bar{\mathbf{B}}$ vs.\ time}
    \label{fig:PC_plots_homo_B}
  \end{subfigure}\hfill
  \begin{subfigure}[t]{0.32\linewidth}
    \centering
    \includegraphics[width=\linewidth]{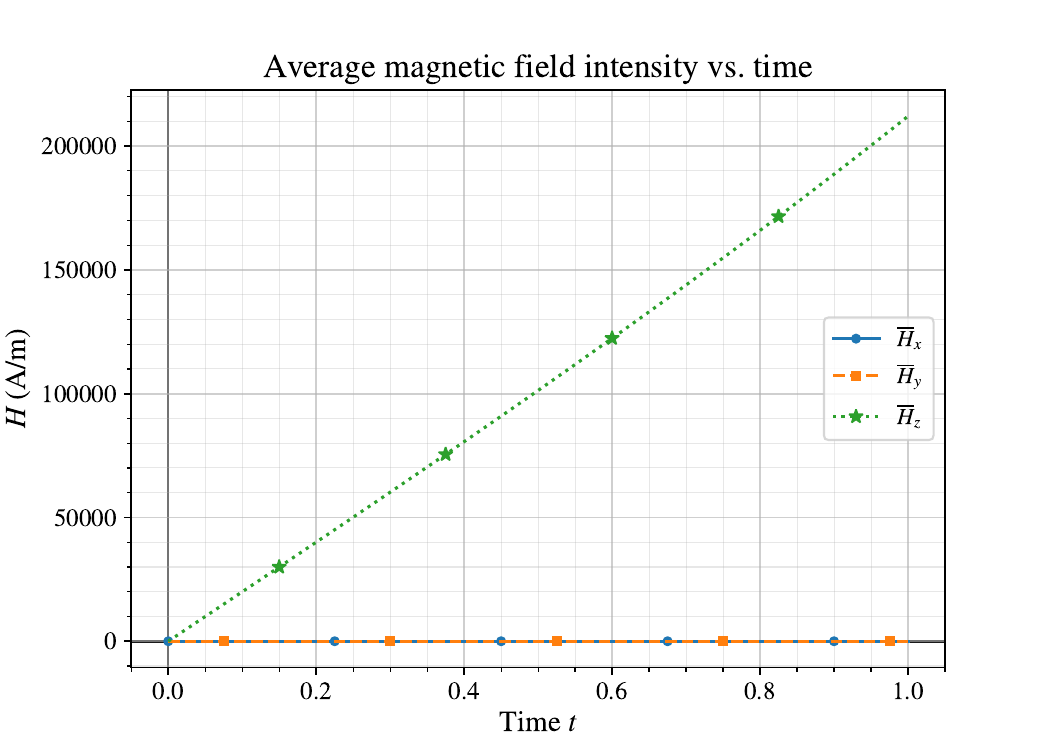}
    \caption{Homogenized magnetic field $\bar{\mathbf{H}}$ vs.\ time}
    \label{fig:PC_plots_homo_H}
  \end{subfigure}

  \caption{Homogenized quantities for the particle-chain RVE under the prescribed magneto-mechanical loading.}
  \label{fig:PC_homogenized_quantities}
\end{figure}

Below are representative visualizations from the particle-chain simulation, including the full-field distributions and planar cut views of the deformation and magnetic quantities.

\begin{figure}[htbp]
    \centering
    \begin{subfigure}[b]{0.32\linewidth}
        \centering
        \includegraphics[width=\linewidth]{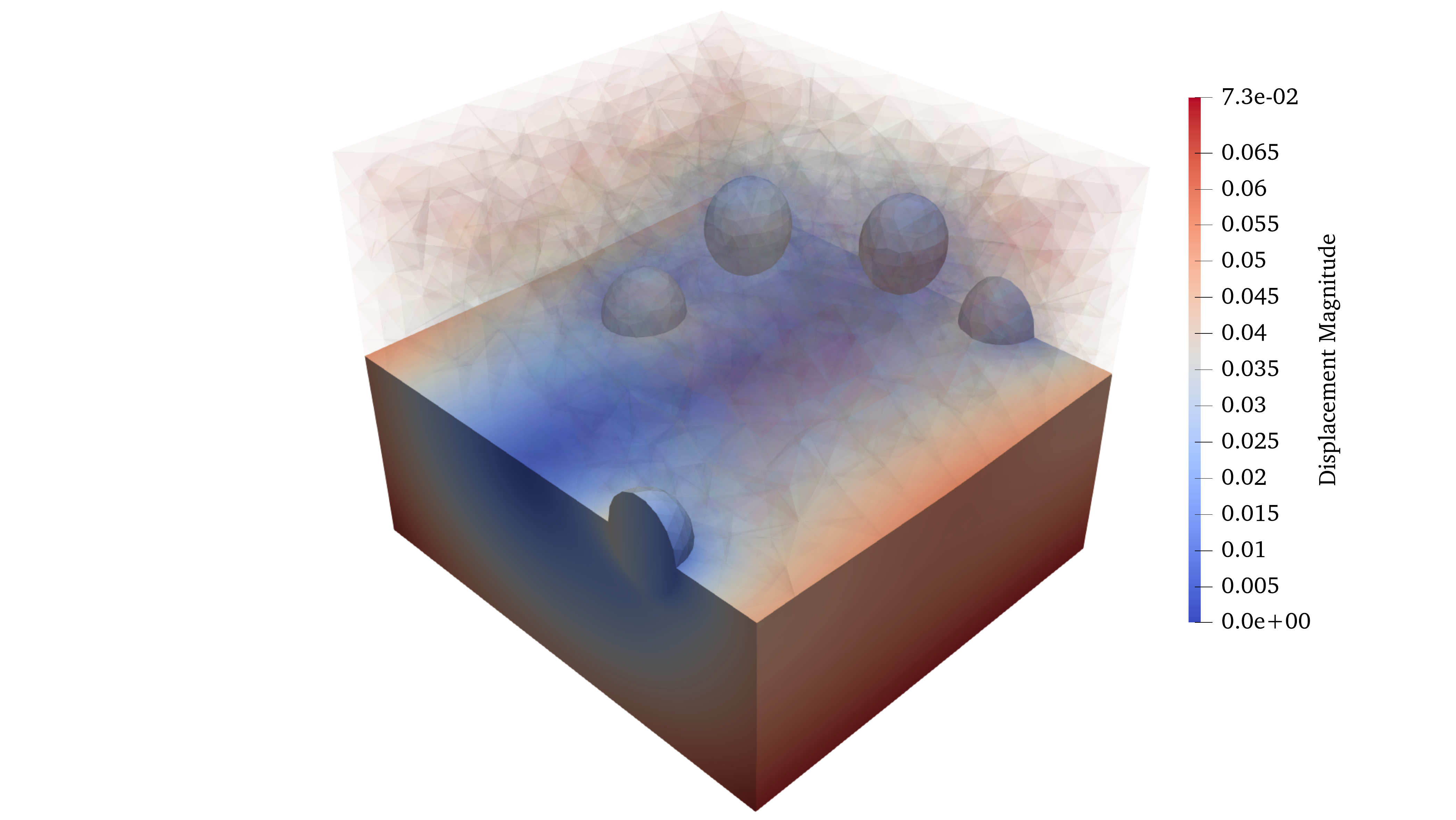}
        \caption{Combined displacement magnitude}
        \label{fig:pc_both_disp_final}
    \end{subfigure}\hfill
    \begin{subfigure}[b]{0.32\linewidth}
        \centering
        \includegraphics[width=\linewidth]{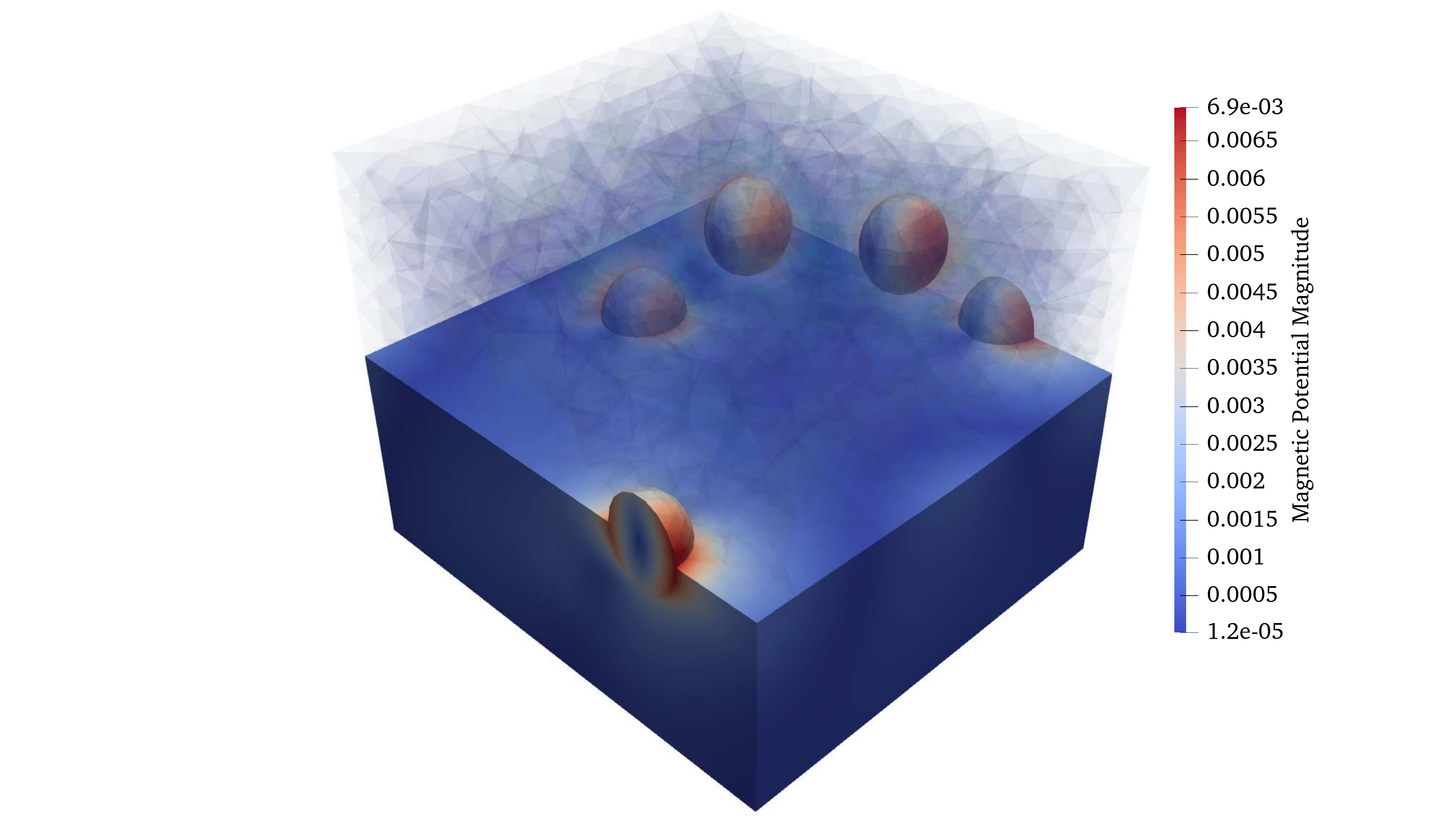}
        \caption{Combined magnetic potential}
        \label{fig:pc_both_magpot_final}
    \end{subfigure}\hfill
    \begin{subfigure}[b]{0.32\linewidth}
        \centering
        \includegraphics[width=\linewidth]{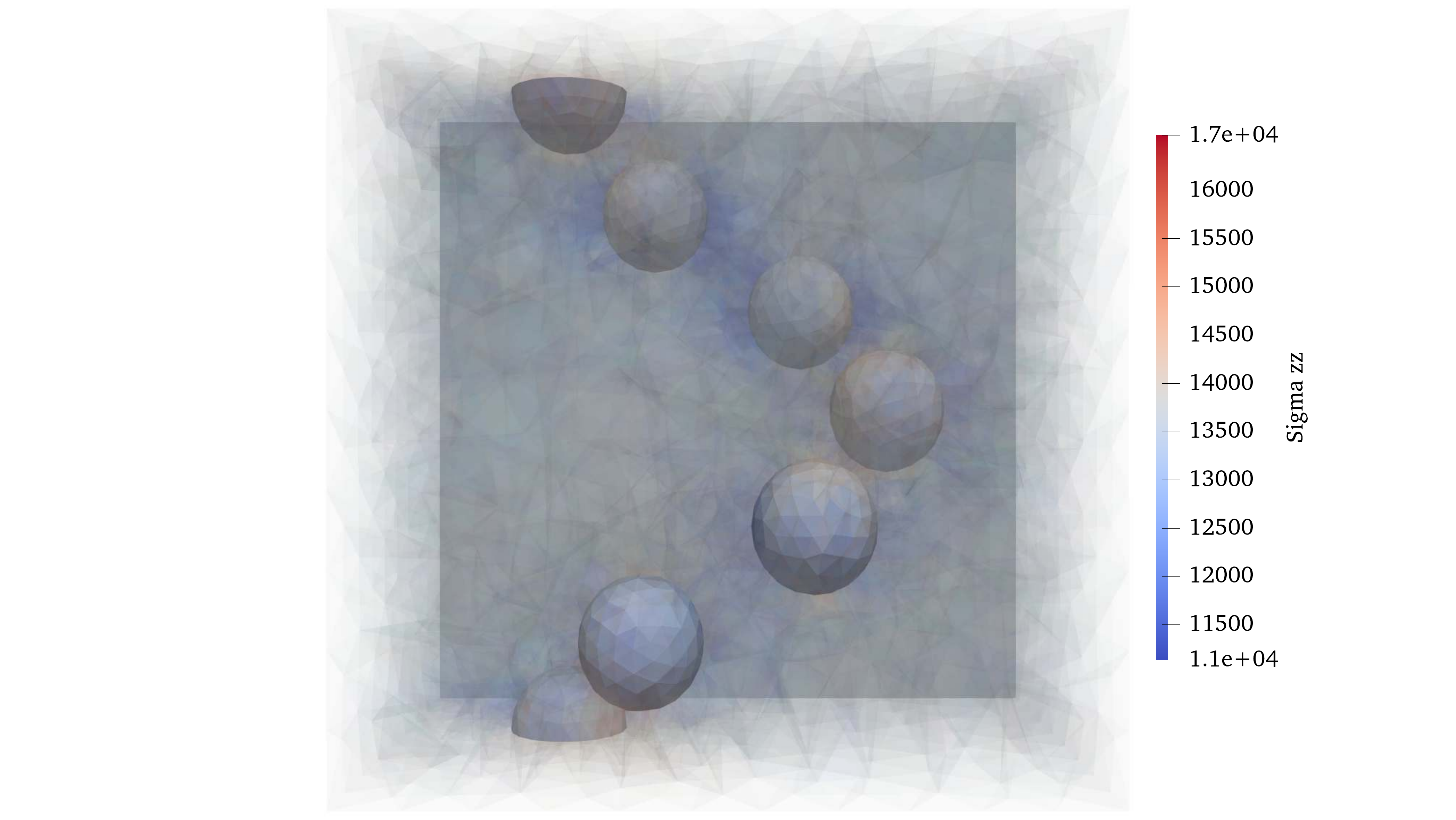}
        \caption{Final $\sigma_{zz}$}
        \label{fig:pc_both_sigmazz_final}
    \end{subfigure}

    \caption{Particle-chain RVE (tet mesh): displacement magnitude, magnetic potential, and final $\sigma_{zz}$ under the prescribed magneto-mechanical loading.}
    \label{fig:pc_both_images}
\end{figure}

\subsection{Random Particles}
\label{sec:random_particles}

As a final demonstration, we consider RVEs containing a large number of randomly distributed spherical particles. In contrast to the single-inclusion and particle-chain benchmarks, this setting is intended to represent a more realistic heterogeneous microstructure and to test the robustness of the coupled solver in the presence of strong phase contrast, complex particle arrangements, and localized interactions across many inclusion--matrix interfaces. All simulations in this section are performed on unstructured tetrahedral meshes to accommodate the geometric complexity and to enable local refinement near particle surfaces and within narrow inter-particle gaps.

The particle distributions are generated using a periodic placement strategy to avoid boundary artifacts and to ensure compatibility with the periodic boundary constraints imposed on the RVE faces. The particle volume fraction is controlled by the prescribed number of inclusions and their radii, and multiple random realizations (seeds) can be considered to assess the variability of the homogenized response for a fixed volume fraction. Periodic boundary conditions are enforced for both the displacement fluctuation field and the magnetic vector potential, and the resulting homogenized quantities $(\overline{\mathbf{\sigma}},\overline{\mathbf{H}})$ are extracted via the averaging procedure described in Section~\ref{sec:homogenization}.

\subsubsection{Meshes and microstructure generation}
\label{sec:random_particles_mesh}

The mesh containing random particles was generated by placing 100 particles with radius 0.2 randomly inside a 6x6x6 cell. The mesh uses tetrahedral elements and is periodic, similar to the previous meshes. Despite the larger unit cell, homogenized quantities are independent of volume and have no effect on the outcome.

\begin{figure}[htbp]
  \centering
  \includegraphics[width=0.48\linewidth]{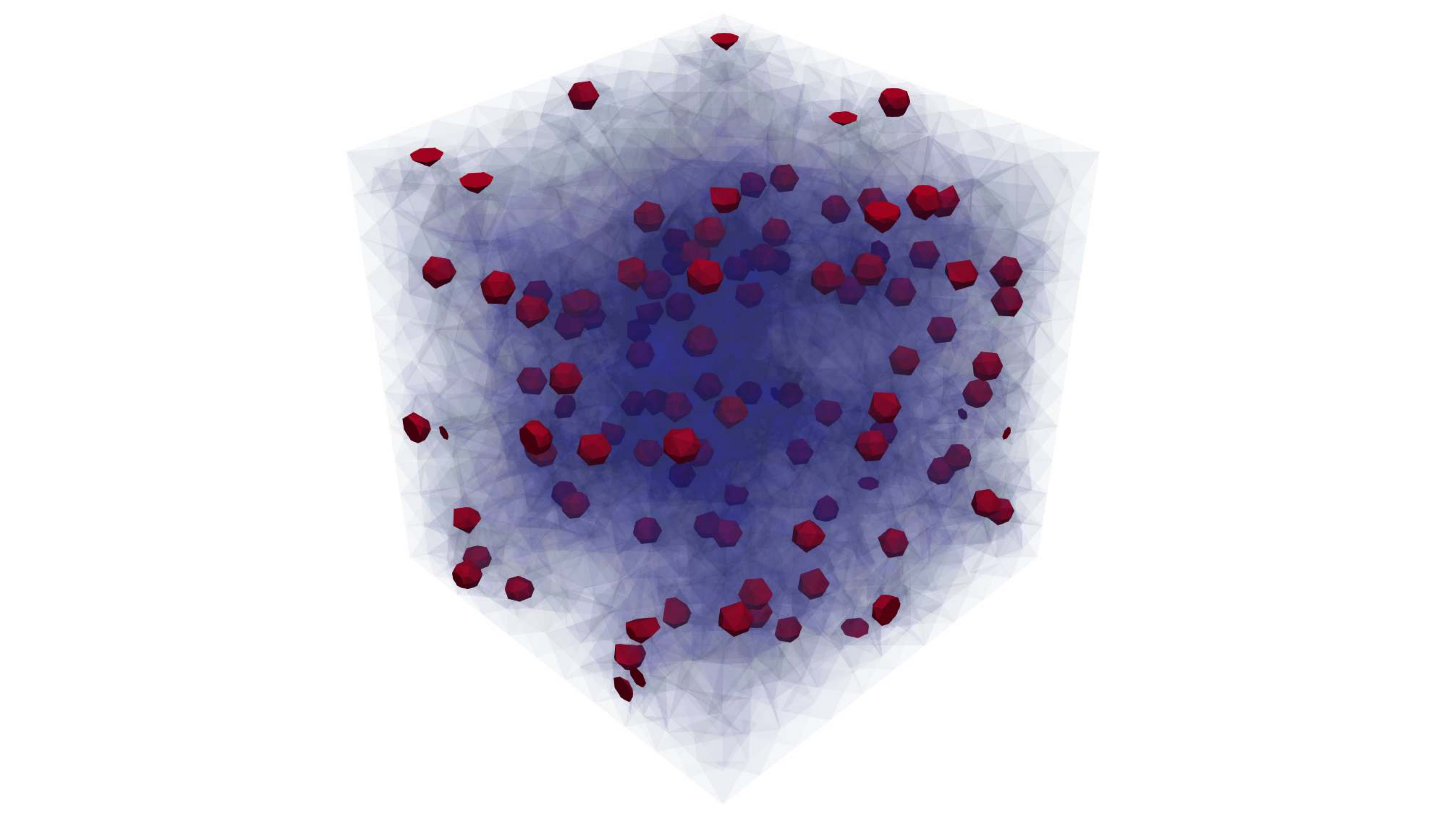}
  \caption{Particle Mesh}
  \label{fig:Particle_mesh}
\end{figure}

\subsubsection{Homogenized Responses}
\label{sec:random_particles_homo_responses}

\paragraph{Mechanical loading:}
The random-particle RVE is first subjected to the same purely mechanical loading path introduced in Paragraph~\ref{par:sphere_mech_loading}; however, due to the associated computational expense, the loading is applied only up to 37\% of the maximum prescribed deformation. In this case, the macroscopic deformation gradient $\overline{\mathbf{F}}(t)$ is prescribed, while the macroscopic magnetic induction is set to $\overline{\mathbf{B}}=\mathbf{0}$. The resulting response is therefore driven entirely by the imposed mechanical deformation, and the homogenized behavior is characterized primarily through the macroscopic Cauchy stress $\overline{\boldsymbol{\sigma}}$. Relative to the single-inclusion and particle-chain cases, the random particle distribution introduces a more heterogeneous microstructure, leading to a more spatially dispersed pattern of local deformation and stress concentrations throughout the RVE.

\begin{figure}[htbp]
  \centering

  \begin{subfigure}[t]{0.32\linewidth}
    \centering
    \includegraphics[width=\linewidth]{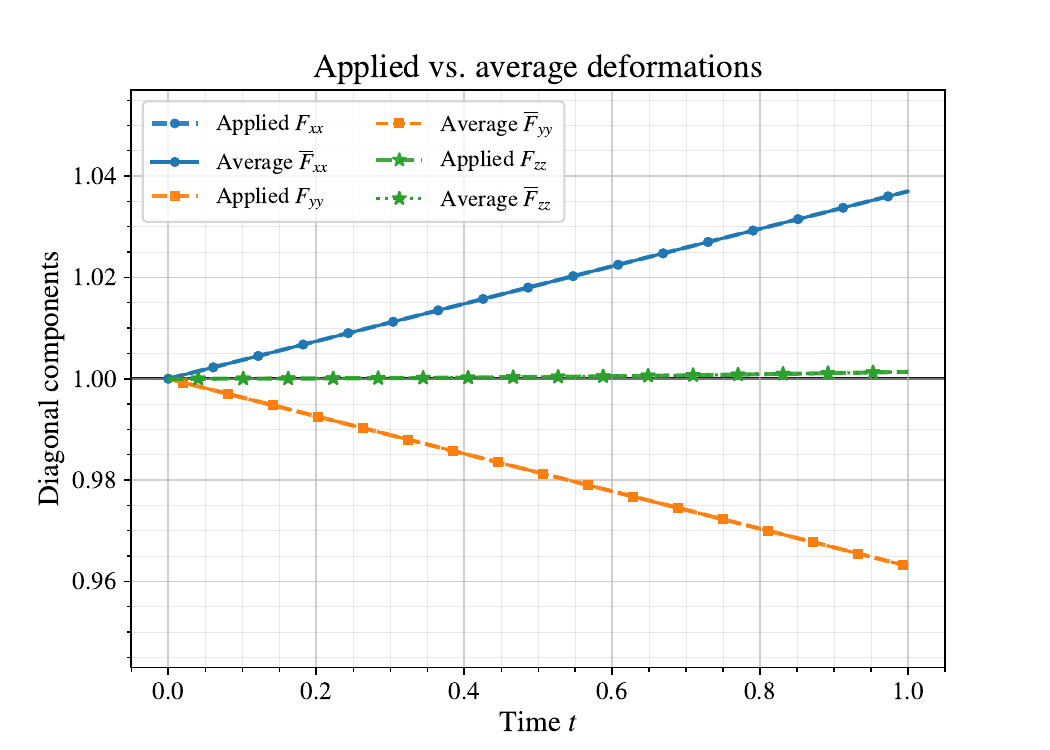}
    \caption{Homogenized deformation (average vs.\ applied)}
    \label{fig:particle_js_plots_homo_def}
  \end{subfigure}\hfill
  \begin{subfigure}[t]{0.32\linewidth}
    \centering
    \includegraphics[width=\linewidth]{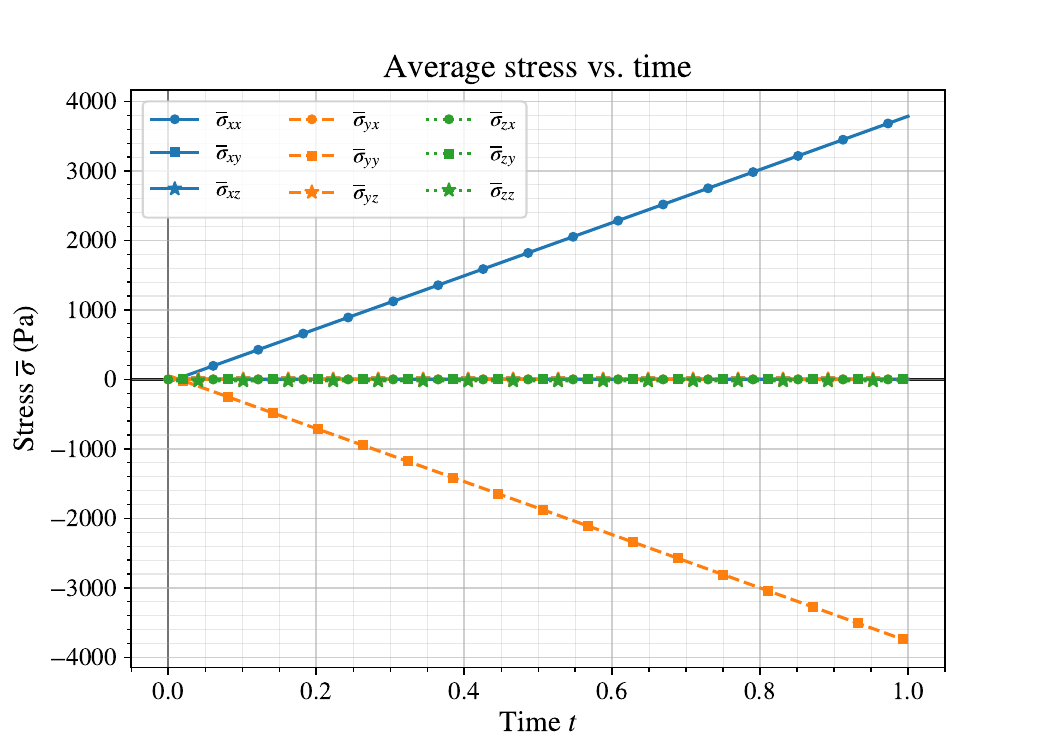}
    \caption{Homogenized stress response}
    \label{fig:particle_js_plots_homo_stress}
  \end{subfigure}\hfill
  \begin{subfigure}[t]{0.32\linewidth}
    \centering
    \includegraphics[width=\linewidth]{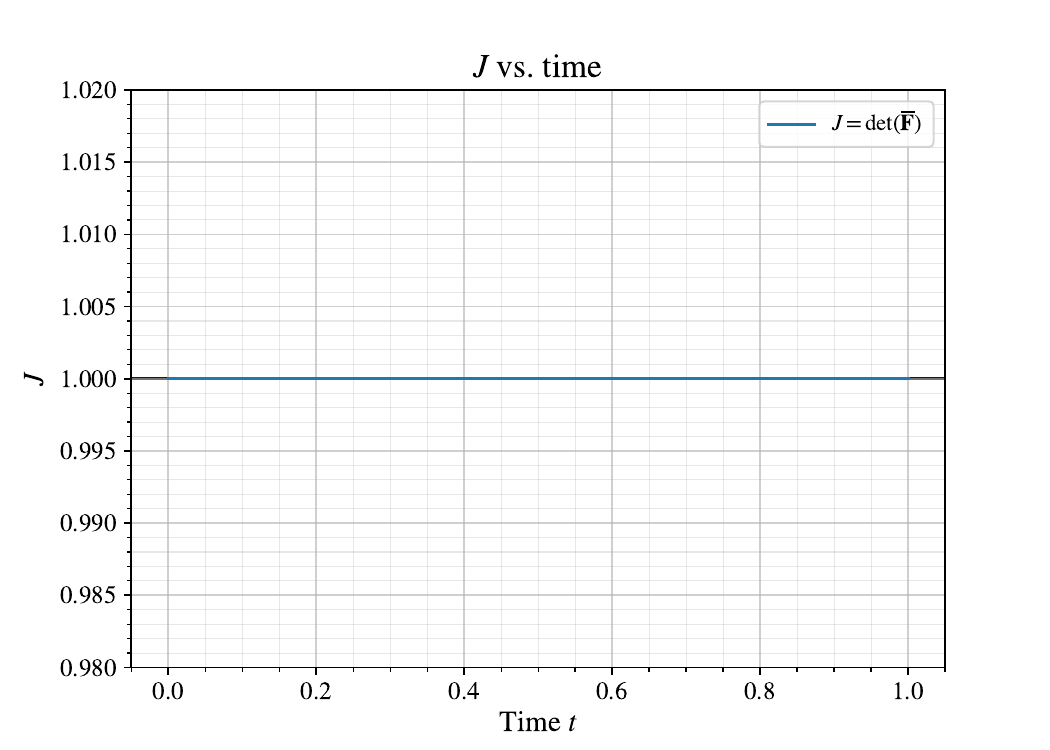}
    \caption{Homogenized Jacobian $J$ (volume change)}
    \label{fig:particle_js_plots_homo_J}
  \end{subfigure}

  \caption{Homogenized quantities for the random particle under strain loading.}
  \label{fig:particle_js_homogenized_quantities}
\end{figure}

\begin{figure}[htbp]
    \centering
    \begin{subfigure}[b]{0.49\linewidth}
        \centering
        \includegraphics[width=\linewidth]{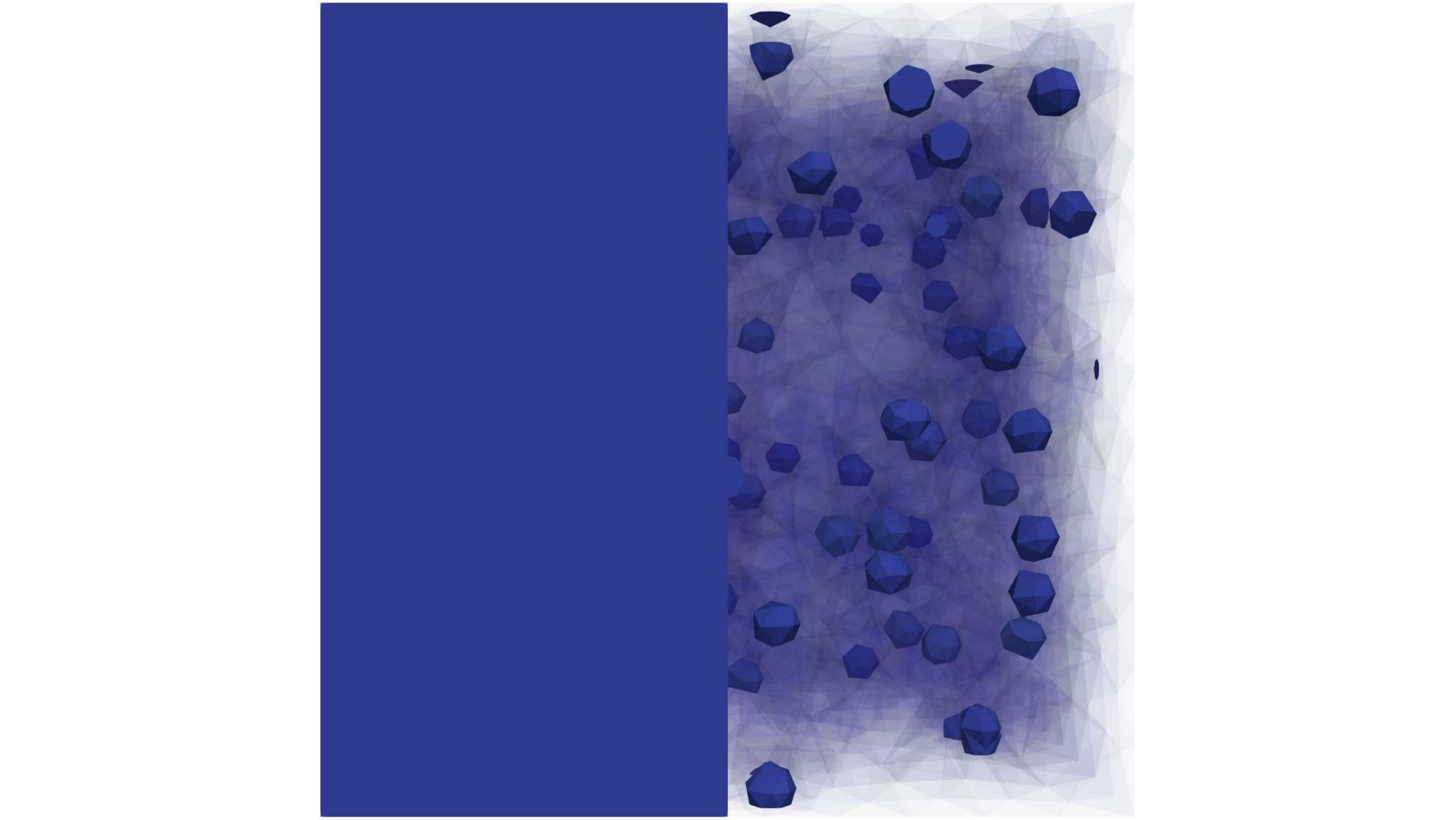}
        \caption{Initial state.}
        \label{fig:particles_js_initial}
    \end{subfigure}\hfill
    \begin{subfigure}[b]{0.49\linewidth}
        \centering
        \includegraphics[width=\linewidth]{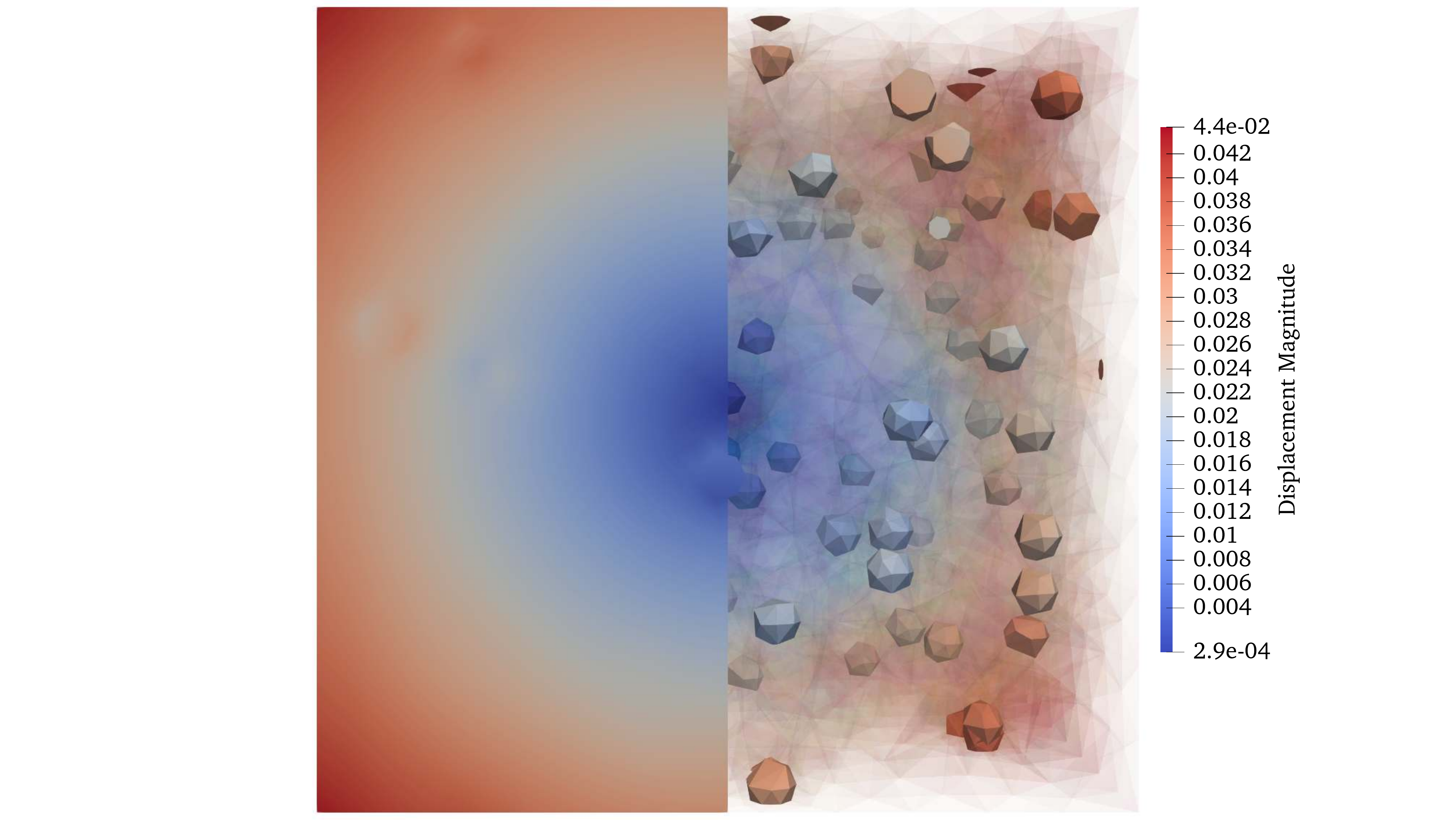}
        \caption{Final state.}
        \label{fig:particles_js_final}
    \end{subfigure}

    \caption{Random Particles: initial and final configurations under the prescribed deformation.}
    \label{fig:particles_js_def}
\end{figure}

The random-particle RVE exhibits a macroscopic strain response that is very similar to that of the single spherical-inclusion case. This similarity is consistent with the fact that both configurations share the same constituent material properties and approximately the same particle volume fraction, which largely governs the overall effective stiffness. However, although the averaged response remains similar, the random particle morphology produces a more spatially heterogeneous local response, with deformation and stress concentrations distributed across many inclusion--matrix interfaces. This behavior is clearly observed in Figure~\ref{fig:particles_js_def}.

\paragraph{Magnetic loading:}
The same purely magnetic loading path used for the single-inclusion case is applied to the random-particle RVE. Specifically, the macroscopic magnetic induction $\overline{\mathbf{B}}(t)$ is prescribed while the macroscopic deformation is held fixed at $\overline{\mathbf{F}}=\mathbf{I}$. The homogenized response is reported in terms of the macroscopic magnetic field $\overline{\mathbf{H}}$ and the induced macroscopic stress $\overline{\boldsymbol{\sigma}}$. Compared with the single-inclusion benchmark, the random particle morphology is expected to generate a more heterogeneous microscopic response, with localized magnetic and mechanical fields arising throughout the RVE rather than being concentrated around a single inclusion.

\begin{figure}[htbp]
  \centering

  \begin{subfigure}[t]{0.32\linewidth}
    \centering
    \includegraphics[width=\linewidth]{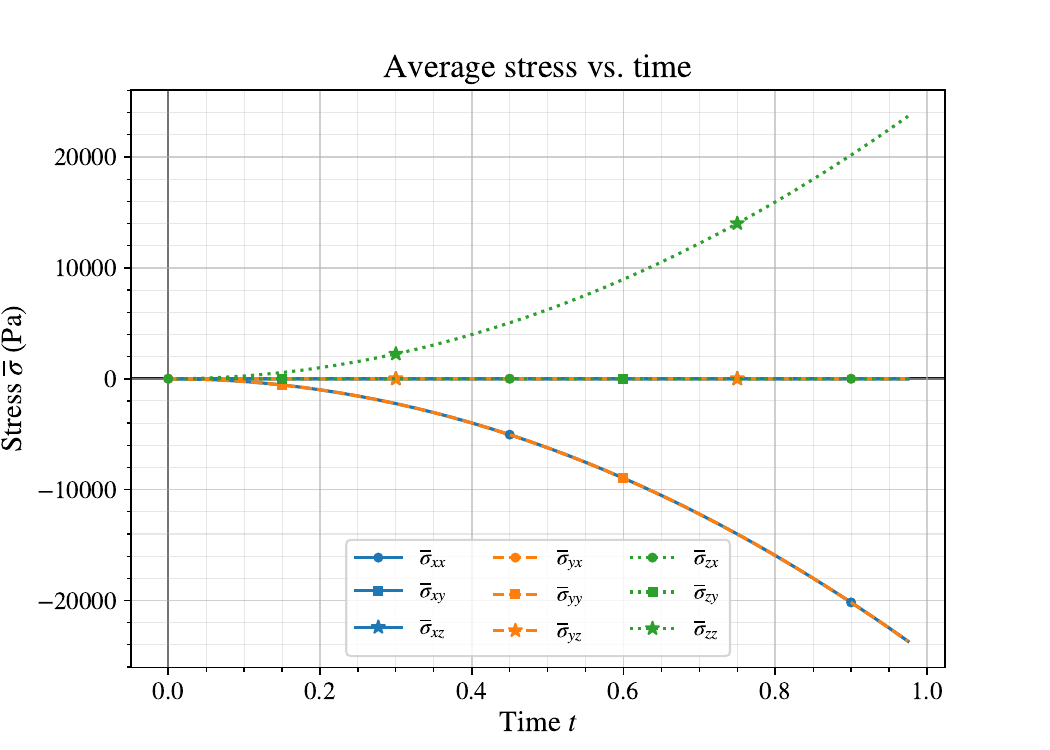}
    \caption{Homogenized stress response}
    \label{fig:particle_jm_plots_homo_stress}
  \end{subfigure}\hfill
  \begin{subfigure}[t]{0.32\linewidth}
    \centering
    \includegraphics[width=\linewidth]{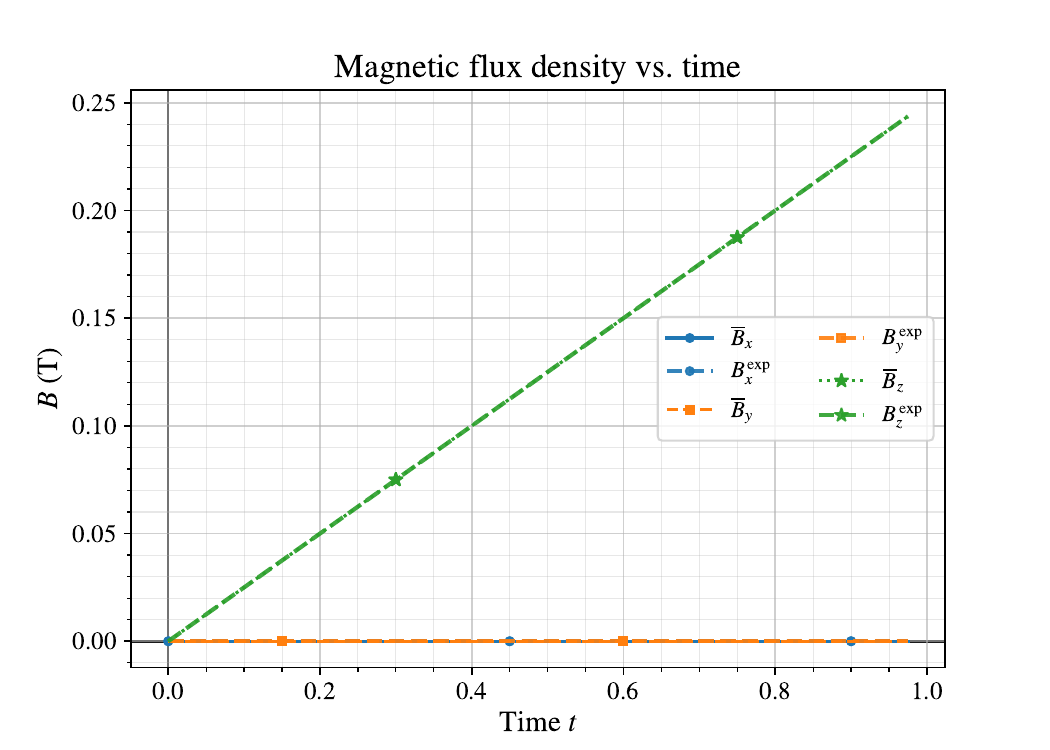}
    \caption{Homogenized magnetic induction $\bar{\mathbf{B}}$ vs.\ time}
    \label{fig:particle_jm_plots_homo_B}
  \end{subfigure}\hfill
  \begin{subfigure}[t]{0.32\linewidth}
    \centering
    \includegraphics[width=\linewidth]{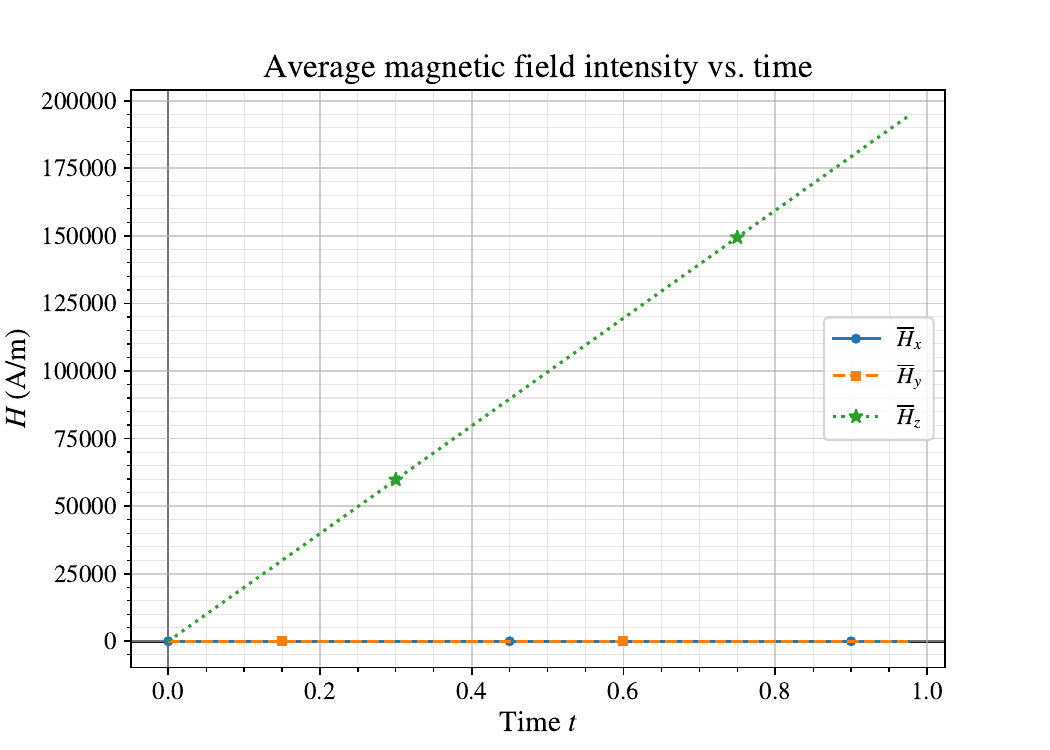}
    \caption{Homogenized magnetic field $\bar{\mathbf{H}}$ vs.\ time}
    \label{fig:particle_jm_plots_homo_H}
  \end{subfigure}

  \caption{Homogenized quantities for the random particle under magnetic loading.}
  \label{fig:particle_jm_homogenized_quantities}
\end{figure}

\begin{figure}[htbp]
    \centering
    \begin{subfigure}[b]{0.49\linewidth}
        \centering
        \includegraphics[width=\linewidth]{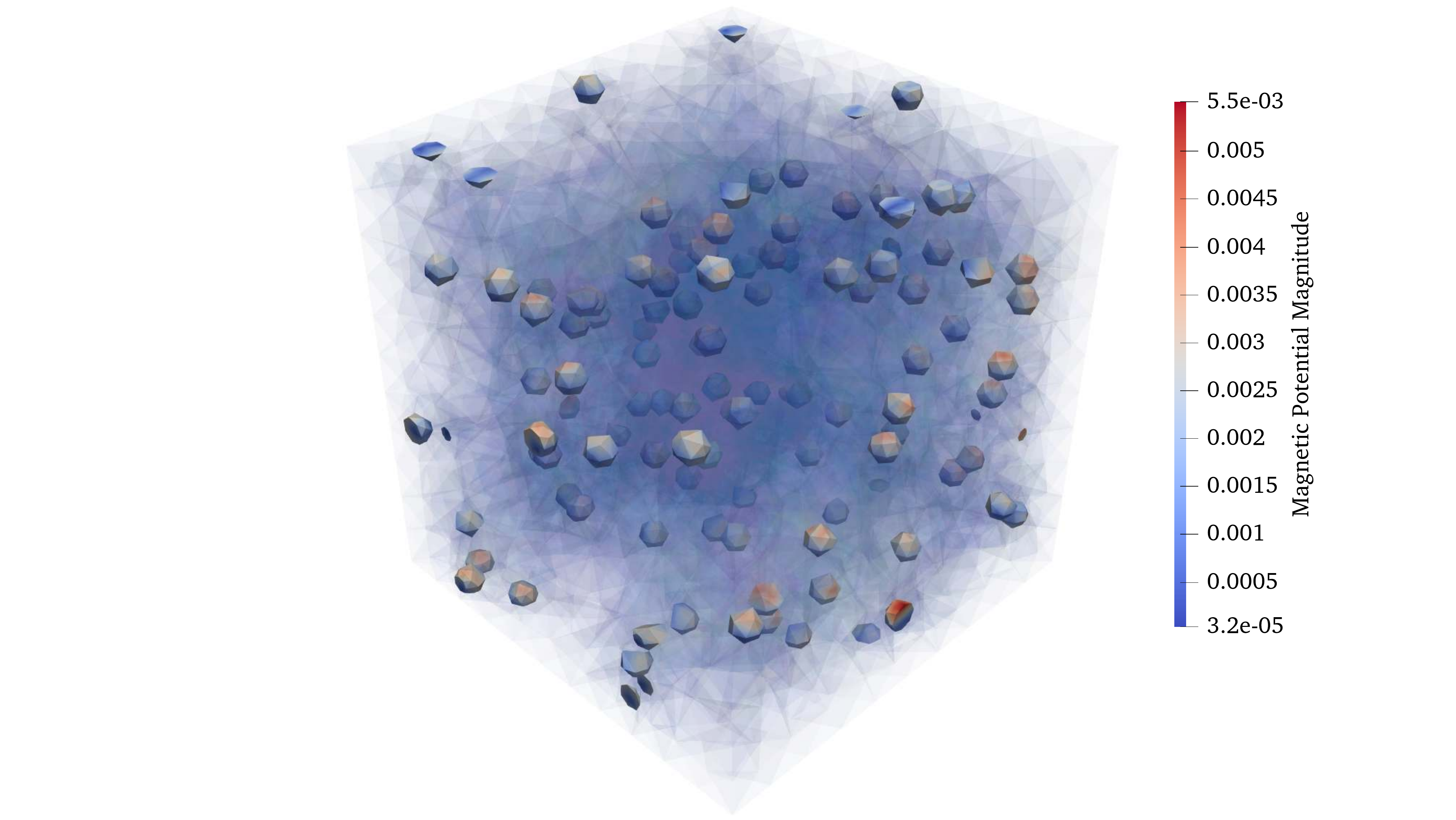}
        \caption{Magnetic Potential}
        \label{fig:particle_jm_mag_pot}
    \end{subfigure}\hfill
    \begin{subfigure}[b]{0.49\linewidth}
        \centering
        \includegraphics[width=\linewidth]{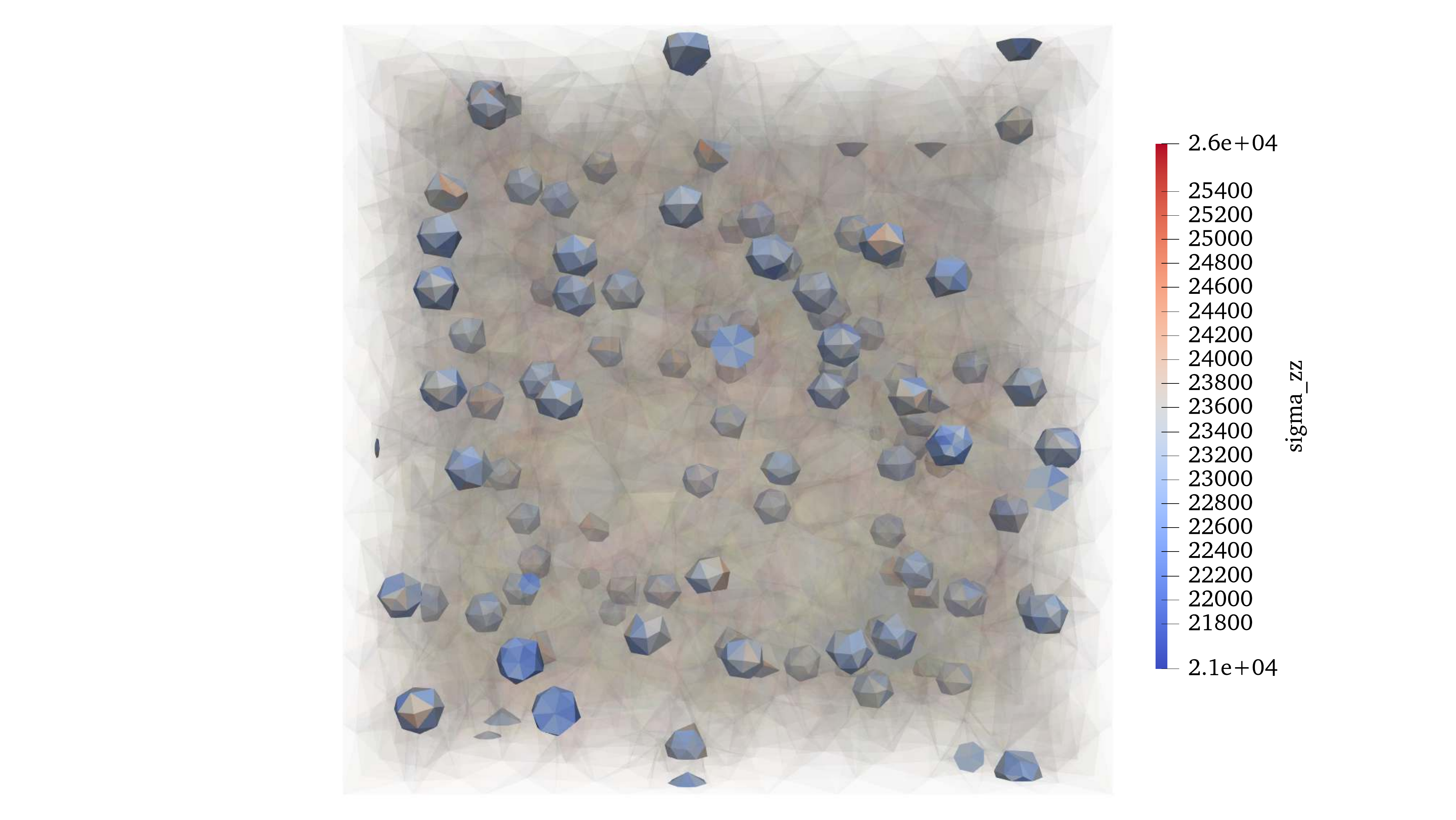}
        \caption{$\sigma_{zz}$}
        \label{fig:particle_js_sig_zz}
    \end{subfigure}

    \caption{Random Particles: Magnetic Potential and $\sigma_{zz}$ under magnetic loading.}
    \label{fig:particle_jm_figs}
\end{figure}

Like the strain response, the macroscopic magnetic response of the random-particle RVE is very similar to that of the spherical-inclusion case. This similarity is expected because both configurations share the same constituent material properties and nearly the same particle volume fraction. However, the microscopic response differs substantially. In particular, the random particle arrangement produces a more heterogeneous spatial distribution of magnetic potential and stress throughout the RVE, as shown in Figure~\ref{fig:particle_jm_figs}.

\begin{figure}[htbp]
  \centering

  \begin{subfigure}[b]{0.40\linewidth}
    \centering
    \includegraphics[width=\linewidth,height=5cm,keepaspectratio]{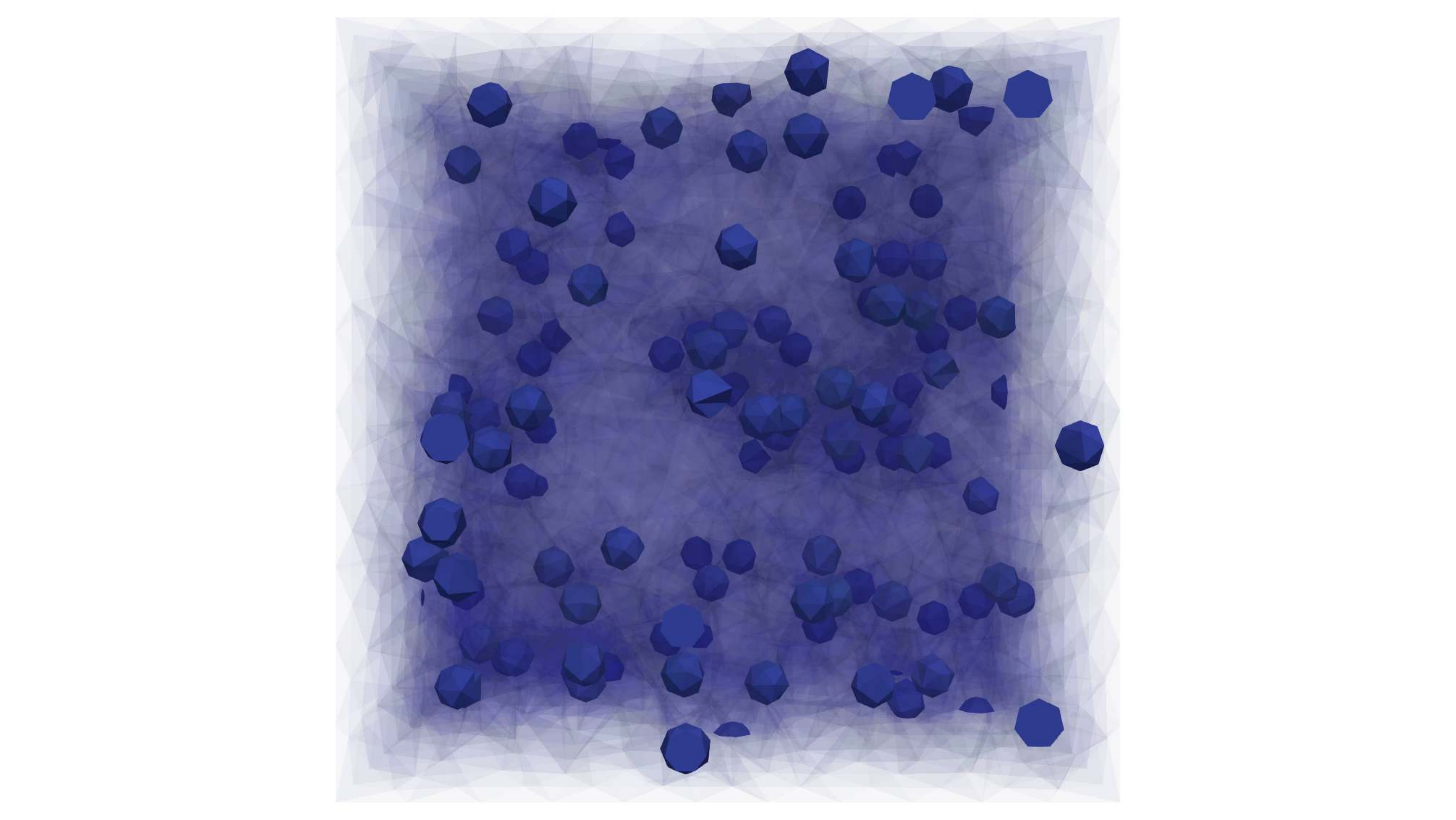}
    \caption{Initial configuration.}
    \label{fig:particle_move_t0}
  \end{subfigure}
  \hspace{0.02\linewidth}
  \begin{subfigure}[b]{0.40\linewidth}
    \centering
    \includegraphics[width=\linewidth,height=5cm,keepaspectratio]{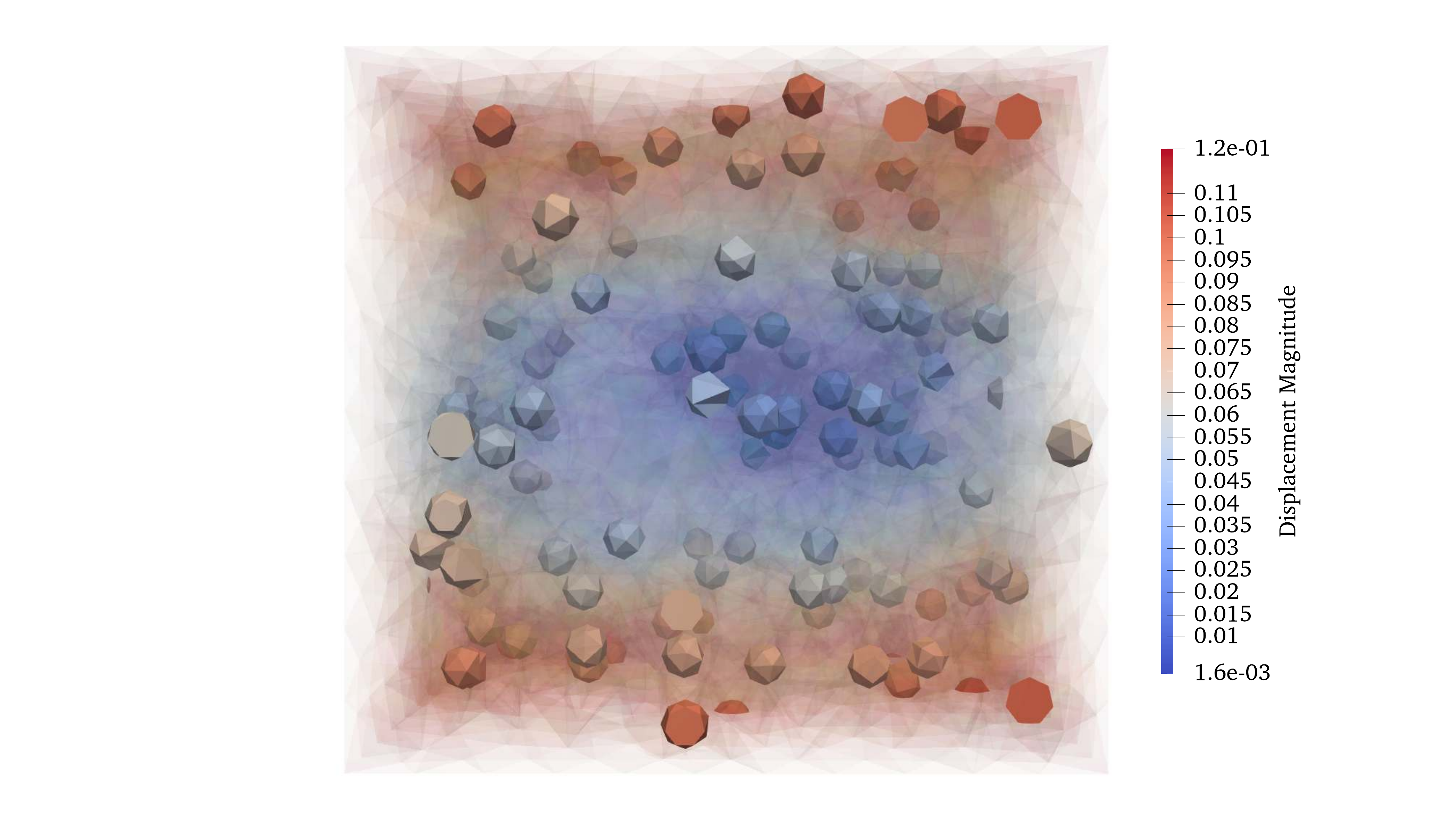}
    \caption{Final configuration.}
    \label{fig:particle_move_tfinal}
  \end{subfigure}

  \caption{Initial and final random-particle configurations under stress-relaxed boundary conditions. The particle assembly undergoes a cooperative but spatially heterogeneous rearrangement toward the central plane of the cell.}
  \label{fig:particle_move_displacement}
\end{figure}

\begin{figure}[htbp]
    \centering
    \includegraphics[width=0.95\linewidth]{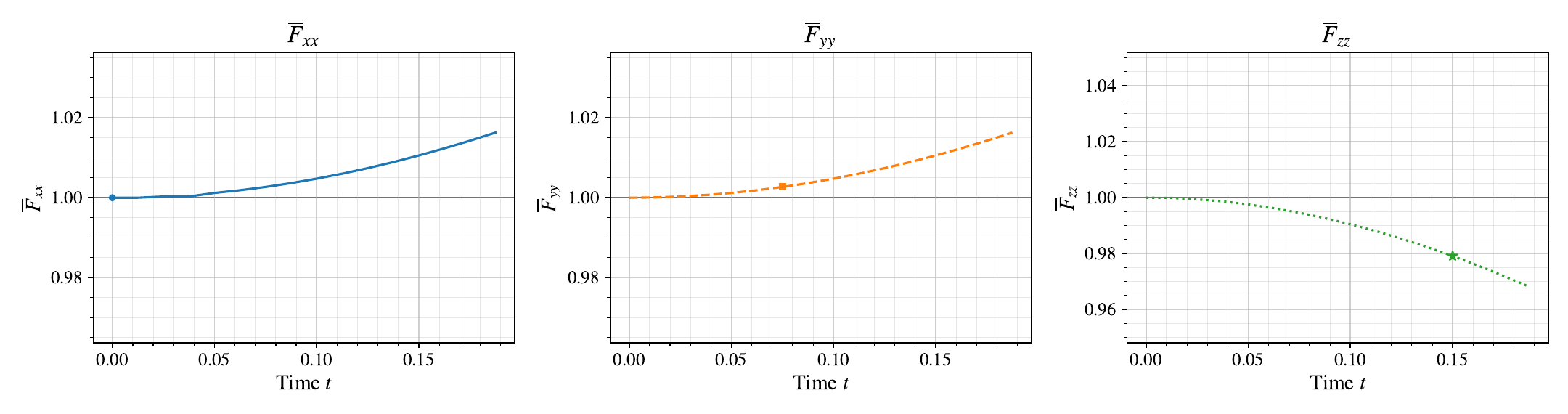}
    \caption{Homogenized deformation response under stress-relaxed boundary conditions, with the residual stress tolerance set to $10^{-3}$ and the loading applied up to approximately 18\% of the full prescribed value.}
    \label{fig:particle_ns_def}
\end{figure}

The microscopic response becomes even more pronounced when particle motion is permitted through the stress-relaxed boundary conditions. The random particle assembly undergoes a collective but spatially heterogeneous rearrangement under the applied magnetic loading. The dominant motion is aligned with the direction of the applied magnetic induction, with particles above and below the mid-plane of the cell moving toward one another in the \(z\)-direction. This produces an overall contraction toward the central plane of the microstructure, while small accompanying motions in the \(x\)- and \(y\)-directions remain secondary and are best interpreted as compatibility-driven transverse adjustments caused by the irregular local particle neighborhood. Thus, the dominant response is a cooperative \(z\)-directed rearrangement of the random particle assembly, modulated locally by the disordered microstructure. This behavior is illustrated in Figure~\ref{fig:particle_move_displacement}. As in the previous cases, the corresponding homogenized deformation response is reported in Figure~\ref{fig:particle_ns_def}. Similar to the particle-chain case, these simulations were performed only up to approximately 18\% of the full prescribed loading owing to their computational cost.

\begin{figure}[htbp]
    \centering
    \includegraphics[width=0.5\linewidth]{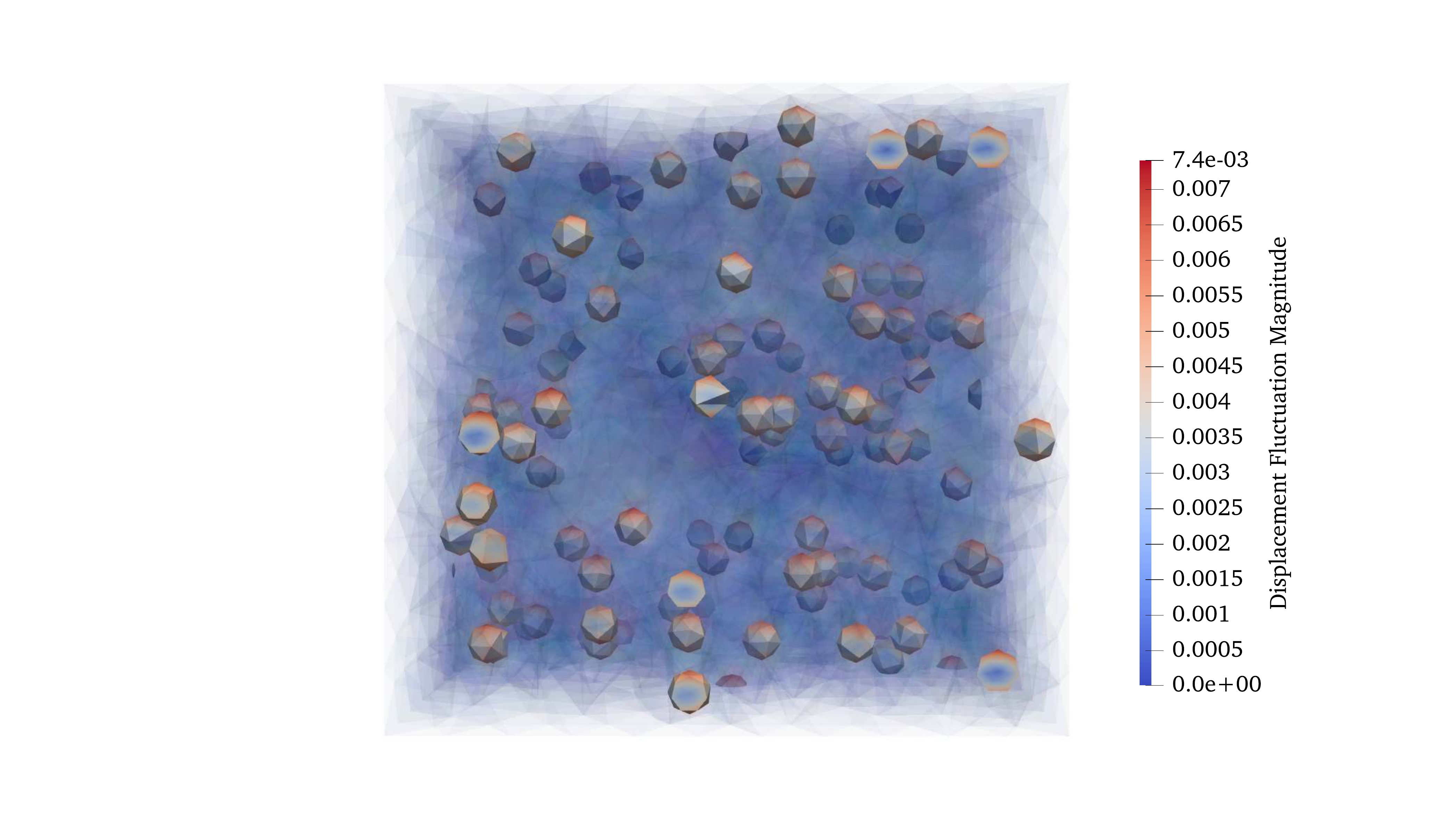}
    \caption{Displacement fluctuation field for the random-particle RVE under stress-relaxed boundary conditions.}
    \label{fig:particle_move_fluctuation}
\end{figure}

Because the RVE is periodic, this motion should not be interpreted as the behavior of isolated particles in free space. Rather, it represents the local kinematics of a repeating disordered microstructure, in which particles near the cell boundaries interact with neighboring periodic images. In that sense, the observed response may be interpreted on a larger scale as a relative approach of neighboring inclusions along the loading direction, while the random arrangement of the particles introduces the additional local variability visible in the deformation field. To better highlight the role of the surrounding inclusions, Figure~\ref{fig:particle_move_fluctuation} shows the displacement fluctuation field. This quantity more clearly exposes the spatial variation induced by particle--particle interaction and periodic compatibility. In particular, it reveals how the local motion varies throughout the microstructure and how the disordered neighborhood of each inclusion perturbs the surrounding response. This interpretation is also consistent with the homogenized deformation response shown later in Figure~\ref{fig:particle_ns_def}, which reflects the same field-direction contraction at the macroscale.

\paragraph{Coupled magneto--mechanical loading:}
The random-particle RVE is finally subjected to a coupled loading path in which the macroscopic deformation gradient $\overline{\mathbf{F}}(t)$ and the macroscopic magnetic induction $\overline{\mathbf{B}}(t)$ are prescribed simultaneously; however, as in the strain-loading case, the loading is applied only up to 29\% of the maximum prescribed level. Under these conditions, the homogenized response is reported in terms of both the macroscopic Cauchy stress $\overline{\boldsymbol{\sigma}}$ and the macroscopic magnetic field $\overline{\mathbf{H}}$. Relative to the purely mechanical and purely magnetic cases, this loading path highlights the nonlinear interaction between deformation and magnetic effects. In particular, the random particle arrangement is expected to generate a more heterogeneous coupled response throughout the RVE, as the combined loading activates many localized particle--particle and particle--matrix interactions rather than a single preferred microstructural direction.

\begin{figure}[htbp]
  \centering

  \begin{subfigure}[t]{0.32\linewidth}
    \centering
    \includegraphics[width=\linewidth]{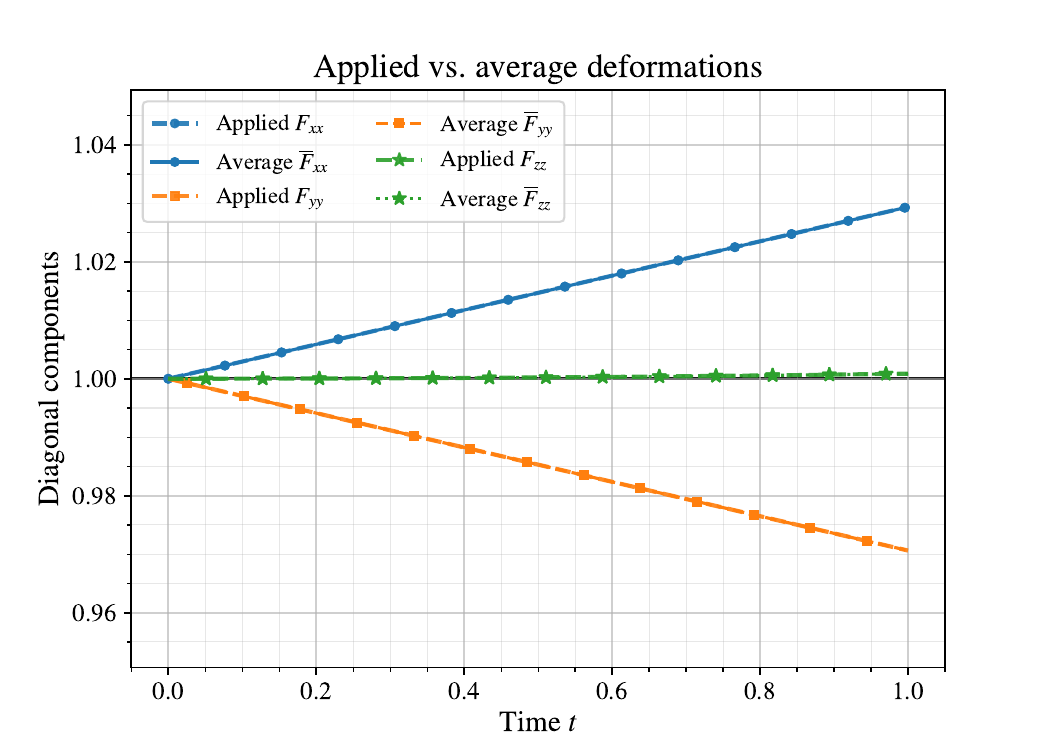}
    \caption{Homogenized deformation (average vs.\ applied)}
    \label{fig:Particle_plots_homo_def}
  \end{subfigure}\hfill
  \begin{subfigure}[t]{0.32\linewidth}
    \centering
    \includegraphics[width=\linewidth]{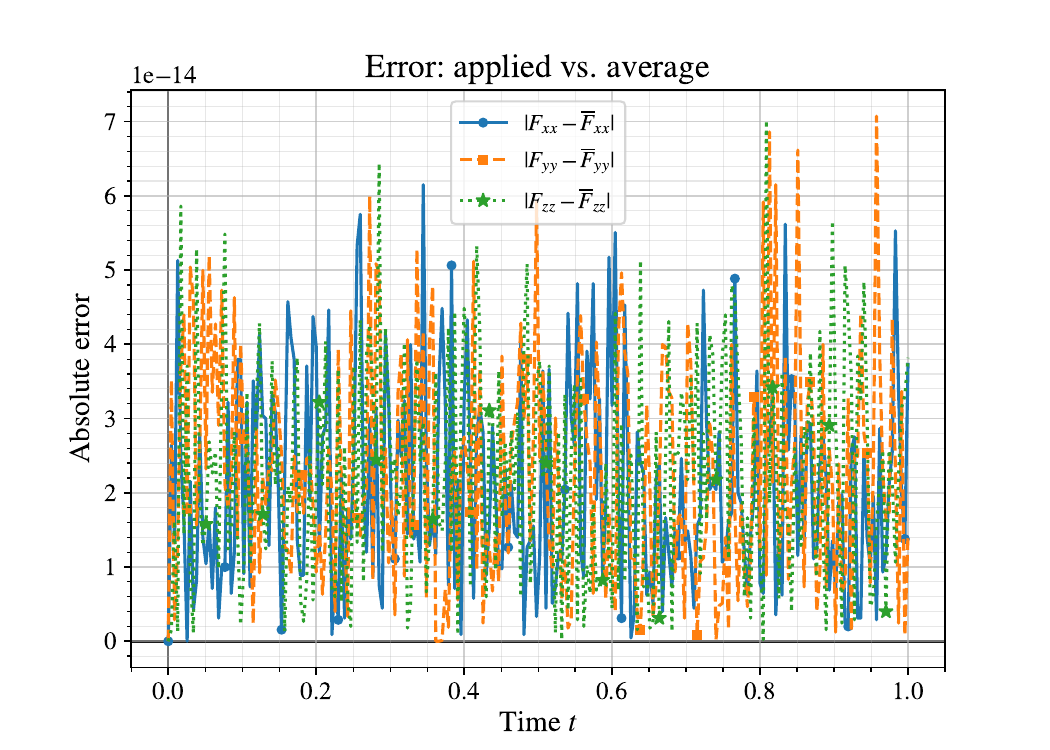}
    \caption{Deformation mismatch (average $-$ applied)}
    \label{fig:Particle_plots_homo_def_error}
  \end{subfigure}\hfill
  \begin{subfigure}[t]{0.32\linewidth}
    \centering
    \includegraphics[width=\linewidth]{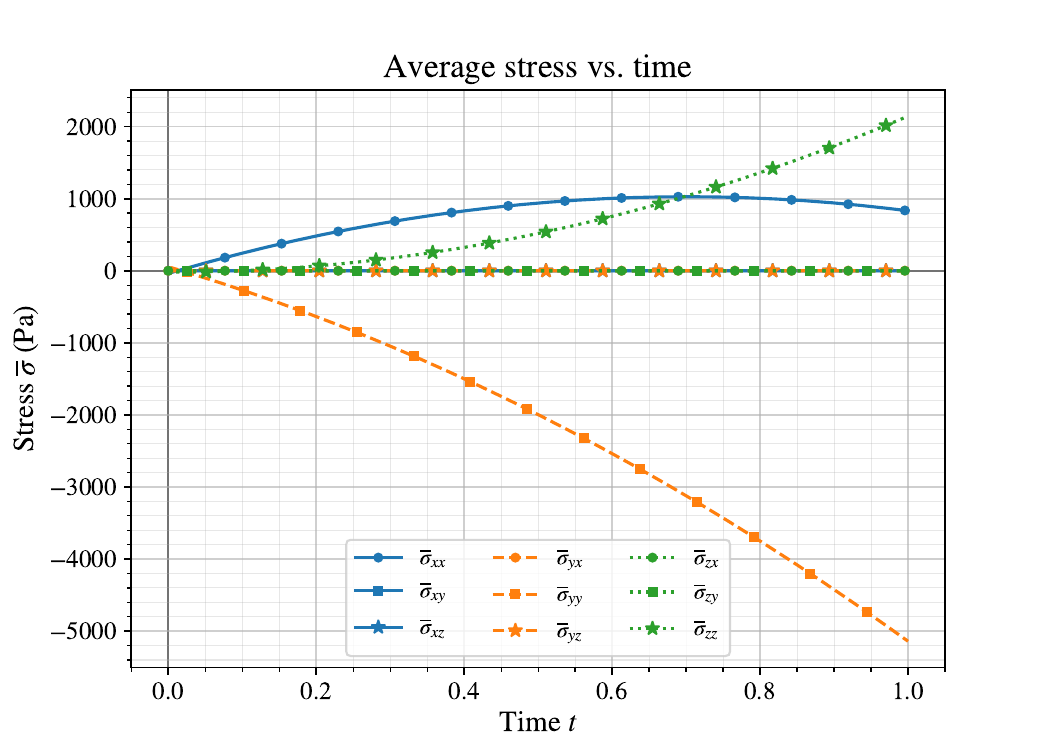}
    \caption{Homogenized stress response}
    \label{fig:Particle_plots_homo_stress}
  \end{subfigure}

  \vspace{0.6em}

  \begin{subfigure}[t]{0.32\linewidth}
    \centering
    \includegraphics[width=\linewidth]{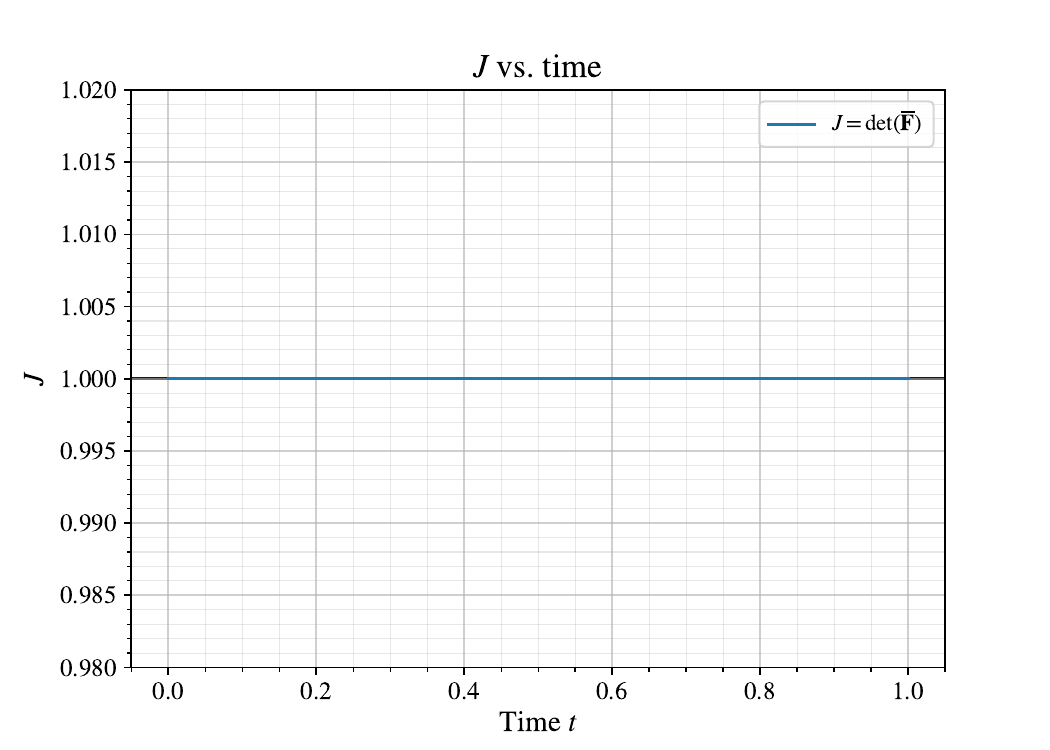}
    \caption{Homogenized Jacobian $J$ (volume change)}
    \label{fig:Particle_plots_homo_J}
  \end{subfigure}\hfill
  \begin{subfigure}[t]{0.32\linewidth}
    \centering
    \includegraphics[width=\linewidth]{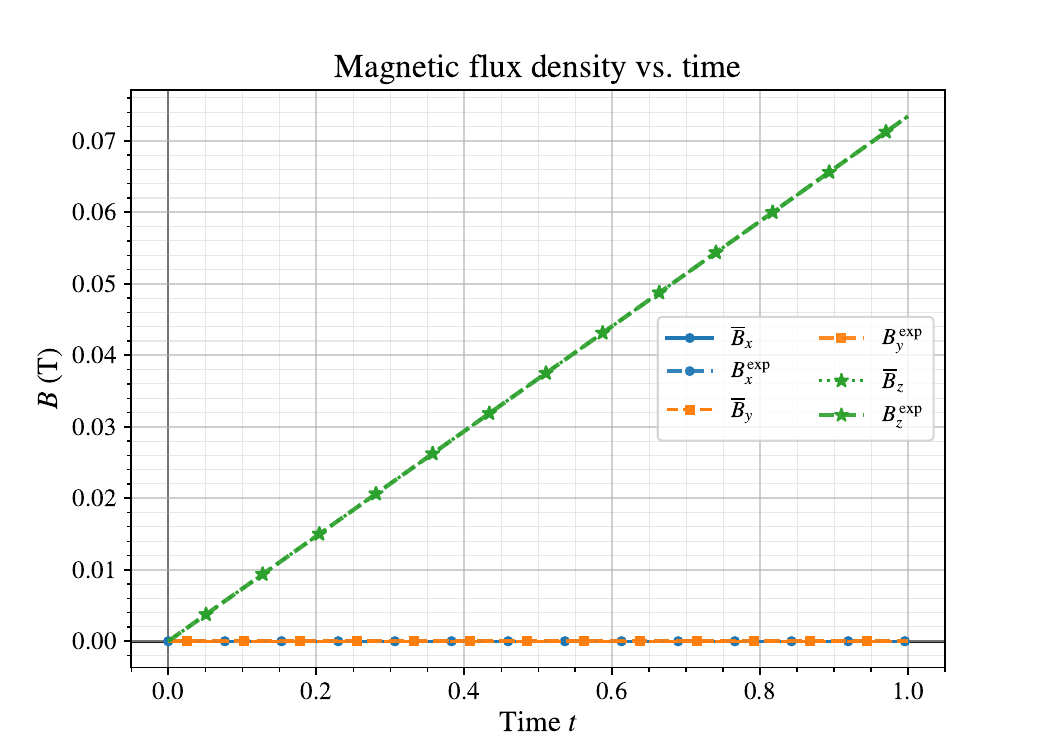}
    \caption{Homogenized magnetic induction $\bar{\mathbf{B}}$ vs.\ time}
    \label{fig:Particle_plots_homo_B}
  \end{subfigure}\hfill
  \begin{subfigure}[t]{0.32\linewidth}
    \centering
    \includegraphics[width=\linewidth]{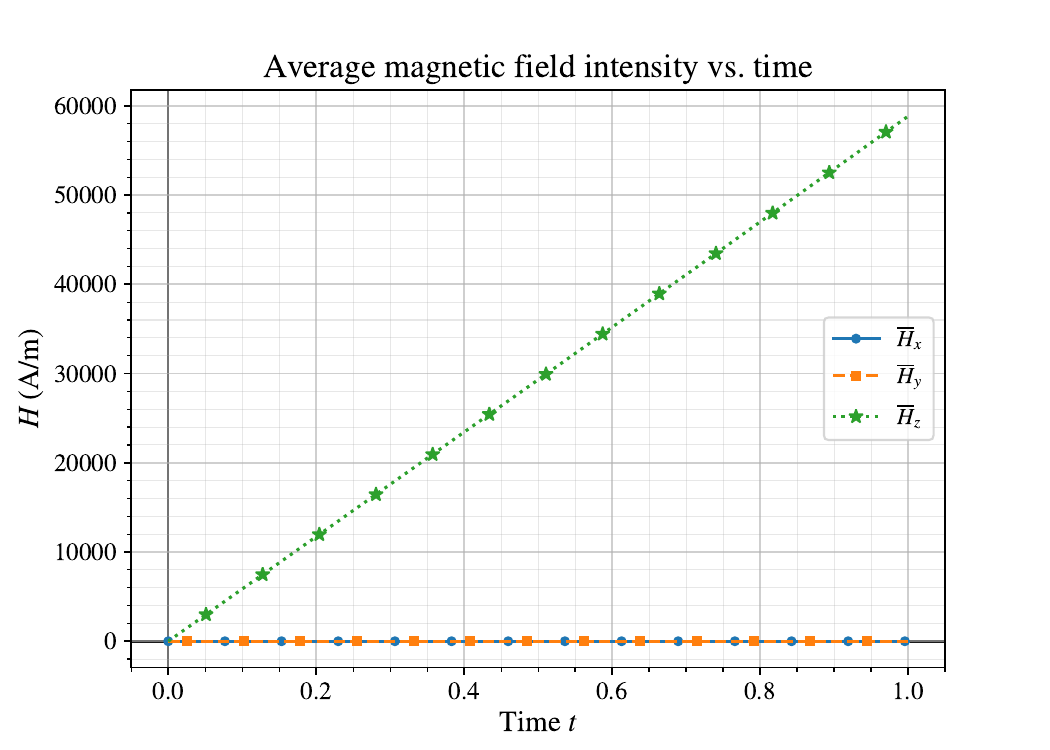}
    \caption{Homogenized magnetic field $\bar{\mathbf{H}}$ vs.\ time}
    \label{fig:Particle_plots_homo_H}
  \end{subfigure}

  \caption{Homogenized quantities for the particle RVE under the prescribed magneto-mechanical loading.}
  \label{fig:Particle_homogenized_quantities}
\end{figure}

\begin{figure}[htbp]
    \centering
    \begin{subfigure}[b]{0.32\linewidth}
        \centering
        \includegraphics[width=\linewidth]{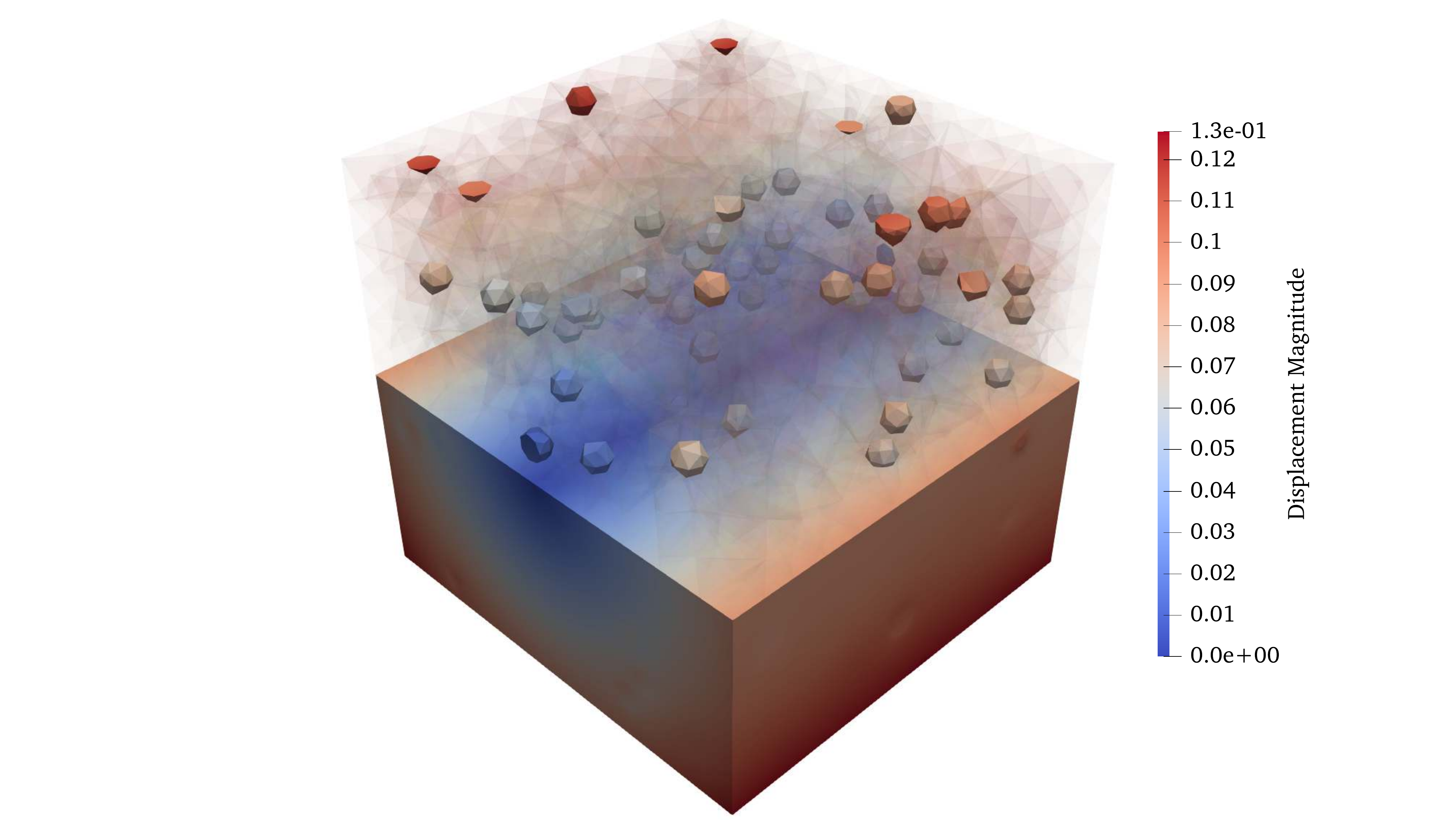}
        \caption{Displacement magnitude}
        \label{fig:particle_both_disp_final}
    \end{subfigure}\hfill
    \begin{subfigure}[b]{0.32\linewidth}
        \centering
        \includegraphics[width=\linewidth]{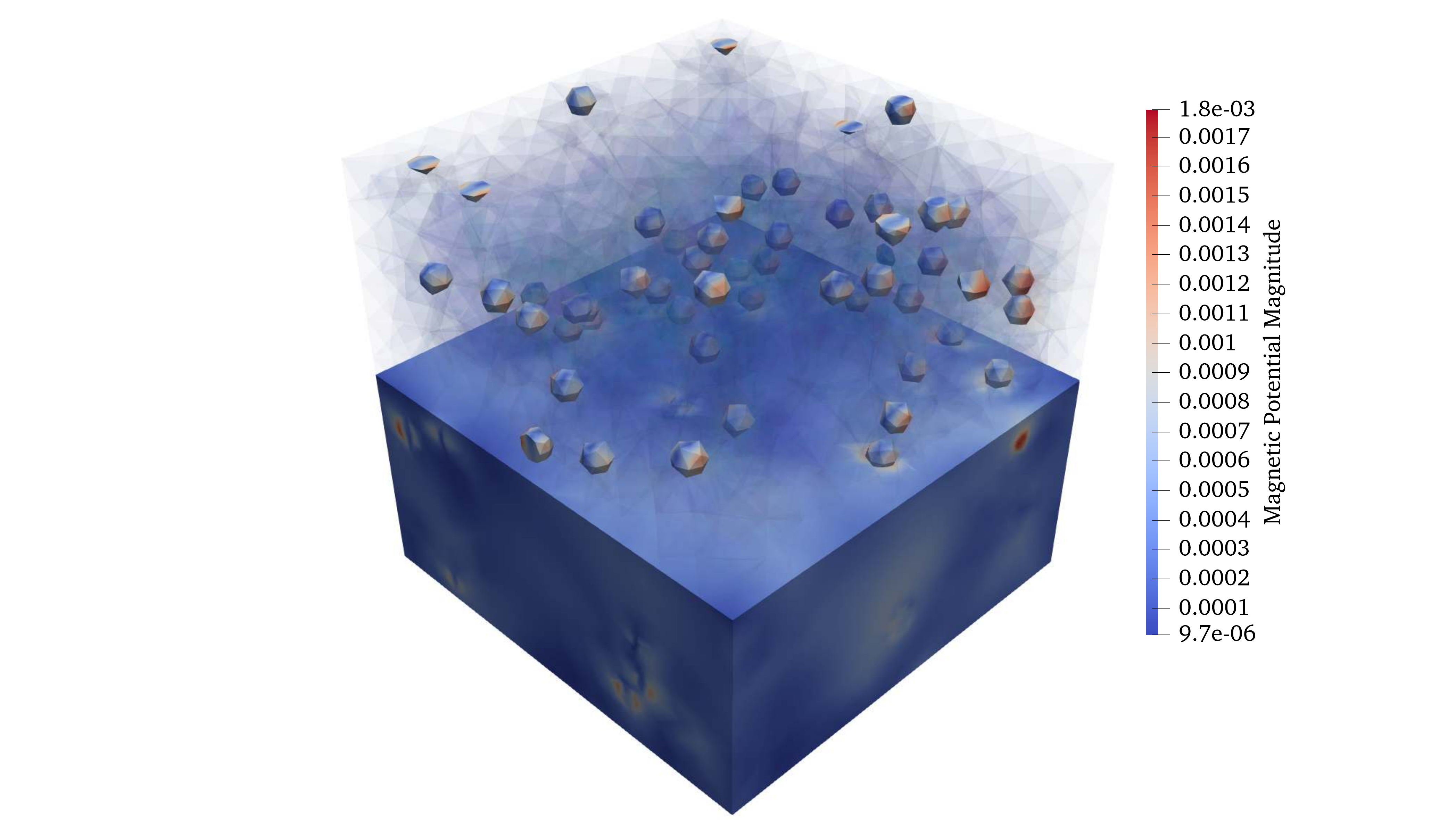}
        \caption{Magnetic potential}
        \label{fig:particle_both_magpot_final}
    \end{subfigure}\hfill
    \begin{subfigure}[b]{0.32\linewidth}
        \centering
        \includegraphics[width=\linewidth]{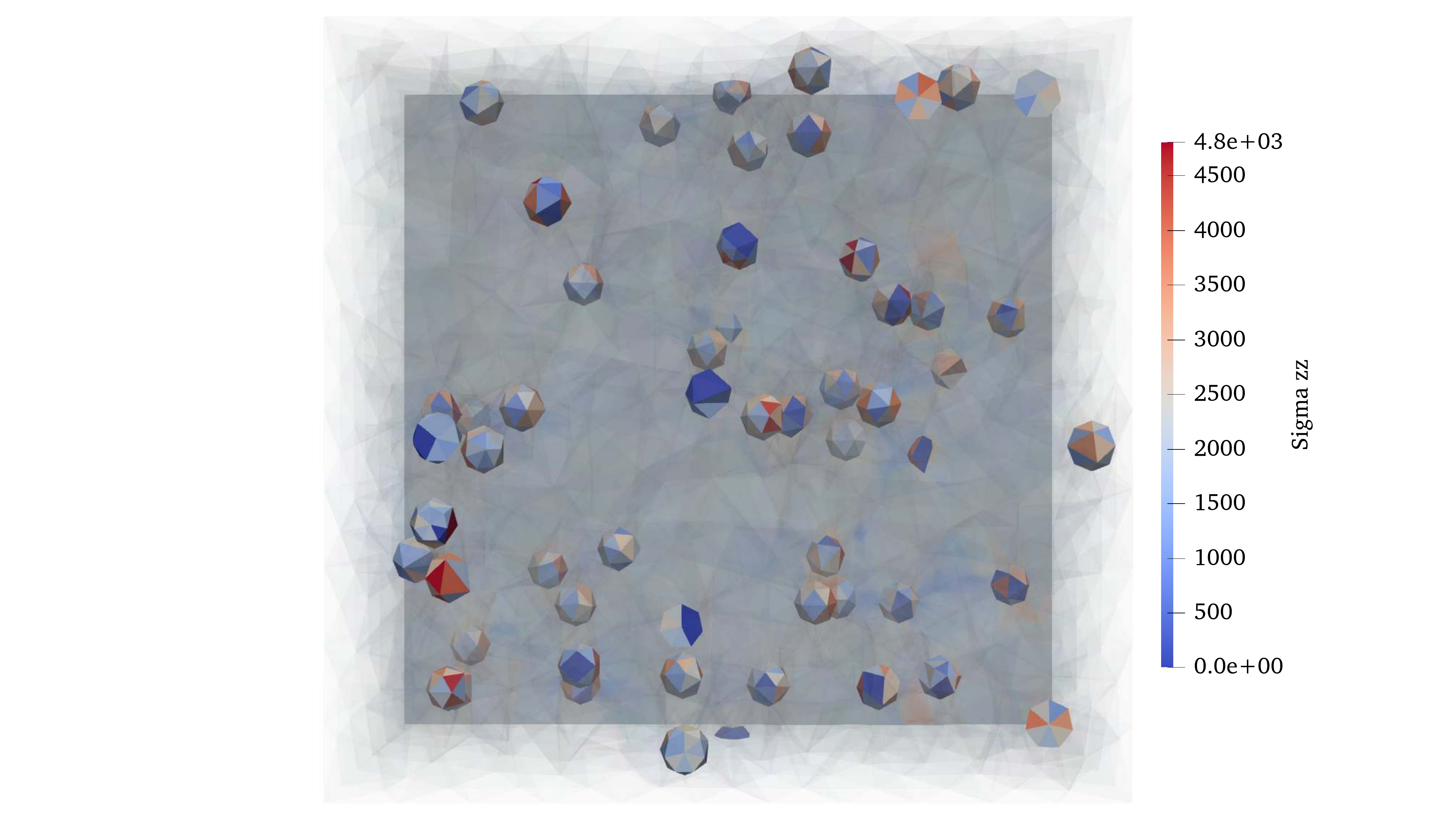}
        \caption{Final $\sigma_{zz}$}
        \label{fig:particle_both_sigmazz_final}
    \end{subfigure}

    \caption{Random-particle RVE (tet mesh): displacement magnitude, magnetic potential, and final $\sigma_{zz}$ under the prescribed magneto-mechanical loading.}
    \label{fig:particle_both_figs}
\end{figure}

As expected, the random-particle RVE exhibits a macroscopic response similar to that observed in the previous examples. However, as shown in Figure~\ref{fig:particle_both_figs}, the microscopic response is considerably more heterogeneous. In particular, the random particle distribution produces higher local stress concentrations together with more spatially variable displacement and magnetic potential fields.

\subsection{Comparison of effective magnetostriction across particle arrangements}
\label{sec:magnetostriction_compare}

In magneto-active polymers, magnetically induced deformation is commonly referred to as the magnetostrictive effect. In the present computational homogenization setting, the appropriate measure of this response is the homogenized deformation obtained under purely magnetic loading and macroscopically stress-relaxed periodic boundary conditions. That is, the RVE is allowed to deform in response to the applied magnetic induction while the macroscopic deformation is adjusted such that the targeted homogenized stress components remain approximately zero, rather than being prescribed a priori.

To quantify this effect, we define the effective magnetostriction in the Cartesian directions as
\begin{equation}
\lambda_{ii}^{\mathrm{mag}} := \bar{F}_{ii} - 1,
\qquad i \in \{x,y,z\},
\label{eq:lambda_diag_mag}
\end{equation}
where $\bar{F}_{ii}$ denotes the corresponding homogenized deformation component. More generally, the magnetostriction along a unit direction $\mathbf{N}$ may be written as
\begin{equation}
\lambda^{\mathrm{mag}}(\mathbf{N}) = \sqrt{\mathbf{N}\cdot\bar{\mathbf{C}}\mathbf{N}} - 1,
\qquad
\bar{\mathbf{C}} = \bar{\mathbf{F}}^T \bar{\mathbf{F}},
\label{eq:lambda_dir_mag}
\end{equation}
but in the present study, the primary quantities of interest are the Cartesian components $\lambda_{xx}^{\mathrm{mag}}$, $\lambda_{yy}^{\mathrm{mag}}$, and $\lambda_{zz}^{\mathrm{mag}}$.

To relate the observed field-induced deformation to the constitutive structure of the free energy, we consider a small-strain, low-field expansion of the same Helmholtz free energy used in the full formulation,
\begin{equation}
\psi(F,B)
=
\psi_{\mathrm{mech}}(F,\bar{J}^{(e)})
+
\psi_{\mathrm{vac}}(F,B)
+
\psi_{\mathrm{mag}}(F,B).
\label{eq:psi_total_magstr}
\end{equation}
Let
\begin{equation}
F = I + H,
\qquad
\varepsilon = \mathrm{sym}\,H,
\qquad
\|H\| \ll 1.
\label{eq:small_strain_ansatz_magstr}
\end{equation}
Then, to first order in strain,
\begin{equation}
J = \det F = 1 + \mathrm{tr}\,\varepsilon,
\qquad
J^{-1} = 1 - \mathrm{tr}\,\varepsilon,
\label{eq:J_small_strain_magstr}
\end{equation}
and
\begin{equation}
(FB)\cdot(FB)
=
B\cdot B + 2\,\varepsilon:(B\otimes B).
\label{eq:FB_expand_magstr}
\end{equation}

The mechanical contribution reduces to the usual small-strain elastic form,
\begin{equation}
\psi_{\mathrm{mech}}
=
\frac{1}{2}\,C_{ijkl}\,\varepsilon_{ij}\varepsilon_{kl},
\label{eq:psi_mech_small_magstr}
\end{equation}
where $C_{ijkl}$ is the linearized elasticity tensor. For the present Yeoh model, the nearly incompressible small-strain limit gives
\begin{equation}
E = 6 C_1,
\label{eq:E_from_C1_magstr}
\end{equation}
and the corresponding compliance tensor $S_{ijkl}$ satisfies
\begin{equation}
S_{ijpq} C_{pqkl} = \delta_{ik}\delta_{jl}.
\label{eq:compliance_def_magstr}
\end{equation}

We next expand the two magnetic contributions. The vacuum term is
\begin{equation}
\psi_{\mathrm{vac}}(F,B)
=
\frac{1}{2\mu_0 J}(FB)\cdot(FB).
\label{eq:psi_vac_magstr}
\end{equation}
Using \eqref{eq:J_small_strain_magstr}--\eqref{eq:FB_expand_magstr}, this becomes
\begin{equation}
\psi_{\mathrm{vac}}
=
\frac{1}{2\mu_0} B^2
+
\frac{1}{\mu_0}\,\varepsilon:(B\otimes B)
-
\frac{1}{2\mu_0}(\mathrm{tr}\,\varepsilon) B^2.
\label{eq:psi_vac_expand_magstr}
\end{equation}

For the particle magnetization term,
\begin{equation}
\psi_{\mathrm{mag}}(F,B)
=
\eta\,J\,w_{\mathrm{mag}}(b),
\qquad
b = \|J^{-1}FB\|,
\label{eq:psi_particle_magstr}
\end{equation}
we use the low-field expansion of the Langevin law,
\begin{equation}
L(x) = \frac{x}{3},
\qquad
\chi_L := \frac{m_s^{\mathrm{leg}}\alpha^{\mathrm{leg}}}{3},
\qquad
\mu_{\mathrm{eff}} := \mu_0(1+\chi_L),
\label{eq:chi_mu_eff_magstr}
\end{equation}
so that, to quadratic order in $b$,
\begin{equation}
w_{\mathrm{mag}}(b)
=
\frac{b^2}{2\mu_{\mathrm{eff}}}.
\label{eq:wmag_lowfield_magstr}
\end{equation}
Hence,
\begin{equation}
\psi_{\mathrm{mag}}
=
\frac{\eta}{2\mu_{\mathrm{eff}}J}(FB)\cdot(FB)
=
\frac{\eta}{2\mu_{\mathrm{eff}}} B^2
+
\frac{\eta}{\mu_{\mathrm{eff}}}\,\varepsilon:(B\otimes B)
-
\frac{\eta}{2\mu_{\mathrm{eff}}}(\mathrm{tr}\,\varepsilon) B^2.
\label{eq:psi_mag_expand_magstr}
\end{equation}

Combining \eqref{eq:psi_mech_small_magstr}, \eqref{eq:psi_vac_expand_magstr}, and \eqref{eq:psi_mag_expand_magstr}, the total small-strain free energy may be written as
\begin{equation}
\psi(\varepsilon,B)
=
\frac{1}{2} C_{ijkl}\varepsilon_{ij}\varepsilon_{kl}
+
\frac{1}{2}\nu^{-1}_{mn}B_mB_n
-
\Gamma_{ijmn}\,\varepsilon_{ij}B_mB_n,
\label{eq:gamma_form_magstr}
\end{equation}
where the magnetoelastic coupling tensor is
\begin{equation}
\Gamma_{ijmn}
=
-\frac{1}{2}
\left(
\frac{1}{\mu_0}
+
\frac{\eta}{\mu_{\mathrm{eff}}}
\right)
\left(
\delta_{im}\delta_{jn}
+
\delta_{in}\delta_{jm}
-
\delta_{ij}\delta_{mn}
\right).
\label{eq:gamma_tensor_magstr}
\end{equation}
The corresponding small-strain stress is therefore
\begin{equation}
\sigma_{ij}
=
\frac{\partial \psi}{\partial \varepsilon_{ij}}
=
C_{ijkl}\varepsilon_{kl}
-
\Gamma_{ijmn}B_mB_n.
\label{eq:sigma_gamma_magstr}
\end{equation}
Under the stress-relaxed conditions considered here, the targeted homogenized stress is approximately zero, so the induced strain is
\begin{equation}
\varepsilon_{ij}
=
S_{ijkl}\Gamma_{klmn}B_mB_n
=:
\Lambda_{ijmn} B_m B_n,
\label{eq:Lambda_def_magstr}
\end{equation}
with
\begin{equation}
\Lambda_{ijmn}
=
S_{ijkl}\Gamma_{klmn}.
\label{eq:Lambda_tensor_magstr}
\end{equation}

For a field applied in the $z$-direction, $B = B\,e_z$, one obtains
\begin{equation}
\varepsilon_{11} = \Lambda_{1133} B^2,
\qquad
\varepsilon_{22} = \Lambda_{2233} B^2,
\qquad
\varepsilon_{33} = \Lambda_{3333} B^2.
\label{eq:eps_components_Bz_magstr}
\end{equation}
In the isotropic nearly incompressible limit, using $E = 6C_1$, these components reduce to
\begin{equation}
\Lambda_{3333}
=
-\frac{1}{6C_1}
\left(
\frac{1}{\mu_0}
+
\frac{\eta}{\mu_{\mathrm{eff}}}
\right),
\qquad
\Lambda_{1133}
=
\Lambda_{2233}
=
\frac{1}{12C_1}
\left(
\frac{1}{\mu_0}
+
\frac{\eta}{\mu_{\mathrm{eff}}}
\right).
\label{eq:Lambda_components_magstr}
\end{equation}
Using the particle-phase material parameters from Table~\ref{tab:sef_parameters},
\begin{equation}
\begin{aligned}
C_1 &= 1.0\times 10^7\,\mathrm{Pa},
\qquad
m_s^{\mathrm{leg}} = 8.41\times 10^5\,\mathrm{A\,m^{-1}}, \\
\alpha^{\mathrm{leg}} &= 2.18\times 10^{-5}\,\mathrm{m\,A^{-1}},
\qquad
\mu_0 = 1.2564\times 10^{-6}\,\mathrm{H\,m^{-1}}.
\end{aligned}
\label{eq:particle_params_magstr}
\end{equation}
gives
\begin{equation}
\chi_L = 6.11,
\qquad
\mu_{\mathrm{eff}} = 8.93\times 10^{-6}\,\mathrm{H\,m^{-1}}.
\label{eq:chi_mu_num_magstr}
\end{equation}
Taking $\eta = 1$ in the particle phase then yields
\begin{equation}
\Lambda_{3333} = -1.51\times 10^{-2}\,\mathrm{T}^{-2},
\qquad
\Lambda_{1133} = \Lambda_{2233} = 7.56\times 10^{-3}\,\mathrm{T}^{-2}.
\label{eq:Lambda_num_magstr}
\end{equation}

\begin{figure}[htbp]
    \centering
    \includegraphics[width=0.95\linewidth]{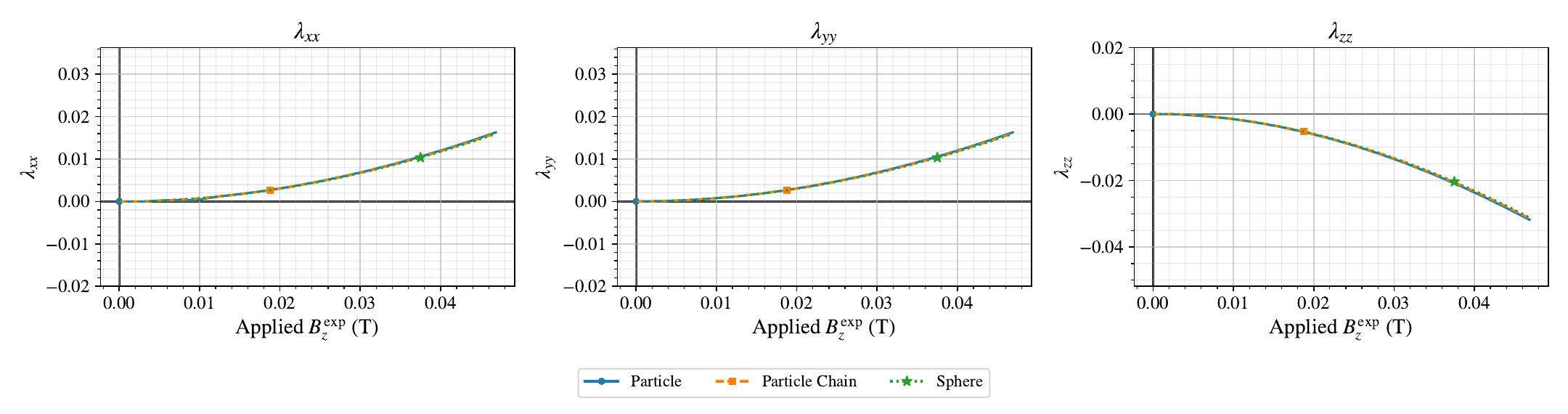}
    \caption{Effective magnetostrictive strain components $\lambda_{ii}^{\mathrm{mag}}=\bar{F}_{ii}-1$ for the sphere, spiral, and random-particle RVEs under purely magnetic loading.}
    \label{fig:magnetostriction_compare}
\end{figure}

Thus, the small-strain particle model predicts a dominant field-aligned contraction together with transverse expansion, with the axial component having twice the magnitude of each transverse component. This sign pattern is consistent with the present homogenized results, where the applied magnetic loading produces $\lambda_{zz}^{\mathrm{mag}} < 0$ and $\lambda_{xx}^{\mathrm{mag}},\lambda_{yy}^{\mathrm{mag}} > 0$. Figure~\ref{fig:magnetostriction_compare} shows the full evolution of the diagonal effective magnetostriction components during loading.

To facilitate comparison with the analytical coefficients in \eqref{eq:Lambda_num_magstr}, it is useful to define effective RVE-scale magnetostrictive coefficients by normalizing the final homogenized deformation by a common final applied magnetic induction. In the present comparison, we use the same final loading level for all three RVEs, namely
\begin{equation}
B_{z,\mathrm{final}}^{\mathrm{app}} = 0.18\times 0.25 = 0.045\ \mathrm{T}.
\label{eq:Bfinal_common}
\end{equation}
We therefore define
\begin{equation}
\Lambda_{ii33}^{\mathrm{RVE}}
=
\frac{\lambda_{ii}^{\mathrm{mag}}}{\left(B_{z,\mathrm{final}}^{\mathrm{app}}\right)^2},
\qquad i\in\{x,y,z\}.
\label{eq:Lambda_RVE_def_commonB}
\end{equation}
The resulting values are reported in Table~\ref{tab:magnetostriction_lambda_compare_commonB}. These effective coefficients preserve the same sign structure as the analytical estimate, with positive transverse components and a negative axial component, and provide a consistent RVE-scale measure of the magnetostrictive response across all three microstructures.

The difference in magnitude between the particle magnetostriction and the effective RVE-scale values should be interpreted in light of the underlying assumptions. The particle coefficients are derived from a homogeneous constitutive model in which the entire material is treated as magnetically active. By contrast, the numerical RVEs are two-phase composites in which only the particle phase possess magnetostrictive coupling, while the surrounding matrix is mechanically compliant but non-magnetostrictive. The homogenized response therefore reflects not only the intrinsic particle-level magnetoelastic coupling, but also phase contrast, particle--matrix interaction, and the geometric constraints imposed by the heterogeneous microstructure. Moreover, the normalized RVE-scale magnetostriction is much larger because the observed deformation is not simply the constitutive response of a homogeneous magnetic medium, but a particle-driven composite rearrangement in which local inclusion motion and interaction are transmitted through the surrounding matrix and then amplified at the structural level of the RVE. In fact, this rearrangement dominates the magnetostriction of the RVE in that its magnitude is on the order $10^{4}$ times that of the inclusions themselves.

Across all three microstructures, the response is qualitatively similar, with a dominant field-aligned contraction and compensating transverse expansion. Among the three cases, the random-particle arrangement exhibits the largest magnitude response, followed closely by the spiral configuration, whereas the spherical-inclusion case yields the smallest overall deformation. This ordering suggests that increased microscale freedom for rearrangement and interaction enhances the net magnetostrictive deformation captured at the homogenized level, while remaining consistent with the constitutive sign structure predicted by the small-strain analytical expansion.

\begin{table}[htbp]
\centering
\caption{Effective RVE-scale magnetostrictive coefficients obtained by dividing the final magnetostrictive strains by the square of the final applied magnetic induction $B_{z,\mathrm{final}}^{\mathrm{app}}=0.045\,\mathrm{T}$.}
\label{tab:magnetostriction_lambda_compare_commonB}
\begin{tabular}{l c c c}
\hline
Arrangement & $\Lambda_{1133}^{\mathrm{RVE}}\,[\mathrm{T}^{-2}]$ & $\Lambda_{2233}^{\mathrm{RVE}}\,[\mathrm{T}^{-2}]$ & $\Lambda_{3333}^{\mathrm{RVE}}\,[\mathrm{T}^{-2}]$ \\
\hline
Spherical Inclusion & $7.8706$ & $7.8706$ & $-15.3723$ \\
Spiral              & $7.9343$ & $7.9348$ & $-15.4943$ \\
Random Particles    & $8.0405$ & $8.0469$ & $-15.7121$ \\
\hline
\end{tabular}
\end{table}

\subsection{Incompressibility Study}
\label{sec:incomp_study}

To further investigate the role of incompressibility, a parameter study was performed in which the Poisson’s ratio of the magnetic inclusion was varied. Two loading cases were examined: a purely magnetic loading case, intended to illustrate differences in particle motion, and a uniaxial-strain case under the combined influence of mechanical loading and a magnetic field. Both cases were analyzed using a single spherical inclusion. The volume fraction, however, differed between the two studies: the purely magnetic case employed the same volume fraction as the example in Section~\ref{sec:simple_inclusion}, while the uniaxial-strain case used a volume fraction of 37\%.

For the compressible material, the following Neo-Hookean strain-energy function, shown in Eq.~\ref{eq:neohookean_formulation}, was adopted:
\begin{equation}
\label{eq:neohookean_formulation}
\psi_{\mathrm{mech}}^{\mathrm{comp}}
=
\frac{\mu}{2}\left(\bar{I}_1 - 3\right)
+
\frac{\kappa}{2}\left(\ln J\right)^2.
\end{equation}

Additionally, to permit compressible material behavior, the volumetric contribution of the compressible phase was excluded from the $\bar{J}$ averaging procedure and treated separately.

\subsubsection{Purely Magnetic Loading}

For this study, the Poisson's ratio of the inclusion was varied over the values \(\nu = 0.40\), \(0.45\), \(0.48\), \(0.49\), \(0.495\), and \(0.50\), while the Young's modulus was held fixed at \(E = 5.6\times10^7\). In this way, the effect of incompressibility was isolated by varying only \(\nu\). For each case, the corresponding shear and bulk moduli were computed from the standard isotropic relations. The resulting material parameters are summarized in Table~\ref{tab:incomp_study_mag_loading}.

\begin{table*}[htbp]
\centering
\caption{Material parameters used in the purely magnetic incompressibility study, with Young's modulus held constant at \(E = 5.6\times10^7\).}
\label{tab:incomp_study_mag_loading}
\begin{tabular*}{\textwidth}{@{\extracolsep{\fill}} c c c c @{}}
\hline
\(\nu\) & \(E\) & \(\mu\) & \(\kappa\) \\
\hline
0.40  & \(5.6\times10^7\) & \(2.000\times10^7\) & \(9.333\times10^7\) \\
0.45  & \(5.6\times10^7\) & \(1.931\times10^7\) & \(1.867\times10^8\) \\
0.48  & \(5.6\times10^7\) & \(1.892\times10^7\) & \(4.667\times10^8\) \\
0.49  & \(5.6\times10^7\) & \(1.879\times10^7\) & \(9.333\times10^8\) \\
0.495 & \(5.6\times10^7\) & \(1.873\times10^7\) & \(1.867\times10^9\) \\
0.50  & \(5.6\times10^7\) & \(1.867\times10^7\) & \(\infty\) \\
\hline
\end{tabular*}
\end{table*}

\begin{figure}[htbp]
    \centering
    \includegraphics[width=0.95\linewidth]{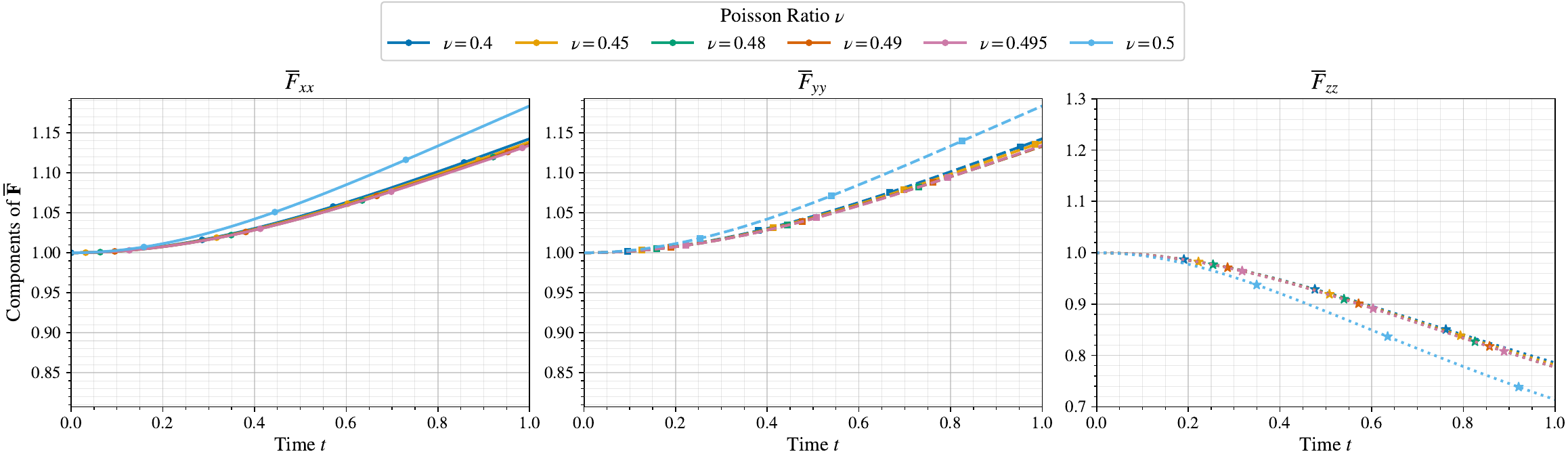}
    \caption{Homogenized deformation response for each value of \(\nu\).}
    \label{fig:incomp_def_plot}
\end{figure}

The corresponding homogenized deformation responses are shown in Figure~\ref{fig:incomp_def_plot}. The results demonstrate that the inclusion compressibility influences the overall deformation response, with increasing \(\nu\) leading to a progressively more constrained macroscopic deformation. In particular, the diagonal components show the expected suppression of volumetric change as \(\nu\) increases. Overall, these results indicate that even under purely magnetic loading, the mechanical compressibility of the particle phase significantly affects the effective deformation response.

\begin{figure}[htbp]
    \centering
    \includegraphics[width=0.5\linewidth]{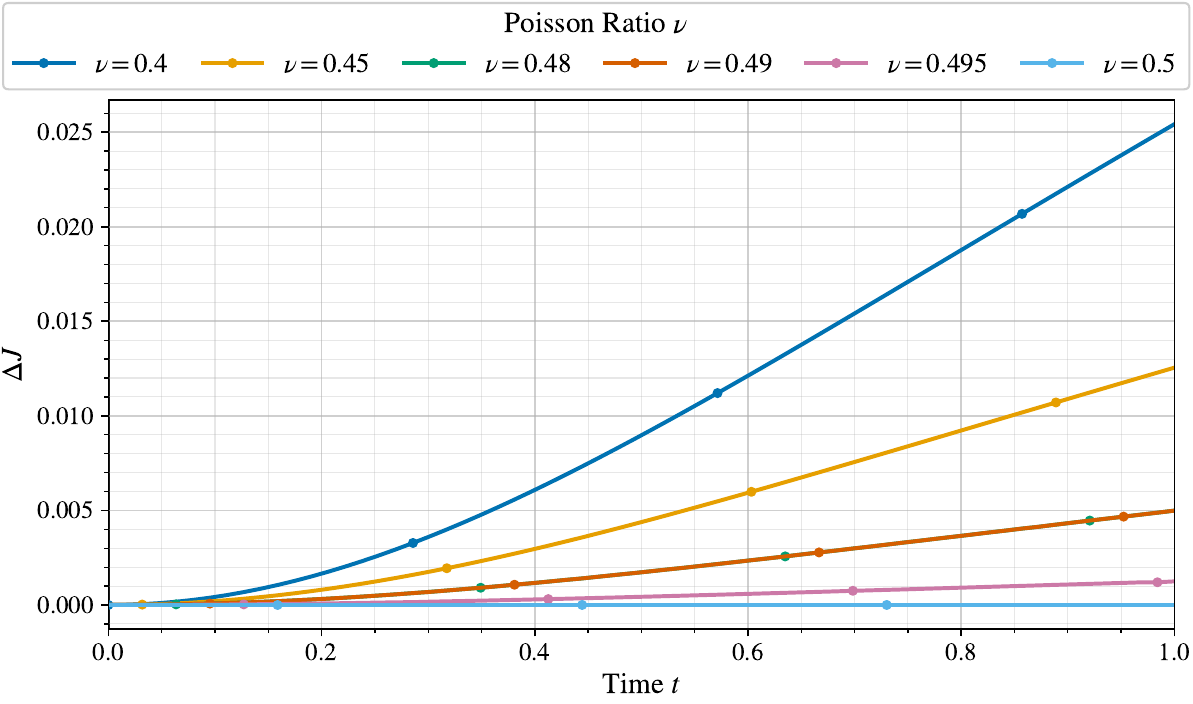}
    \caption{Deviation of the homogenized Jacobian from unity for each value of \(\nu\).}
    \label{fig:incomp_J_deviation}
\end{figure}

Figure~\ref{fig:incomp_J_deviation} shows the corresponding deviation of the homogenized Jacobian from unity. As expected, lower values of \(\nu\) result in larger volumetric changes, while the nearly incompressible and incompressible cases remain close to \(J = 1\). This confirms that volumetric deformation is increasingly suppressed as \(\nu\) approaches the incompressible limit.

\begin{figure}[htbp]
    \centering
    \includegraphics[width=0.95\linewidth]{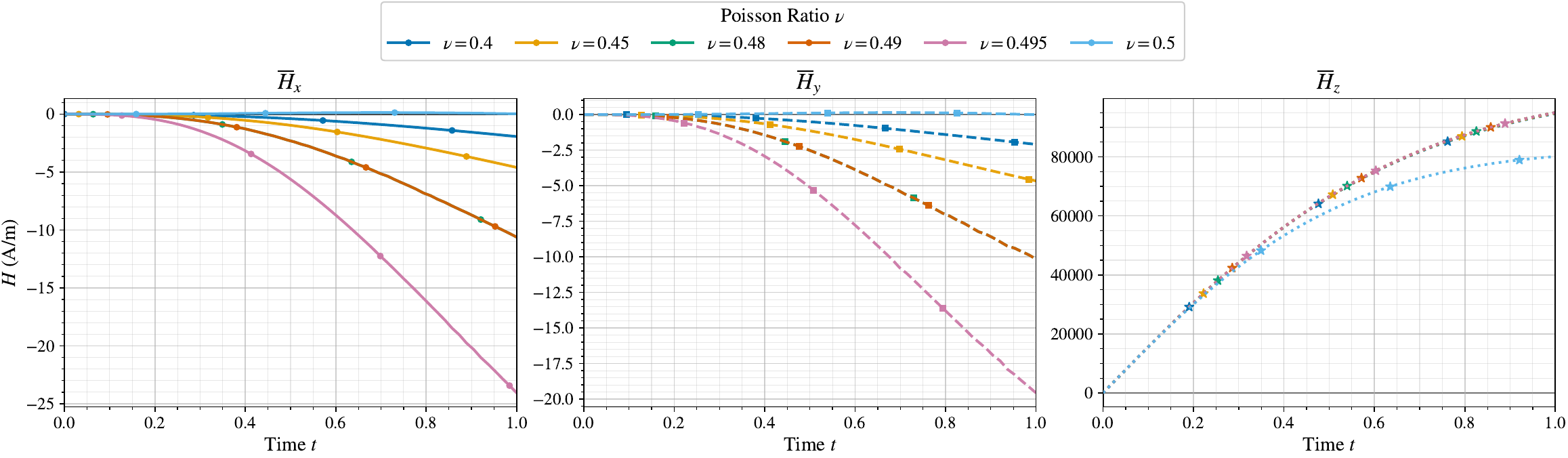}
    \caption{Homogenized magnetic field response for each value of \(\nu\).}
    \label{fig:incomp_H_plots}
\end{figure}

Figure~\ref{fig:incomp_H_plots} shows the homogenized magnetic field response \(\overline{\mathbf{H}}\) for each value of \(\nu\). Variations in \(\nu\) lead to noticeable differences in the magnetic response, indicating that compressibility influences not only the mechanical deformation but also the coupled magnetic behavior. This highlights the role of magneto-mechanical coupling, where the mechanical constraint imposed by the inclusion phase directly impacts the effective magnetic response.

Overall, these results demonstrate that inclusion compressibility has a non-negligible effect on both the mechanical and magnetic homogenized responses, even in a loading case driven purely by the applied magnetic induction.

\subsubsection{Uniaxial Strain with Magnetic Field}

As in the previous study, the Young’s modulus was held constant while the Poisson’s ratio $\nu$ was varied. The primary objective of this case was to examine the particle stress response under combined mechanical and magnetic loading. In contrast to the purely magnetic study, the particle--matrix stiffness ratio was reduced to 50, and the particle volume fraction was increased in order to amplify the influence of the inclusion response on the homogenized behavior. The values of $\nu$ considered in this case were $0.20$, $0.45$, and the incompressible limit. The corresponding material parameters are summarized in Table~\ref{tab:incomp_study_strain_mag}.

\begin{table}[htbp]
\centering
\caption{Material parameters used in the uniaxial-strain study with magnetic loading, with Young's modulus held constant at $E = 3.7785\times10^6$.}
\label{tab:incomp_study_strain_mag}
\begin{tabular}{c c c c}
\hline
$\nu$ & $E$ & $\mu$ & $\kappa$ \\
\hline
0.20 & $3.779\times10^6$ & $1.574\times10^6$ & $2.099\times10^6$ \\
0.45 & $3.779\times10^6$ & $1.302\times10^6$ & $1.255\times10^7$ \\
0.50 & $3.779\times10^6$ & $1.259\times10^6$ & incompressible limit \\
\hline
\end{tabular}
\end{table}

In this case, a uniaxial stretch of 15\% was prescribed in the $x$-direction, while the stretches in the $y$- and $z$-directions were adjusted such that the overall deformation remained isochoric. In addition, a magnetic induction of $250\,\mathrm{mT}$ was applied in the $z$-direction. Consistent with the preceding studies, both loads were applied incrementally over time using a linear ramp. This loading scenario was chosen to assess the influence of inclusion compressibility on the particle stress response under coupled magneto-mechanical loading.

\begin{figure}[htbp]
    \centering
    \includegraphics[width=0.8\linewidth]{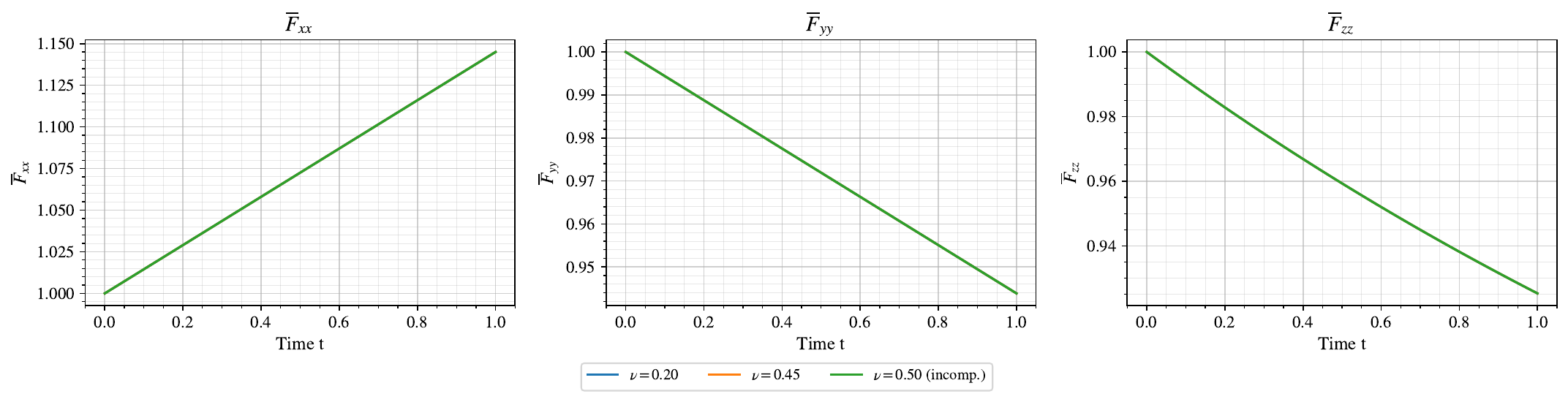}
    \caption{Homogenized deformation response for the different values of the inclusion Poisson's ratio under uniaxial strain with an applied magnetic field.}
    \label{fig:incomp_strain_deformation}
\end{figure}

\begin{figure}[htbp]
    \centering
    \includegraphics[width=0.8\linewidth]{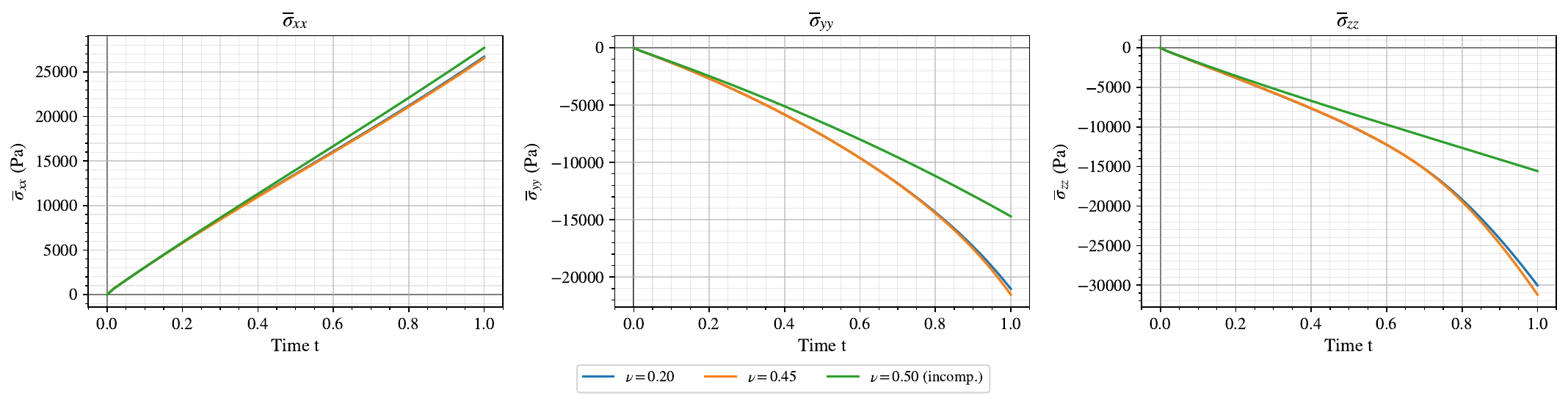}
    \caption{Homogenized stress response for the different values of the inclusion Poisson's ratio under uniaxial strain with an applied magnetic field.}
    \label{fig:incomp_stress_response}
\end{figure}

\begin{figure}[htbp]
    \centering
    \includegraphics[width=0.5\linewidth]{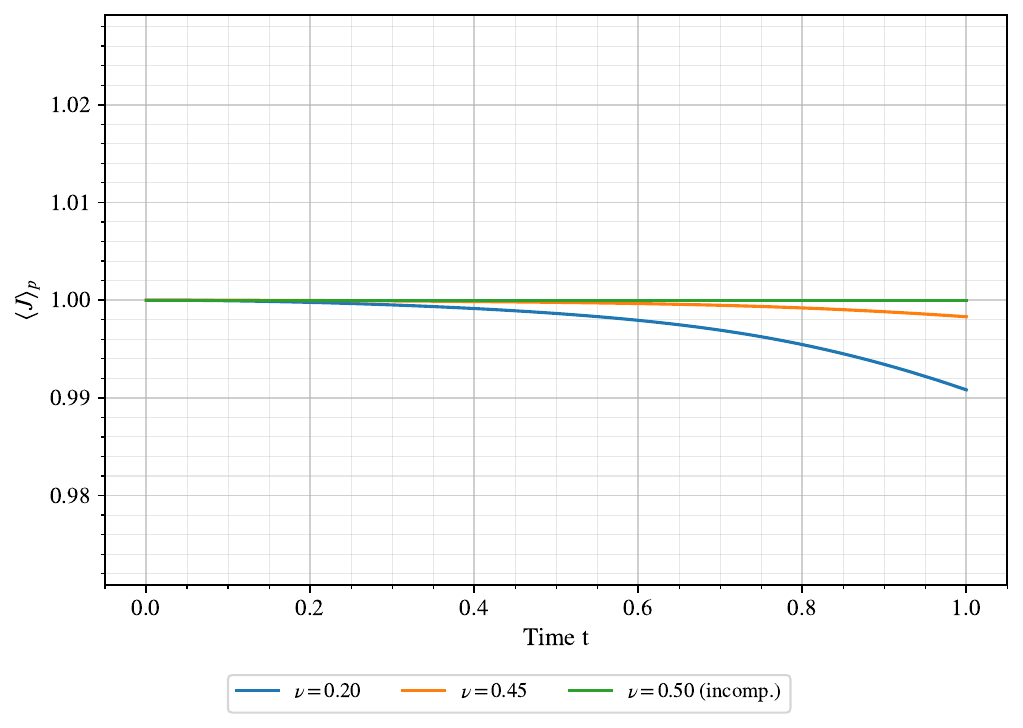}
    \caption{Evolution of the particle Jacobian for the different values of the inclusion Poisson's ratio under uniaxial strain with an applied magnetic field.}
    \label{fig:incomp_particle_J}
\end{figure}

Figure~\ref{fig:incomp_strain_deformation} shows the resulting homogenized deformation response for each case, while Figure~\ref{fig:incomp_stress_response} presents the corresponding homogenized stress response. Additionally, Figure~\ref{fig:incomp_particle_J} illustrates the evolution of the particle Jacobian for each case, highlighting differences in the volumetric response as the inclusion compressibility is varied.

From these results, the difference in particle behavior is apparent, with the compressible case producing a higher stress under this deformation. This is consistent with the fact that the applied deformation is isochoric, so the volumetric contribution to the strain energy does not contribute, and the stress response is governed primarily by the shear modulus. Since Young’s modulus is held fixed while Poisson’s ratio is increased toward incompressibility, the bulk modulus increases but the shear modulus decreases, leading the more incompressible case to exhibit a lower stress response.

\begin{figure}[htbp]
    \centering
    \begin{subfigure}[b]{0.32\linewidth}
        \centering
        \includegraphics[width=\linewidth]{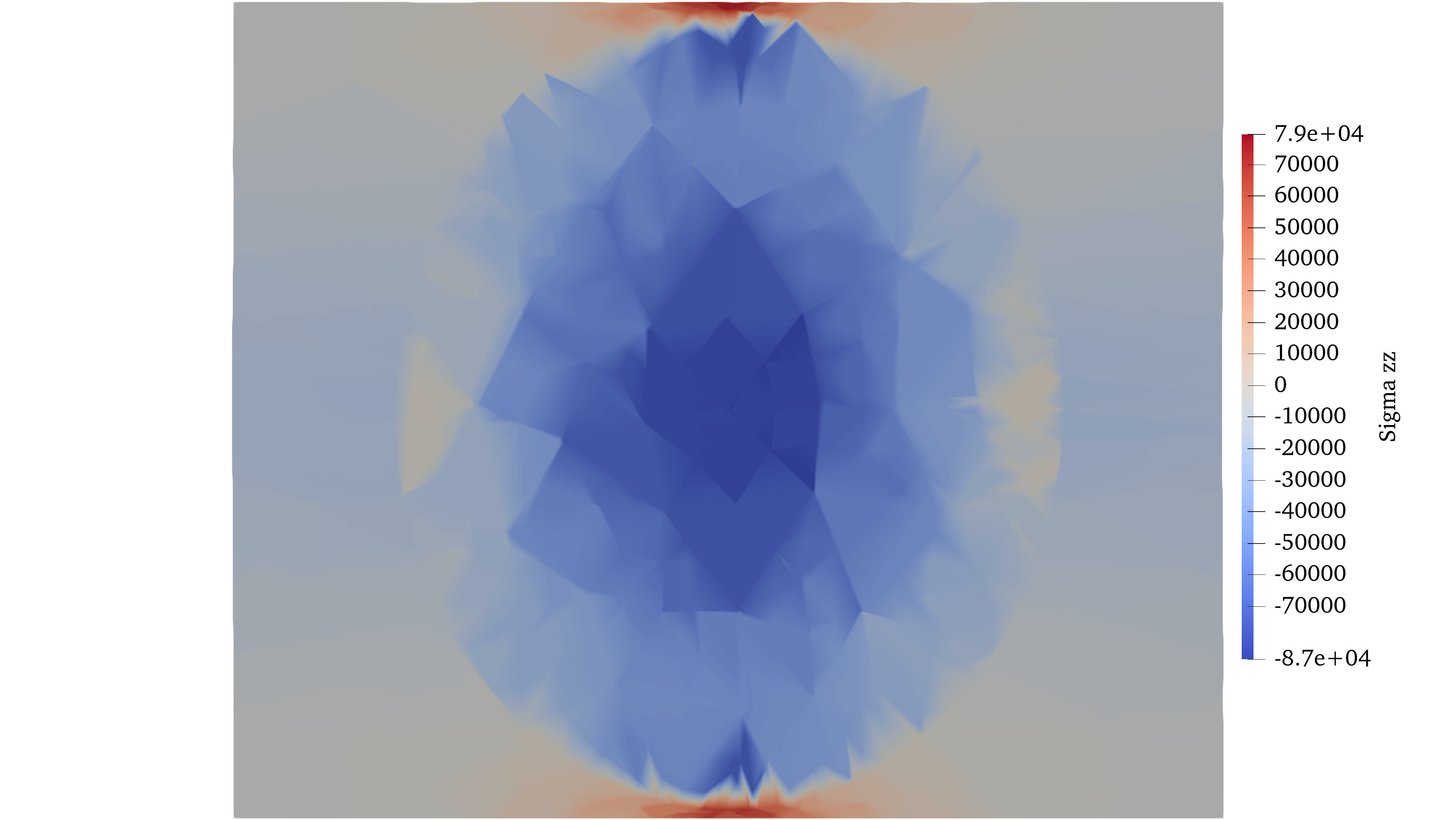}
        \caption{$\nu = 0.20$}
        \label{fig:incomp_local_dif_02}
    \end{subfigure}\hfill
    \begin{subfigure}[b]{0.32\linewidth}
        \centering
        \includegraphics[width=\linewidth]{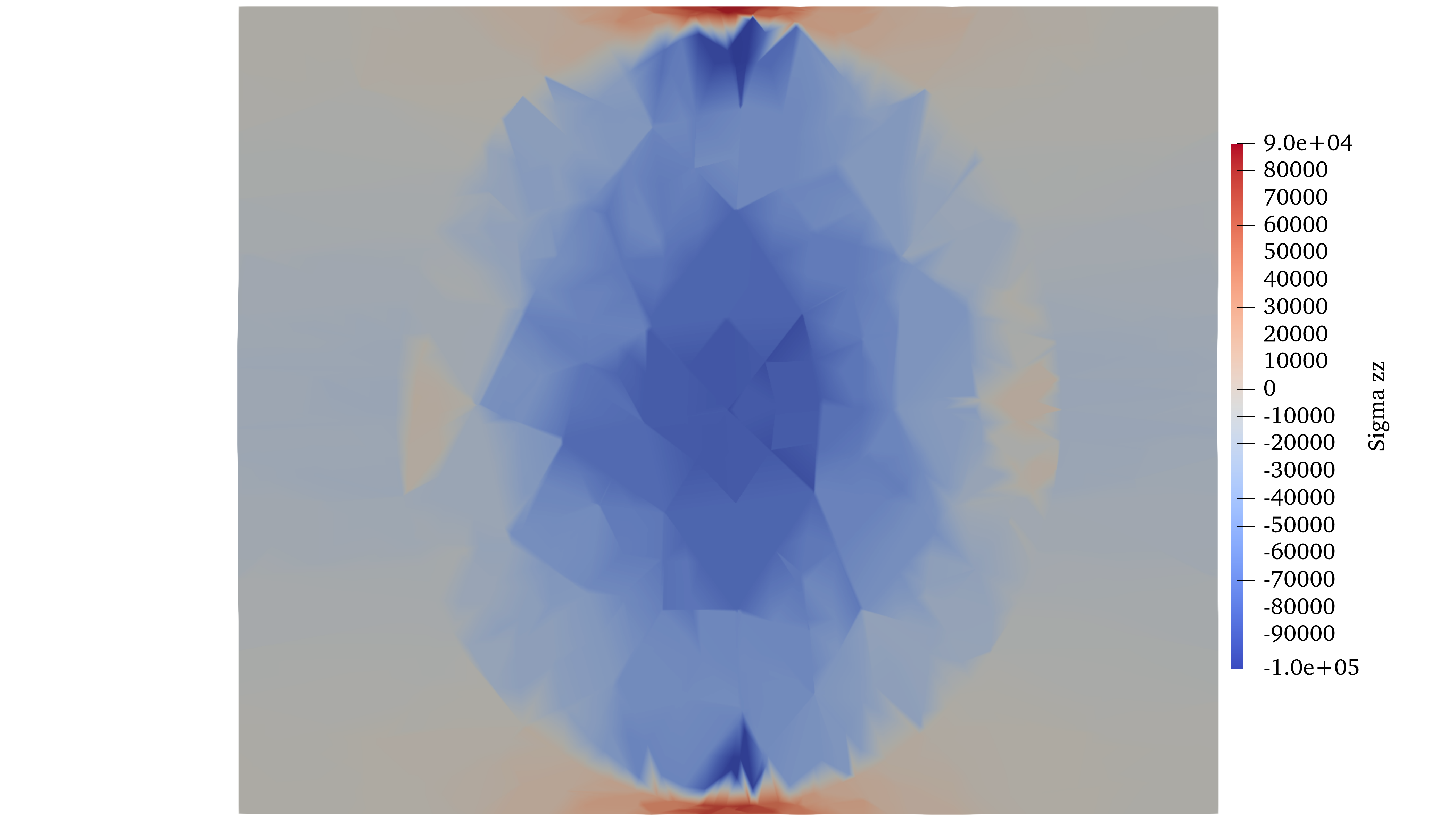}
        \caption{$\nu = 0.45$}
        \label{fig:incomp_local_dif_045}
    \end{subfigure}\hfill
    \begin{subfigure}[b]{0.32\linewidth}
        \centering
        \includegraphics[width=\linewidth]{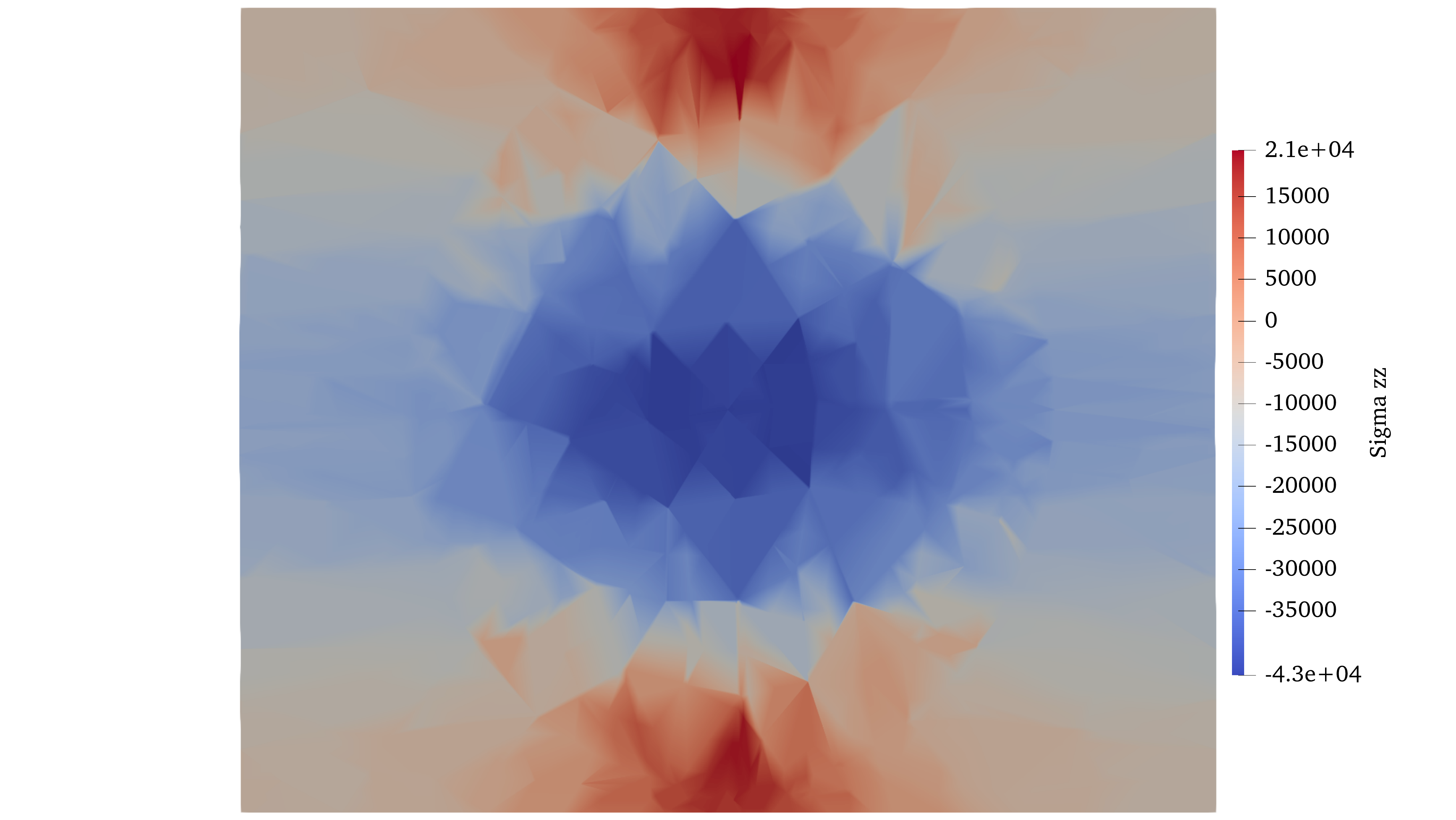}
        \caption{Incompressible}
        \label{fig:incomp_local_dif_inc}
    \end{subfigure}
    \caption{Comparison of the local stress field $\sigma_{zz}$ for the three inclusion compressibility cases under uniaxial strain with an applied magnetic field.}
    \label{fig:incomp_local_dif}
\end{figure}

Additionally, Figure~\ref{fig:incomp_local_dif} compares the differences in the local stress field $\sigma_{zz}$ among the three cases.

\section{Conclusion}
\label{sec:conclusion}

In this work, we developed a three-dimensional computational homogenization framework for magnetically active elastomers that enables fully coupled magneto-mechanical simulations at the RVE level. The formulation combines a $J$-bar-based treatment of near-incompressibility, periodic boundary conditions for consistent multiscale modeling, and a magnetic vector-potential description of the magnetic field.
The framework was used to analyze a range of microscale configurations exhibiting overall incompressible behavior with heterogeneous inclusions. Both tetrahedral and hexahedral discretizations were successfully employed, and consistent homogenized responses were obtained under periodic boundary conditions. The results highlight the ability of the formulation to capture nontrivial particle kinematics and local field variations while preserving macroscopic incompressibility.
Furthermore, the effect of inclusion compressibility was investigated, showing that variations in Poisson’s ratio influence both the local deformation patterns and the homogenized mechanical and magnetic responses. While these effects remain modest for low volume fractions and high stiffness contrasts, they become more pronounced as the relative stiffness and volumetric compliance of the inclusion phase change, particularly under coupled magneto--mechanical loading.
Despite the robustness of the proposed formulation, several limitations should be noted. First, the use of the $J$-bar stabilization introduces an element-wise averaged volumetric response, which may smear localized volumetric effects and limit accuracy in problems where strong local dilatation gradients are important. Second, the present study is restricted to quasi-static magnetostatics and does not account for dynamic electromagnetic effects, eddy currents, or rate-dependent material behavior, which may become relevant in high-frequency or transient applications. Third, while the tetrahedral discretization enables complex geometries, the use of an $H^1$-conforming approximation for the magnetic vector potential does not enforce strict $H(\mathrm{curl})$ conformity, which may impact accuracy for problems involving sharp magnetic field discontinuities.
In addition, the computational cost of fully coupled three-dimensional RVE simulations remains significant, particularly for high particle volume fractions or when fine mesh resolution is required to capture strong localization effects. Finally, the material models employed here are based on idealized hyperelastic and saturation-type magnetization laws; extensions to more complex constitutive behaviors, including hysteresis, viscoelasticity, and rate-dependent magnetization, remain an open area for future work.
In addition, several constitutive assumptions adopted in the present study remain to be validated experimentally. In particular, the use of idealized hyperelastic and saturation-type magnetic laws assumes a largely reversible, rate-independent response and does not yet resolve possible hysteretic, viscoelastic, or other dissipative effects in the composite. Moreover, while the simulations capture particle rotation, translation, and interaction-driven rearrangement at the microscale, these mechanisms have not yet been directly compared against experimental observations. Experimental validation through free-magnetostriction tests, constrained or blocked-force actuation tests, cyclic magnetic loading, and time-dependent stress-relaxation or creep experiments, ideally combined with in situ imaging of particle motion, would provide a direct basis for assessing and refining these constitutive assumptions \cite{schumann2017situ}.
Several directions for future work follow naturally from the present formulation. A key extension is the integration of data-driven and machine learning techniques to accelerate the computational homogenization process. In particular, surrogate models trained on RVE simulations could be used to approximate the homogenized constitutive response, enabling efficient multiscale simulations and opening the door to inverse design and parameter identification \cite{kalina2026physics, roth2025data}.
Another important direction is the extension of the framework to more general coupled multiphysics settings, including electro--magneto--mechanical behavior. Incorporating electric fields alongside magnetic and mechanical effects would allow the modeling of a broader class of smart materials, such as magneto-electro-active composites and multifunctional metamaterials \cite{pradhan2021applications, cai2025ferroelectricity}.
Finally, further developments could include the incorporation of more complex material behavior, such as hysteresis, rate dependence, and nonlinear magnetic saturation, as well as the exploration of higher inclusion volume fractions and more complex microstructural geometries. These extensions would enhance both the predictive capability and the practical applicability of the proposed multiscale framework \cite{kalina2017modeling}. Ultimately, these developments aim to bridge high-fidelity multiscale modeling and real-time predictive capabilities, enabling the design and control of advanced multifunctional materials.

\section*{Acknowledgements}
This material is based upon work partially supported by the U.S. National Science Foundation under award No. 2452029 and partially under award No. 2235856 (MKR).  
The opinions, findings, and conclusions, or recommendations expressed are those of the authors and do not necessarily reflect the views of the NSF.
\\
The authors acknowledge the Texas Advanced Computing Center (TACC) at The University of Texas at Austin for providing computational resources that have contributed to the research results reported within this paper. URL: \url{http://www.tacc.utexas.edu}

\bibliographystyle{unsrt}  
\bibliography{references}  






\end{document}